\documentclass[aps,prd,preprint,onecolumn,notitlepage,superscriptaddress,nofootinbib,longbibliography]{revtex4-2}
\usepackage[utf8]{inputenc}
\usepackage[T1]{fontenc}
\usepackage[english]{babel}
\usepackage{graphicx}
\usepackage[dvipsnames]{xcolor}
\usepackage{tikz}
\usetikzlibrary{arrows,scopes,shapes.misc}
\usepackage{amsmath,amsfonts,amssymb}
\usepackage{mathtools}
\usepackage{indentfirst}
\usepackage{hyperref}
\usepackage{subcaption}
\usepackage[export]{adjustbox}

\renewcommand{\theequation}{\arabic{section}.\arabic{equation}}
\renewcommand{\thesection}{\arabic{section}}

\catcode`@=11
\@addtoreset{equation}{section}
\catcode`@=12

\usepackage{hyperref}
\hypersetup{
	colorlinks,%
	citecolor=RoyalPurple,%
	filecolor=blue,%
	linkcolor=OliveGreen,%
	urlcolor=purple
}

\DeclareMathOperator{\Det}{Det}
\DeclareMathOperator{\Tr}{Tr}

\newcommand{\calD}{{\mathcal D}}

\begin{document}

\title{Real-time diagram technique for instantonic systems}	
\author{Nikita Kolganov}
\email{nikita.kolganov@phystech.edu}
\affiliation{Moscow Institute of Physics and Technology, 141700, Institutskiy pereulok, 9, Dolgoprudny, Russia}
\affiliation{Institute for Theoretical and Mathematical Physics, Moscow State University, 119991, Leninskie Gory, GSP-1, Moscow, Russia}
\affiliation{Institute for Theoretical and Experimental Physics, 117218, B. Cheremushkinskaya, 25, Moscow, Russia}
\begin{abstract}
	The Schwinger-Keldysh diagram technique is usually involved in the calculation of real-time in-in correlation functions. In the case of a thermal state, one can analytically continue imaginary-time Matsubara correlation functions to real times. Nevertheless, not all real-time correlation functions usually can be obtained by such procedure. Moreover, numerical analytic continuation is an ill-posed problem. Thus, even in the case of a thermal state one may need for the Schwinger-Keldysh formalism. If the potential of a system admits degenerate minima, instantonic effects enter the game, so one should also integrate over the instantonic moduli space, including the one, corresponding to the imaginary time translational invariance. However, the Schwinger-Keldysh closed time contour explicitly breaks such invariance. We argue, that this invariance must be recovered, and show, how it can be done. After that, we construct an extension of the Schwinger-Keldysh diagram technique to instantonic systems and demonstrate it on the example of the first few-point correlation functions.
\end{abstract}
\maketitle
\newpage
\section{Introduction}
	Quantum field theory (QFT) provides a framework with the greatest predictability power in a broad range of areas, including particle physics, condensed matter, and quantum cosmology. The main quantities, which can be calculated in QFT, are the correlation functions, that encode all the information about a particular physical system. These correlation functions can be further transformed into problem-specific observable quantities such as scattering amplitudes (via reduction formulas), response functions, etc.

	Correlation functions are usually calculated as a perturbative series using an appropriate diagrammatic technique, which significantly depends on a problem under consideration, namely the type of correlations functions and the state the system is in. For example, in the context of the scattering problem, one is usually interested in vacuum time-ordered in-out correlation functions, in studying non-equilibrium phenomena, one deals with in-in correlation functions, which is a product of time and anti-time ordered Heisenberg operators, averaged over some density matrix. At the same time, the investigation of more subtle phenomena, such as quantum chaos, may need out-of-time-ordered correlation functions (OTOCs), where operators may have an arbitrary time ordering. The calculation procedure of in-out correlation functions leads to the Feynman diagram technique, having the time-ordered propagator as a main building block, whereas the calculation of in-in correlation functions needs more complicated techniques such as the Schwinger-Keldysh diagram technique \cite{Keldysh:1964ud,schwinger1961brownian,Arseev:2015}, involving many-component propagators (and vertices), making the calculations rather cumbersome.
	
	A significant simplification may occur in the calculation of real-time in-in correlation functions if the state is a thermal state defined by the time-independent Hamiltonian, while the time evolution is governed by the same Hamiltonian. Namely, imaginary-time correlation functions, calculated with the use of the Matsubara diagram technique \cite{matsubara1955new,abrikosov2012methods}, can be analytically continued to the real times \cite{baym1961determination,Evans:1991ky,Kobes:1990kr,Baier:1993yh}. The latter technique is much simpler than the Schwinger-Keldysh one, so in the case of a thermal state, many-component calculations seem to be avoided. However, there are at least two difficulties, that may arise in this scheme. The first problem is that an analytic continuation of Matsubara correlation functions leads to a very particular type of the Schwinger-Keldysh correlation functions \cite{Evans:1991ky}. Despite the fact that some other can be reconstructed using the generalized fluctuation-dissipation relations \cite{Wang:1998wg}, and some more sophisticated techniques \cite{kugler2021multipoint}, one typically cannot obtain all Schwinger-Keldysh correlation functions by an analytic continuation, except some simple situations like the two-point case. One can encounter another difficulty in an attempt to calculate correlation functions numerically. Specifically, the numerical analytic continuation is an ill-posed problem \cite{Tripolt:2018xeo}, so it is very hard to perform in the case of many-point correlation functions. Thus, one still may need to use the Schwinger-Keldysh diagram technique even for thermal states.
	
	While some physical effects can be studied perturbatively, a vast number of other phenomena require nonperturbative methods. In particular, if the potential energy of the theory possesses degenerate and/or local minima, tunneling effects should be taken into account. These effects are usually handled with the use of instantons. Namely, the tunneling process is identified with the evolution of the system in the imaginary time, and the solution to corresponding Wick rotated equations of motion are referred as instantonic solutions. In common prescription one should take into account not only perturbations of trivial vacuum solutions, but also instantons and the fluctuations about them (working in the path integral formalism). In such a way one may calculate, e.g. a partition function (0-point correlation function), and find the corrections to the energy spectrum due to tunneling effects \cite{Lowe:1978ug, Escobar-Ruiz:2015nsa, *Escobar-Ruiz:2015rfa}. This method has broader applications in various topics, from the structure of QCD vacuum \cite{Callan:1977gz,*Callan:1976je} and false vacuum decay \cite{coleman1977fate,*callan1977fate,Linde:1981zj} to tunneling effects in condensed matter systems \cite{chudnovsky1997first,liang1998periodic}. The calculation of nontrivial imaginary-time correlation functions for the instantonic systems is a more tricky task. In particular, once a correlation function on the fixed instantonic background is calculated, one should average the result over the instantonic moduli space afterwards \cite{Polyakov:1976fu, Callan:1977gz}. The latter space is at least one-dimensional, since the functional integration is performed over periodic (in imaginary time) field configurations, so one always has to integrate over the period to non-perturbatively account for zero mode, corresponding to imaginary time translational symmetry. Imaginary-time correlation functions, calculated in such a way, can (at least in principle) be analytically continued, and hence some part of real-time correlation functions can be reconstructed. However, a direct calculation of real-time correlation functions in the presence of instantons is not known in the literature (see, however, Ref.~\cite{titov2016korshunov} on real-time instantons in a rather different context of the dissipative dynamics).
	
	The aim of this paper is to construct a diagrammatic technique for the systems exhibiting instantonic (tunneling) effects, that allows to calculate correlation functions for thermal states directly in real time. For this purpose, we adopt the Schwinger-Keldysh diagram technique for the instantonic systems. However, the real-time part of the Schwinger-Keldysh contour, having three pieces in the case of thermal states, corresponding to real-rime forward, backward and imaginary-time evolution, (at least naively) breaks the imaginary time translational invariance. At the same time, a zero mode corresponding to the latter (naively broken) symmetry is still present, so it seems that one cannot take into account zero mode non-perturbatively. We develop a procedure, through which the mentioned symmetry can be (and, in fact, must be) recovered. After that, zero mode is taken into account in a usual way, and desired diagrammatic technique for real-time instantonic correlation functions is constructed. As with imaginary-time correlation functions, the results involve integration over the imaginary time period as part of the instantonic moduli space. In our case it also plays the role of the point of gluing of the imaginary and real-time parts of the Schwinger-Keldysh time contour. This means that the path integral is saturated by complex-valued backgrounds, while the real-valued backgrounds correspond to only finite a number of points in the whole space.
	
	This paper is organized as follows. In Section~\ref{sec:imaginary_time_corr} we revisit the procedure of calculating the imaginary-time correlation functions in the presence of instantons, generalizing the logic of \cite{Polyakov:1976fu} to $n$-point functions. In Section~\ref{sec:real_time_corr} we resolve the zero mode issue for real-time correlation functions in instantonic systems, and develop the perturbative expansion. In Section~\ref{sec:diagrams} we propose a diagram technique, corresponding to the perturbation theory constructed and demonstrate it on the first few-point correlation functions. Then, we discuss the results obtained, its potential applications and generalizations in Section~\ref{sec:discussion}. Appendices~\ref{sec:app_zero_mode_gf_eucl},~\ref{sec:app_zero_mode_gf_lor} describe the methods for constructing specific Green's functions, that were used through the paper. In Appendix~\ref{sec:app_keldysh_rot} we recast the previously constructed diagrammatic technique in the Keldysh-rotated form.

\section{Imaginary-time instantonic correlation functions} \label{sec:imaginary_time_corr}
Before proceeding to the study of real-time instantonic correlation functions, we revisit the calculation scheme for its imaginary-time counterparts. The latter procedure has mostly the same subtleties, but in a much reduced form, so it will give us an important intuition for the generalization to the real-time case of our main interest.

Let us consider the thermal partition function for a quantum mechanical system
\begin{equation}
	Z = \Tr e^{-\beta \hat H},
\end{equation}
where $\beta$ stands for an inverse temperature. It can be represented as a path integral over periodic trajectories
\begin{equation}
	Z = \int\limits_{\text{periodic}} \calD x \; e^{-S_e[x]} \label{eq:eucl_pi}, \qquad
	S_e[x] = \int_{0}^{\beta} d\tau \; \left[\frac{\dot x^2}2 + V(x)\right].
\end{equation}
Here $S_e$ is the Euclidean action of the system under consideration. Until the end of this section, we will omit the subscript $e$ of $S_e$.

Calculation of the partition function with the use of perturbation theory about nontrivial saddle point has the well-known zero mode issue. To see this, let us expand the action about $\bar{x}$, obeying the classical equation of motion
\begin{equation}
	\ddot{\bar{x}}(\tau) - V'(\bar{x}) = 0, \qquad \bar{x}(0) = \bar{x}(\beta), \quad \dot{\bar{x}}(0) = \dot{\bar{x}}(\beta) \label{eucl_bg_eom}
\end{equation}
and go from the path integration over $x$ to the integration over a perturbation $\eta$ about $\bar{x}$, i.e. $x(\tau) = \bar{x}(\tau) + \eta(\tau)$. Thus, the partition function takes the form
\begin{equation}
	Z = e^{-S[\bar{x}]}\int\limits_{\text{periodic}} \calD\eta \; \exp \Bigl( -S^{\scriptscriptstyle (2)}[\bar{x}, \eta] - S^{\text{int}}[\bar{x}, \eta] \Bigr)
\end{equation}
where $S^{\scriptscriptstyle(2)}[\bar{x}, \eta]$ is the quadratic part of $S[\bar{x}+\eta]$
\begin{equation}
	S^{\scriptscriptstyle (2)}[\bar{x}, \eta] = \frac12 \int d\tau\,d\tau' \; \eta(\tau) \, K(\tau, \tau') \, \eta(\tau'), \qquad K(\tau, \tau') = [-\partial_\tau^2 + V''(\bar{x})] \, \delta(\tau-\tau'),
\end{equation}
and $S^\text{int}$ contains interaction terms
\begin{equation}
	S^\text{int}[\bar x, \eta] = \sum_{k\ge3} \frac1{k!} \int_{0}^{\beta} d\tau \, V^{(k)}(\bar{x}(\tau)) \, \eta^k(\tau). \label{eq:action_int_eucl}
\end{equation}
The operator $K$, defined by its kernel $K(\tau, \tau')$ is diagonalised by the set of eigenfunctions $\eta_n$ subject to the periodic boundary conditions
\begin{equation}
	(K\eta_n)(\tau) = \int_0^\beta d\tau \; K(\tau, \tau') \eta_n(\tau') = \lambda_n \eta_n(\tau), 
\end{equation}
Now, taking derivative of (\ref{eucl_bg_eom}), we obtain
\begin{equation}
	[-\partial_\tau^2 + V''(\bar{x})] \dot{\bar{x}}(\tau) = 0,
\end{equation}
and conclude that $K$ has a zero mode 
\begin{equation}
	\eta_0(\tau) = \frac1{\|\dot{\bar{x}}\|} \dot{\bar{x}}(\tau), 	\qquad
	\|\dot{\bar{x}} \|^2 = \int_0^\beta d\tau \, \bigl(\dot{\bar{x}}(\tau)\bigr)^2,
\end{equation}
originating from the invariance of the action under Euclidean time translations
\begin{equation}
	S[x^{\tau_0}] = S[x], \qquad x^{\tau_0} (\tau) = x(\tau + \tau_0).
\end{equation}
Thus, the one-loop partition function is proportional to $(\Det K)^{-1/2}$ hence naively diverges, so we need to perform a fixing of the translation ``gauge'' invariance. For this purpose, we perform a partition of unity
\begin{equation}
	1 = \frac1{\sqrt{\xi}} \int_0^\beta d\tau_0 \; \frac{d\chi[x^{\tau_0}]}{d\tau_0} \; \delta\bigl(\tfrac1{\sqrt{\xi}}\chi[x^{\tau_0}] - \lambda\bigr), \label{unity_part}
\end{equation}
where $\chi$ is a gauge-fixing functional
\begin{equation}
	\chi[x] = \frac1{\|\dot{\bar{x}} \|}\int_0^\beta d\tau \, \dot{\bar{x}}(\tau) x(\tau), \qquad \lambda, \xi = \text{const}.
\end{equation}
Delta-function fixes the projection of $x(t)$ on the zero-mode, and $\lambda$, $\xi$ are the parameters, whose role will become clear in a while. Inserting the partition of unity into the path integral (\ref{eq:eucl_pi}), and passing to integration over $x^{\tau_0}$ in the functional measure, we obtain the following expression for the partition function
\begin{equation}
	Z = \frac{\beta}{\sqrt{\xi}} \int\limits_{\text{periodic}} \calD x \; J[x] \; \delta\bigl(\tfrac1{\sqrt{\xi}}\chi[x] - \lambda\bigr) \; e^{-S[x]}, \qquad J[x] = \frac{d\chi[x^{\tau_0}]}{d\tau_0}\biggl|_{\tau_0=0} = \frac1{\| \dot{\bar{x}}\|}\int_{0}^\beta d\tau \, \dot{\bar{x}}(\tau) \dot x(\tau),
\end{equation}
where the integration over $\tau_0$ decouples and we replace the integral by its value $\beta$.
Since $Z$ does not depend on $\lambda$, we can average it with the Gaussian weight $e^{-\lambda^2/2} d\lambda / \sqrt{2\pi}$, so the partition function takes the form
\begin{equation}
	Z = \frac{\beta}{\sqrt{2\pi\xi}} \int\limits_{\text{periodic}} \calD x \; J[x] \; e^{-S_\xi[x]}, \qquad S_\xi[x] = S[x] + \frac1{2\xi} \chi[x]^2.
\end{equation}
Note that $\xi$ should tend to zero so that the equation, enforced by delta-function, has solution for an arbitrary large $\lambda$. Here $S_\xi$ is called gauge-fixed action. Finally, expanding about $\bar{x}$, we obtain the perturbative form of the partition function\footnote{In fact, there may be a whole set of nonequivalent periodic saddle point solutions, so one needs to perform a summation over them. To shorten the notations, we will keep only one such saddle.}
\begin{equation}
	Z = \frac{\beta \|\dot{\bar{x}} \|}{\sqrt{2\pi\xi}} e^{-S[\bar{x}]} \int\limits_{\text{periodic}} \calD \eta \; \Bigl[1 + \frac1{\|\dot{\bar{x}}\|} \int_{0}^\beta d\tau \, \eta_0(\tau) \dot \eta(\tau)\Bigr] \; e^{-S_\xi^{\scriptscriptstyle(2)}[\bar{x}, \eta] - S^{\text{int}}[\bar{x}, \eta]}.
\end{equation}
The quadratic part of the gauge-fixed action now reads
\begin{align}
	S_\xi^{\scriptscriptstyle (2)}[\bar{x}, \eta] &= \frac12 \int d\tau\,d\tau' \; \eta(\tau) \, K_\xi(\tau, \tau') \, \eta(\tau'), \qquad \\ K_\xi(\tau, \tau') &= [-\partial_\tau^2 + V''(\bar{x})] \, \delta(\tau-\tau') + \frac1{\xi} \eta_0(\tau) \eta_0(\tau'). \label{eq:diff_op_eucl_reg}
\end{align}
Thus, the zero-mode $\eta_0$ acquires a non-zero eigenvalue $\xi^{-1}$.
The Green's function of the operator $K_\xi$ can be found with the use of spectral decomposition
\begin{equation}
	(K_\xi \eta_n)(\tau) = \lambda_n \eta_n(\tau),
\end{equation}
where the eigenfunctions $\eta_n(\tau)$ are periodic and normalized, and $\eta_0(\tau) = \dot{\bar{x}}(\tau) / \| \dot{\bar{x}}\|$, as before. Thus, the Green's function, satisfying
\begin{equation}
	(K_\xi G_\xi)(\tau, \tau') = \delta(\tau-\tau') \label{eq:gf_eq_eucl_reg}
\end{equation}
has the following explicit form
\begin{equation}
	G_\xi(\tau, \tau') = \xi \, \eta_0(\tau) \eta_0(\tau') + \sum_{n\ne0} \lambda_n^{-1} \, \eta_n(\tau) \eta_n(\tau').
\end{equation}
In the case $\xi \to 0$, the corresponding Green's function $G_0$ satisfies the following equation
\begin{equation}
	(-\partial_\tau^2 + V''(\bar{x})) G_0(\tau,\tau') = \delta(\tau-\tau') - \eta_0(\tau) \eta_0(\tau'), \qquad \int_0^\beta d\tau  \,\eta_0(\tau) G_0(\tau,\tau') = 0,
\end{equation}
whose explicit solution is found in Appendix \ref{sec:app_zero_mode_gf_eucl}.

Now, let us generalize the partition function $Z$ to the generating functional of imaginary time ordered (Euclidean) correlation functions
\begin{equation}
	Z[j] = \Tr\Bigl[e^{-\beta \hat H} T_\tau \exp \Bigl( \int_0^\beta d\tau \, j(\tau) \hat x(\tau) \Bigr) \Bigr], \qquad \hat x(\tau) = e^{\tau \hat H} \, \hat x \, e^{-\tau \hat H},
\end{equation}
in terms of which the $n$-point correlation function reads
\begin{equation}
	D_n(\tau_1, \ldots \tau_n) = \Tr\Bigl[e^{-\beta \hat H} T_\tau \Bigl( \hat x(\tau_1) \ldots \hat x(\tau_n) \Bigr) \Bigr] = \frac1Z\frac{\delta^n Z[j]}{\delta j(\tau_1) \ldots \delta j(\tau_n)}\biggl|_{j=0}. \label{eq:corr_fun_eucl}
\end{equation}
Here $T_\tau$ denotes the ordering of the operators by imaginary time.
Path integral representation of $Z[j]$ has the form
\begin{equation}
	Z[j] = \int\limits_{\text{periodic}} \calD x \; e^{-S[x] + \int_0^\beta d\tau \, j(\tau) x(\tau)} 
\end{equation}
The calculation of $Z[j]$ repeats those of $Z$, except that the integration over $\tau_0$, originating from the partition of unity (\ref{unity_part}) does not decouple, but leads to averaging of the exponential of source term over gauge group of Euclidean time translations
\begin{equation}
	Z[j] = \frac{1}{\sqrt{2\pi\xi}\|\dot{\bar{x}} \|} \int\limits_{\text{periodic}} \calD x \; \Bigl[\int_{0}^\beta d\tau \, \dot{\bar{x}}(\tau) \dot x(\tau)\Bigr] \; e^{-S_\xi[x]} \int_0^\beta d\tau_0 \, e^{\int_0^\beta d\tau \, j(\tau) x(\tau-\tau_0)}, \label{eq:gen_fun_eucl}
\end{equation}
or, after expansion about the saddle point solution $\bar{x}$
\begin{multline}
	Z[j] = \frac{ \|\dot{\bar{x}} \|}{\sqrt{2\pi\xi}} e^{-S[\bar{x}]} \int\limits_{\text{periodic}} \calD \eta \, \Bigl[1 + \frac1{\|\dot{\bar{x}}\|} \int_{0}^\beta d\tau \, \eta_0(\tau) \dot \eta(\tau)\Bigr] \, e^{-S_\xi^{\scriptscriptstyle(2)}[\bar{x}, \eta] - S^{\text{int}}[\bar{x}, \eta]} \\ \times \int_0^\beta d\tau_0 \; e^{\int_0^\beta d\tau \, j(\tau) (\bar{x}(\tau-\tau_0) + \eta(\tau-\tau_0))}. \label{eq:gen_fun_eucl_pert}
\end{multline}
The averaging over $\tau_0$ emphasizes the fact that observables should be gauge-invariant. Namely, substituting the generating functional (\ref{eq:gen_fun_eucl}) to (\ref{eq:corr_fun_eucl}), we find that $n$-point function has the form
\begin{equation}
	D_n(\tau_1, \ldots \tau_n) \propto \int \calD x \; \Bigl[\frac1\beta \int_0^\beta d\tau_0 \, x(\tau_1-\tau_0) \ldots x(\tau_n-\tau_0) \Bigr] \; \Bigl[\int_{0}^\beta d\tau \, \dot{\bar{x}}(\tau) \dot x(\tau)\Bigr] \; e^{-S_\xi[x]}.
\end{equation}
Here the quantity to be path integral averaged
\begin{eqnarray}
	\tilde D_n[x](\tau_1, \ldots \tau_n) = \frac1\beta \int_0^\beta d\tau_0 \; x(\tau_1-\tau_0) \ldots x(\tau_n-\tau_0),
\end{eqnarray}
is indeed a gauge invariant, i.e. $\tilde D_n[x^{\tau'}] = \tilde D_n[x]$, where $x^{\tau'}(\tau) = x(\tau+\tau')$ as before.

The generating functional of imaginary-time correlation functions (\ref{eq:gen_fun_eucl}) and its perturbative counterpart (\ref{eq:gen_fun_eucl_pert}) can be rewritten in a slightly different form, having nonetheless far-reaching implications. Specifically, if one performs a substitution $\tau_0 \mapsto -\tau_0$, and goes from the integration over $x(\tau)$ to the integration over $x(\tau+\tau_0)$, the generating functional acquires the following form
\begin{multline}
	Z[j] = \frac{1}{\sqrt{2\pi\xi}\|\dot{\bar{x}} \|} \int_0^\beta d\tau_0 \int\limits_{\text{periodic}} \calD x \; \Bigl[\int_{0}^\beta d\tau \, \dot{\bar{x}}^{\tau_0}(\tau) \dot x(\tau)\Bigr] \\ \times \exp\biggl\{-S[x]-\frac1{2\xi}\bigl(\chi[\bar x^{\tau_0}; x]\bigr)^2 + \int_0^\beta d\tau \, j(\tau) x(\tau)\biggr\}, \label{eq:gen_fun_eucl_bg_av}
\end{multline}
where we introduce the notation
\begin{equation}
	\chi[\bar{x};x] = \frac1{\|\dot{\bar{x}} \|}\int_0^\beta d\tau \, \dot{\bar{x}}(\tau) x(\tau),
\end{equation}
for the gauge-fixing functional, emphasizing its dependence on the saddle-point solution $\bar{x}$. Similarly, the perturbative from (\ref{eq:gen_fun_eucl_pert}) of the generating functional can be rewritten as
\begin{multline}
	Z[j] = \frac{ \|\dot{\bar{x}} \|}{\sqrt{2\pi\xi}} e^{-S[\bar{x}]} \int_0^\beta d\tau_0 \int\limits_{\text{periodic}} \calD \eta \, \Bigl[1 + \frac1{\|\dot{\bar{x}}\|} \int_{0}^\beta d\tau \, \eta_0^{\tau_0}(\tau) \dot \eta(\tau)\Bigr] \\ \times \exp\biggl\{-S_\xi^{\scriptscriptstyle(2)}[\bar{x}^{\tau_0}, \eta] - S^{\text{int}}[\bar{x}^{\tau_0}, \eta] + \int_0^\beta d\tau \, j(\tau) \bigl(\bar{x}^{\tau_0}(\tau) + \eta(\tau)\bigr)\biggr\},
\end{multline}
where $\eta_0^{\tau_0}(\tau) = \eta_0(\tau+\tau_0)$. This form of the generating functional has the following important interpretation. In the process of calculating the correlation function one first fixes the background $\bar{x}^{\tau_0}$ and calculates the latter correlation function on this background. After that, one integrates over $\tau_0$, parameterizing imaginary time shift of the background $\bar{x}$. Thus, in some sense, the procedure of the correlation function computation includes an ``averaging over backgrounds'' as a final step. 

For the further development of the diagram technique, it will be useful to rewrite the generating functional $Z[j]$ in a ``Wick theorem manner'', namely
\begin{multline} 
	Z[j] = Z_{\text{1-loop}} \,\frac1\beta \int_0^\beta d\tau_0 \; \exp\biggl\{\frac12 \int d\tau\, d\tau' \, \frac{\delta}{\delta \eta(\tau)} G^{\tau_0}_\xi(\tau,\tau') \frac{\delta}{\delta \eta(\tau')}\biggr\} \\ \times \Bigl[1 + \frac1{\|\dot{\bar{x}}\|} \int_{0}^\beta d\tau \, \eta_0^{\tau_0}(\tau) \dot \eta(\tau)\Bigr] e^{- S^{\text{int}}[\bar{x}^{\tau_0}, \eta] + \int_0^\beta d\tau \, j(\tau) (\bar{x}^{\tau_0}(\tau) + \eta(\tau))}\Bigr|_{\eta=0}, \label{eq:gen_fun_eucl_wick}
\end{multline}
where we denote
\begin{equation}
	Z_{\text{1-loop}} = \frac{\beta \|\dot{\bar{x}} \|}{\sqrt{2\pi\xi}} e^{-S[\bar{x}]} \int\limits_{\text{periodic}} \calD \eta \; e^{-S_\xi^{\scriptscriptstyle(2)}[\bar{x}, \eta]}, \label{eq:part_fun_1_loop}
\end{equation}
and $G^{\tau_0}_\xi(\tau,\tau')$ is the Green's function of the operator $K_\xi$, defined in (\ref{eq:diff_op_eucl_reg}), but for the transformed background $\bar{x} \mapsto \bar{x}^{\tau_0}$ (consequently, $\eta_0 \mapsto \eta_0^{\tau_0}$).

\section{Instantonic correlation functions in real time} \label{sec:real_time_corr}
	\begin{figure}[h]
		\centering
		\begin{subfigure}[b]{0.49\textwidth}
			\includegraphics[height=4.5cm]{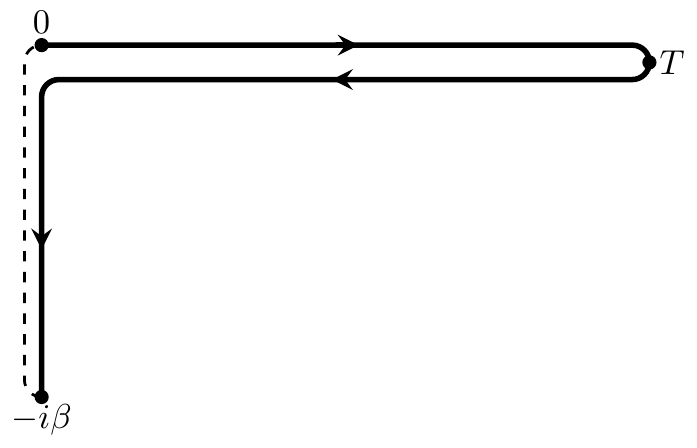}
			\caption{}
			\label{fig:conotur_conv}
		\end{subfigure}
		\begin{subfigure}[b]{0.49\textwidth}
			\includegraphics[height=4.5cm]{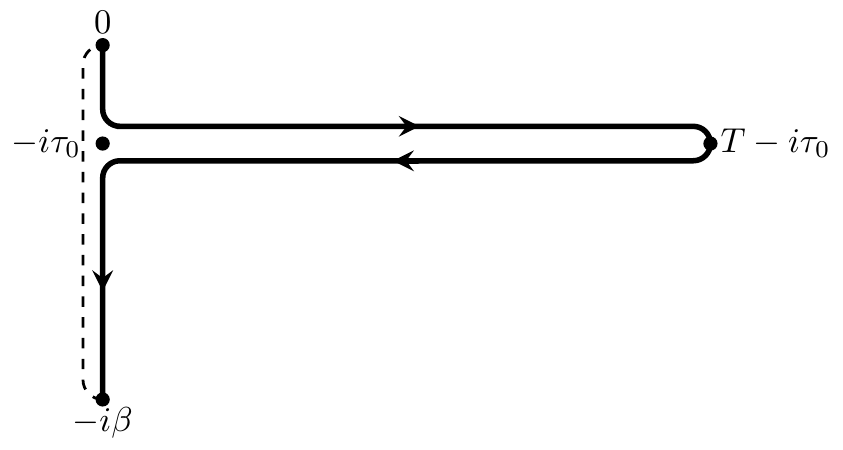}
			\caption{}
			\label{fig:contour_shifted}
		\end{subfigure}
		\caption{(a)~Conventional Schwinger-Keldysh contour $C$ of complex time $z = t - i\tau$. (b)~Schwinger-Keldysh contour $C^{\tau_0}$, shifted by $-i\tau_0$.}
		\label{fig:contour}
	\end{figure}
	Now, we are ready to examine the real-time correlation functions, defined by the generating functional
	\begin{gather}
		Z[j_+,j_-] = \Tr\Bigl[e^{-\beta \hat H} \, \bar T \exp \Bigl( -i\int_0^T dt \, j_-(t) \hat x(t) \Bigr) \, T \exp \Bigl(i \int_0^T dt \, j_+(t) \hat x(t) \Bigr) \Bigr], \label{eq:gen_fun_lor_op}\\ \hat x(t) = e^{i \hat H t} \, \hat x \, e^{-i \hat H t}, \nonumber
	\end{gather}
	so that the variations with respect to $j_\pm$ give the correlation functions of the following form
	\begin{align}
		D_{nm}(t_{1}^+, \ldots t_{n}^+;t_{1}^-, \ldots t_{m}^-) &=
		\Tr\Bigl[e^{-\beta \hat H} \, \bar T\bigl(\hat x(t_1^-) \ldots \hat x(t_m^-) \bigr) \, T \bigl(\hat x(t_1^+) \ldots \hat x(t_n^+) \bigr) \Bigr] \nonumber \\ &= i^{m-n}
		\frac{1}Z \frac{\delta^{n+m} Z[j_+,j_-]}{\delta j_+(t_1^+) \ldots \delta j_+(t_n^+) \, \delta j_-(t_1^-) \ldots \delta j_-(t_m^-)}\biggl|_{j_\pm=0} \label{eq:corr_fun_lor}
	\end{align}
	where $T$ and $\bar T$ denote time and anti-time ordering, respectively. 
	In fact, there is an ambiguity in the definition of the partition function (\ref{eq:gen_fun_lor_op}). Namely, one can use the cyclic property of the trace to rewrite it as  
	\begin{equation}
		Z^{\tau_0}[j_+,j_-] = \Tr\Bigl[e^{-(\beta - \tau_0) \hat H} \, \bar T \exp \Bigl(-i\int_0^T dt \, j_-(t) \hat x(t) \Bigr) \, T \exp \Bigl(i \int_0^T dt \, j_+(t) \hat x(t) \Bigr) \, e^{-\tau_0 \hat H}\Bigr],
	\end{equation}
	for some $\tau_0$, satisfying $0 \le \tau_0 \le \beta$. Thus, the time evolution goes along the complex path\footnote{Note that such type of contour first appear in \cite{konstantinov1961diagram} in a non-instantonic context.} $C^{\tau_0}$, represented at Fig.~\ref{fig:contour_shifted}, whereas the variations over $j_+$, and $j_-$ correspond to insertion of the operators at top and the bottom horizontal segments of the path, respectively.
	The case $\tau_0 = 0$ corresponds to the Schwinger-Keldysh time contour $C$, depicted at Fig.~\ref{fig:conotur_conv} (also referred as Kadanoff-Baym contour in the presence of imaginary time segment \cite{kadanoff2018quantum}), whereas nonzero $\tau_0$ leads to the contour $C^{\tau_0}$ having real-time segments, shifted by $-i\tau_0$.
	Path integral form of the generating functional can be written as follows\footnote{One can also add the source for the Euclidean operators, living on the imaginary time segments of the contour $C^{\tau_0}$, for the purposes of generality. However, it leads to the unnecessary complications, while not giving any new insights.}
	\begin{multline}
		Z^{\tau_0}[j_+, j_-] = \int\limits_{x_e(0) = x_e(\beta)} \calD x_e \int\limits_{\substack{x_\pm(0)=x_e(\tau_0\mp0)\\x_+(T)=x_-(T)}} \calD x_+ \, \calD x_-\\ \times \exp\biggl\{ -S_e[x_e] + i S[x_+] - i S[x_-] + i \int_0^T dt\, j_+(t) x_+(t) - i \int_0^T dt \, j_-(t) x_-(t)\biggr\}. \label{eq:gen_fun_lor_pi}
	\end{multline}
	Here, $S$ is the real-time (Lorentzian) action of the system
	\begin{equation}
		S[x] = \int_{0}^{T} d\tau \; \left[\frac{\dot x^2}2 - V(x)\right]
	\end{equation}
	whereas $S_e$ is its Euclidean counterpart, defined in (\ref{eq:eucl_pi}). Note that in this section we keep the subscript of $S_e$, in contrast to previous section.

\subsection{Saddle point and zero mode}
%	Let us examine saddle points of the integrand in (\ref{eq:gen_fun_lor_pi}) for vanishing sources $j_\pm$. for this purpose, let us explicitly write down the variation of the sum of the actions
%	\begin{multline}
%		\delta\bigl(-S_e[x] + i S[x_+] - i S[x_-]\bigr) = -\int_0^\beta d\tau \, \bigl[-\ddot x_e(\tau) + V'(x_e(\tau))\bigr] \delta x_e(\tau) \\ + i\int_0^T dt \, \bigl[-\ddot x_+(t) - V'(x_+(t))\bigr] \delta x_+(t) - i\int_0^T dt \, \bigl[-\ddot x_-(t) - V'(x_-(t))\bigr] \delta x_-(t)  \\
%		+ \delta x_e(\tau) \dot{x}_e(\tau)\Bigr|_0^\beta + i \delta x_+(t) \dot{x}_+(t)\Bigr|_0^T - i \delta x_-(t) \dot{x}_-(t)\Bigr|_0^T ,
%	\end{multline}
%	The first two lines give the equations of motion
%	\begin{equation}
%		\ddot{x}_e(\tau) - V'(x_e) = 0, \qquad
%		\ddot{x}_\pm(t) + V'(x_\pm) = 0, \label{eq:eom_sections}
%	\end{equation}
%	whereas the third one, supplemented by the conditions arising from the path integral measure
%	\begin{equation}
%		x_e(0) = x_e(\beta), \qquad
%		x_e(\tau_0\mp0) = x_\pm(0),\qquad
%		x_+(T)=x_-(T), \label{eq:bdy_cond_val}
%	\end{equation}
%	give the set of boundary conditions on the derivatives
	Let us examine saddle points of the integrand in (\ref{eq:gen_fun_lor_pi}) for vanishing sources $j_\pm$. To make simpler the account of the boundary conditions, sitting in the path integral measure, it is useful to rewrite the sum of the actions in the exponent as a single integral over a natural parameter $\sigma$ on $C^{\tau_0}$ parameterized as $z = z(\sigma)$
	\begin{equation}
		-S_e[x] + i S[x_+] - i S[x_-] = i \int_0^{2T+\beta} d\sigma \, e(\sigma) \, \left(\frac{(\partial_\sigma x(\sigma))^2}{2 e^2(\sigma)} - V(x)\right), \qquad e(\sigma) = \partial_\sigma z(\sigma),
	\end{equation}
	with the following identifications
	\begin{subequations}
		\begin{align}
			x(\sigma) &= x_e(\sigma), & e(\sigma) &= -i, & 0 \le \sigma{}& \le \tau_0, \label{eq:contour_id_first}\\
			x(\sigma) &= x_+(\sigma - \tau_0), & e(\sigma) &= +1, & \tau_0 \le \sigma{}& \le T+\tau_0, \\
			x(\sigma) &= x_-(2T + \tau_0 - \sigma), & e(\sigma) &= -1, & T + \tau_0 \le \sigma{}& \le 2T+\tau_0, \\
			x(\sigma) &= x_e(\sigma - 2T - \tau_0), & e(\sigma) &= -i, & 2T + \tau_0 \le \sigma{}& \le 2T + \beta. \label{eq:contour_id_last}
		\end{align}
	\end{subequations}
%	\textcolor{red}{
%	\begin{itemize}
%		\item Do the same without natural parameter.
%	\end{itemize}
%	}
	Here $e(\sigma)$ plays the role of einbein. The variation of the action above w.r.t. $x(\sigma)$ gives the saddle point equation
	\begin{equation}
		\partial_\sigma( e^{-1} \partial_\sigma x) + e \, V'(x) = 0. \label{eq:eom_natural_par}
	\end{equation}
	Vanishing of the boundary term, arising in the process of the variation, implies, after use of the identifications above
	\begin{equation}
		\partial_\tau x_e(0) = \partial_\tau x_e(\beta). \label{eq:bdy_cond_eucl}
	\end{equation}
	Outside the breaking points $\tau_0$, $T+\tau_0$, $T+2\tau_0$ this equation can be rewritten as a sat of equations on $x_e$,~$x_\pm$
	\begin{equation}
		\ddot{x}_e(\tau) - V'(x_e) = 0, \qquad
		\ddot{x}_\pm(t) + V'(x_\pm) = 0. \label{eq:eom_sections}
	\end{equation}
	To account the breaking points, one should integrate the equation (\ref{eq:eom_natural_par}) over a small vicinity of these points. As a result, one obtains the following conditions on the derivatives of $x_e$,~$x_\pm$ at these points, namely
	\begin{equation}
		\partial_\tau x_e(\tau_0\mp0) = i \, \partial_t x_\pm(0), \qquad \partial_t x_+(T) = \partial_t x_-(T). \label{eq:bdy_cond_deriv}
	\end{equation}
	These conditions are supplemented by those arising from the path integral measure
	\begin{equation}
		x_e(0) = x_e(\beta), \qquad
		x_e(\tau_0\mp0) = x_\pm(0),\qquad
		x_+(T)=x_-(T). \label{eq:bdy_cond_val}
	\end{equation}
	The boundary conditions obtained have the following important consequences. From (\ref{eq:eom_sections}) we see that $x_+(t)$ and $x_-(t)$ satisfy the same second-order differential equation. Moreover, from the second equality in (\ref{eq:bdy_cond_deriv}) and the third equality in (\ref{eq:bdy_cond_val}) we observe that $x_+(t)$ and $x_-(t)$ has the same values and derivatives at the point $t=T$. In fact, this means that these functions coincide, i.e. $x_+(t) = x_-(t)$ for any $t$. In particular, $x_+(0) = x_-(0)$, so, combining it with the first equality in (\ref{eq:bdy_cond_deriv}) and the second one in (\ref{eq:bdy_cond_val}), one concludes that $x_e(\tau)$ is continuous at $\tau_0$ together with its derivative. Supplemented by the condition (\ref{eq:bdy_cond_eucl}) and the first equality in (\ref{eq:bdy_cond_val}), one finds that $x_e(\tau)$ is smooth, $\beta$-periodic function. Furthermore, the first equality in (\ref{eq:bdy_cond_deriv}) is nothing but Cauchy-Riemann condition, so that $x_+(t)=x_-(t)$ is an analytic continuation of $x_e(\tau)$ from the point $-i\tau_0$ along the axis, parallel to the real one. Summarizing all these facts, one conclude that for sufficiently smooth potential $V(x)$, the saddle point solution can be described by the holomorphic function $\bar{x}(z)$, satisfying the equation
	\begin{equation}
		\partial_z^2 \bar{x}(z) + V'(\bar{x}) = 0, \qquad \bar{x}(z - i \beta) = \bar{x}(z), \label{eq:saddle_eq_an}
	\end{equation}
	together with the following identifications
	\begin{equation}
		x_e(\tau) = \bar{x}(-i\tau) = \bar{x}_e(\tau), \qquad x_\pm(t) = \bar{x}(t-i\tau_0) = \bar{x}^{\tau_0}(t). \label{eq:contour_inentif}
	\end{equation}
	
	Let us consider linear perturbations $\eta$ about $\bar{x}$, i.e.
	\begin{equation}
		x_e(\tau) = \bar{x}_e(\tau) + \eta_e(\tau), \qquad
		x_\pm(t) = \bar{x}^{\tau_0}(t) + \eta_\pm(t).
	\end{equation}
	Substitution to the exponent in the integrand (\ref{eq:gen_fun_lor_pi}) gives
	\begin{align}
		-S_e[x] + i S[x_+] - i S[x_-] = -S_e[\bar{x}_e] {}& - S_e^{\scriptscriptstyle (2)}[\bar{x}_e,\eta_e] + i S^{\scriptscriptstyle (2)}[\bar{x}^{\tau_0},\eta_+] - i S^{\scriptscriptstyle (2)}[\bar{x}^{\tau_0},\eta_-] \nonumber \\ {}&- S^{\text{int}}_e[\bar{x}_e,\eta_e] + i S^{\text{int}}[\bar{x}^{\tau_0},\eta_+] - i S^{\text{int}}[\bar{x}^{\tau_0},\eta_-],
	\end{align}
	where the quadratic part of the action has the following explicit form
	\begin{align}
		&S_e^{\scriptscriptstyle (2)}[\bar{x}_e,\eta_e] = \frac12 \int d\tau\,d\tau' \; \eta_e(\tau) \, K_e(\tau, \tau') \, \eta_e(\tau'), \qquad K_e(\tau, \tau') = \bigl[-\partial_\tau^2 + V''(\bar{x}_e(\tau))\bigr] \, \delta(\tau-\tau'),\\
		&S^{\scriptscriptstyle (2)}[\bar{x}^{\tau_0},\eta_\pm] = \frac12 \int dt\,dt' \; \eta_\pm(t) \, K^{\tau_0}(t, t') \, \eta_\pm(t'), \qquad K^{\tau_0}(t, t') = \bigl[-\partial_t^2 - V''(\bar{x}^{\tau_0}(t))\bigr] \, \delta(t-t').
	\end{align}
%	The quadratic part of the sum of the actions can be rewritten in a useful matrix form, namely
%	\begin{align}
%		-&S_e^{\scriptscriptstyle (2)}[\bar{x}_e,\eta_e] + i S^{\scriptscriptstyle (2)}[\bar{x}^{\tau_0},\eta_+] - i S^{\scriptscriptstyle (2)}[\bar{x}^{\tau_0},\eta_-] = -\frac12 \eta_A\, K_{AB} \,\eta_B, \\
%		&K_{AB} = \operatorname{diag} \bigl( K_e(\tau, \tau'), -i K^{\tau_0}(t, t'), i K^{\tau_0}(t, t') \bigr), \\ 
%		&\eta_A = \bigl( \eta_e(\tau), \eta_+(t), \eta_-(t) \bigr),
%	\end{align}
%	where we assume summation over repeated indices together with integration over appropriate variable.
	Perturbative calculation of the generating functional (\ref{eq:gen_fun_lor_pi}) begins with solving the saddle point equation for the quadratic part of the action (including the source terms), which has the following form
	\begin{equation}
		\begin{bmatrix}
			-\partial_\tau^2 + V''(\bar{x}_e) & 0 & 0 \\
			0 & \partial_t^2 + V''(\bar{x}^{\tau_0}) & 0 \\
			0 & 0 & \partial_t^2 + V''(\bar{x}^{\tau_0})
		\end{bmatrix}
		\begin{bmatrix}
			\eta_{e}(\tau) \\ \eta_{+}(t) \\ \eta_{-}(t)
		\end{bmatrix} = 
		\begin{bmatrix*}[r]
			j_{e}(\tau) \\ j_{+}(t) \\ j_{-}(t)
		\end{bmatrix*}, \label{eq:quadratic_eom}
	\end{equation}
	where we introduce the source $j_e$ for $\eta_e$ for the sake of convenience. Here $\eta_e$,~$\eta_\pm$ are subject to the same boundary conditions as those for the saddle point solution $x_e$,~$x_\pm$, i.e.
	\begin{subequations} \label{eq:pert_bc}
	\begin{align}
		&\eta_e(0) = \eta_e(\beta), & &\partial_\tau\eta_e(0) = \partial_\tau\eta_e(\beta), \\
		&\eta_\pm(0) = \eta_e(\tau_0), & &\partial_\tau\eta_e(0) = i\,\partial_t\eta_\pm(0), \\
		&\eta_+(T) = \eta_-(T), & &\partial_t \eta_+(T) = \partial_t \eta_-(T).
	\end{align}
	\end{subequations}
	Thus, one should invert the differential operator on the l.h.s. of (\ref{eq:quadratic_eom}). However, it can be easily seen that this operator is not invertible, since it has zero mode. The latter can be constructed by an analytic continuation of some function $\eta_0(z)$, satisfying the equation
	\begin{equation}
		\bigl[\partial_z^2 + V''(\bar{x}(z))\bigr] \eta_0(z) = 0 \label{eq:zero_mode_eq_compl}
	\end{equation}
	by the same identifications as in (\ref{eq:contour_inentif}), namely
	\begin{equation}
		\eta_{0e}(\tau) = \eta_0(-i\tau), \qquad \eta_{0\pm}(t) = \eta_0(t-i\tau_0), \label{eq:zero_mode cont}
	\end{equation}
	so it satisfies the boundary conditions (\ref{eq:pert_bc}). Taking derivative of (\ref{eq:saddle_eq_an}), we find that $\partial_z \bar{x}(z)$ satisfies the equation (\ref{eq:zero_mode_eq_compl}), and the normalized zero mode reads
	\begin{equation}
		\eta_0(z) = -\frac{i}{\|\dot{\bar{x}}\|} \partial_z \bar{x}(z), \label{eq:zero_mode_full}
	\end{equation}
	where we introduce $-i$ factor for the convenience so that $\eta_{0e}(\tau) = \partial_\tau \bar{x}_e(\tau) / \|\dot{\bar{x}}\|$ as in the Euclidean construction of the previous section.

	Thus, as in the case of generating functional of the imaginary-time correlation functions (\ref{eq:corr_fun_eucl}), we run into a problem of zero mode. But unlike the imaginary time case, there is no invariance of the integrand of (\ref{eq:gen_fun_lor_pi}) under the transformations like $x_e(\tau) \mapsto x_e(\tau+\tau')$ for finite $\tau'$, since the contour $C^{\tau_0}$ (namely, its Lorentzian piece) explicitly breaks it. More specifically, there are no transformations of the fields $x_e$,~$x_\pm$, including imaginary time shifts of $x_e$, preserving the boundary conditions, enforced by the path integral measure in (\ref{eq:gen_fun_lor_pi}). This means, that one cannot account zero mode (\ref{eq:zero_mode_full}) with the use of technique similar to those of Section~\ref{sec:imaginary_time_corr}, so we need to develop more sophisticated method.

\subsection{Recovery of invariance under imaginary time translations}
	To save the day, we restore the invariance of the integral (\ref{eq:gen_fun_lor_pi}) under the translation of imaginary time. For this purpose we use the fact that $Z^{\tau_0}$ is actually independent of $\tau_0$. It allows to average it over $\tau_0$ without any changes in the result and define $Z$ as
	\begin{equation}
		Z[j_+, j_-] = \frac1\beta \int_0^\beta d\tau_0 \, Z^{\tau_0}[j_+, j_-] = Z^{\tau_0}[j_+, j_-],
	\end{equation}
	with the following path integral form
	\begin{equation}
		Z[j_+, j_-] = \frac1\beta\int_0^\beta d\tau_0 \int \calD[x]^{\tau_0} \; \exp\biggl\{ -S_e[x_e] + i S[x_+] - i S[x_-] + i \int_0^T dt\, j_+ x_+ - i\int_0^T dt \, j_- x_-\biggr\}. \label{eq:gen_fun_lor_av}
	\end{equation}
	Here we denote the measure of integration as
	\begin{equation}
		\int \calD[x]^{\tau_0} \; (\ldots) =  \int\limits_{x_e(0) = x_e(\beta)} \calD x_e \int\limits_{\substack{x_\pm(0)=x_e(\tau_0\mp0)\\x_+(T)=x_-(T)}} \calD x_+ \, \calD x_- \; (\ldots).
	\end{equation}
	Treating the integration over $\tau_0$ on the same footing as the functional integration over $x_e$,~$x_+$,~$x_-$, we observe that the integrand together with the boundary conditions in the functional measure are invariant under the transformation
	\begin{equation}
		x_e(\tau) \; \mapsto \; x_e^{\tau_1}(\tau) = x_e(\tau+\tau_1), \qquad x_\pm(t) \mapsto x_\pm(t), \qquad \tau_0 \; \mapsto \; \tau_0^{\tau_1} = \tau_0-\tau_1. \label{eq:gauge_tr} 
	\end{equation}
	Once the invariance is restored, we can apply the same method for account of zero mode, as in imaginary time case. To proceed in such way, let us define a partition of unity
	\begin{equation}
		1 = \frac1{\sqrt{\xi}} \int_0^{\beta} d\tau_1 \; \frac{d \chi^{\tau_1}}{d\tau_1} \; \delta\bigl(\tfrac1{\sqrt{\xi}} \chi^{\tau_1}-\lambda\bigr), \qquad \chi^{\tau_1} = \chi[x_e^{
			\tau_1}, x_\pm; \tau_0^{\tau_1}). \label{eq:unity_part_lor}
	\end{equation}
 	We choose a gauge-fixing  functional $\chi$ as follows
	\begin{align}
		\chi[x_e, x_\pm;\tau_0) &= \frac1{\|\dot{\bar{x}}\|} \int_0^{2T+\beta} dz(\sigma) \, \partial_z \bar{x}(z(\sigma)) \, x(\sigma) \nonumber \\&= 
		\frac1{\|\dot{\bar{x}}\|} \biggl[ \int_0^\beta d\tau \; \partial_\tau \bar{x}_e(\tau) \, x_e(\tau) + \int_0^T dt \; \partial_t \bar{x}^{\tau_0}(t) \,\bigl(x_+(t) - x_-(t)\bigr) \biggr],
	\end{align}
	where we use the identifications (\ref{eq:contour_id_first})--(\ref{eq:contour_id_last}) when moving from the first to the second line.
	Insertion of (\ref{eq:unity_part_lor}) to (\ref{eq:gen_fun_lor_av}) and subsequent transformation, inverse to (\ref{eq:gauge_tr}) gives
	\begin{multline}
		Z[j_+, j_-] = \frac1{\sqrt{\xi}} \int_0^\beta d\tau_0 \int \calD[x]^{\tau_0} \; J[x_e, x_\pm;\tau_0) \; \delta\bigl(\tfrac1{\sqrt{\xi}} \chi-\lambda\bigr)  \\ \times \exp\biggl\{ -S_e[x_e] + i S[x_+] - i S[x_-] + i \int_0^T dt\, j_+(t) x_+(t) - i\int_0^T dt \, j_-(t) x_-(t)\biggr\}.
	\end{multline}
	where the ``Faddeev-Popov determinant'' $J$ reads
	\begin{equation}
		J[x_e, x_\pm;\tau_0) = \frac{d \chi^{\tau_1}}{d\tau_1}\biggr|_{\tau_1=0} = 
		\frac1{\|\dot{\bar{x}}\|} \biggl[\int_0^\beta d\tau \; \partial_\tau{\bar{x}}_e(\tau) \, \dot{x}_e(\tau) +i \int_0^T dt \; \partial_t^2{\bar{x}}^{\tau_0}(t) \,\bigl(x_+(t) - x_-(t)\bigr) \biggr],
	\end{equation}
	and the integration over $\tau_1$ was performed explicitly, since the integrand becomes independent of it after the redefinitions made.
	As in pure Euclidean case, generating functional is independent of $\lambda$, so we can average it with the weight $e^{-\lambda^2/2} d\lambda / \sqrt{2\pi}$, so that we have
	\begin{multline}
		Z[j_+, j_-] = \frac1{\sqrt{2\pi\xi}} \int_0^\beta d\tau_0 \int \calD[x]^{\tau_0} \;J[x_e, x_\pm;\tau_0) \\ \times  \exp\biggl\{ -S_e[x_e] + i S[x_+] - i S[x_-] - \frac1{2\xi} (\chi)^2 + i \int_0^T dt\, j_+(t) x_+(t) - i\int_0^T dt \, j_-(t) x_-(t)\biggr\}. 
	\end{multline}
	
	Now, let us rewrite the generating functional obtained in the form, similar to those of (\ref{eq:gen_fun_eucl_bg_av}). For this purpose, in the functional integral above we go from the integration over $x_e(\tau)$ to the integration over $x_e(\tau+\tau_0)$. The result is as follows
	\begin{multline} \label{eq:gen_fun_lor_bg_av}
		Z[j_+, j_-] = \frac1{\sqrt{2\pi\xi}} \int_0^\beta d\tau_0 \int\limits_{\substack{x_+(0)=x_e(\beta)\\x_-(0)=x_e(0)\\x_+(T)=x_-(T)}} \calD x_e \, \calD x_+ \, \calD x_- \; J[\bar{x}^{\tau_0}; x_e, x_\pm]\\ \times  \exp\biggl\{ -S_e[x_e] + i S[x_+] - i S[x_-] - \frac1{2\xi} \bigl(\chi[\bar{x}^{\tau_0};x_e, x_\pm]\bigr)^2 \\ + i \int_0^T dt\, j_+(t) x_+(t) - i\int_0^T dt \, j_-(t) x_-(t)\biggr\}. 
	\end{multline}
	Here we update the notations for the gauge-fixing functional $\chi$ and corresponding Faddeev-Popov determinant $J$ as
	\begin{align}
		\chi[\bar{x};x_e, x_\pm] &= \frac1{\|\dot{\bar{x}}\|} \biggl[ \int_0^\beta d\tau \; \partial_\tau \bar{x}_e(\tau) \, x_e(\tau) + \int_0^T dt \; \partial_t \bar{x}(t) \,\bigl(x_+(t) - x_-(t)\bigr) \biggr], \label{eq:gf_fun_lor_bg}\\
		J[\bar{x}; x_e, x_\pm] &= 
		\frac1{\|\dot{\bar{x}}\|} \biggl[\int_0^\beta d\tau \; \partial_\tau{\bar{x}}_e(\tau) \, \dot{x}_e(\tau) +i \int_0^T dt \; \partial_t^2{\bar{x}}(t) \,\bigl(x_+(t) - x_-(t)\bigr) \biggr], \label{eq:fp_det_lor_bg}
	\end{align}
	emphasizing its dependence on the background solution $\bar{x}$. The outcome of such a transformation is as follows. The integrand of the generating functional in the form (\ref{eq:gen_fun_lor_bg_av}) has the dependence on $\tau_0$ only as the imaginary time shift of the saddle point solution $\bar{x}$. More specifically, the complex time contour is fixed now (cf. boundary conditions in the functional measure) and corresponds to the contour $C$, represented at Fig.~\ref{fig:conotur_conv}, corresponding to $\tau_0=0$.
	
\subsection{Perturbation theory}
	Now, let us develop a perturbative version of the generating functional (\ref{eq:gen_fun_lor_bg_av}). For this purpose, we expand the integrand about the saddle point solution $\bar{x}^{\tau_0}$
	\begin{equation}
		x_e(\tau) = \bar{x}^{\tau_0}_e(\tau) + \eta_e(\tau), \qquad 
		x_\pm(t) = \bar{x}^{\tau_0}(t) + \eta_\pm(t), \label{eq:pert_def_lor}
	\end{equation}
	and go to the integration over the perturbations $\eta_e$,~$\eta_\pm$. Here we have $\bar{x}^{\tau_0}_e(\tau) = \bar{x}(-i(\tau+\tau_0))$, and $\bar{x}^{\tau_0}(t) = \bar{x}(t-i\tau_0)$ as before. For the sake of further convenience we introduce the ``inner product'' $\langle \bullet, \bullet \rangle$ on the functions, defined on the whole contour $C$, e.g. $\boldsymbol{\varphi} = [\varphi_e(\tau), \varphi_+(t), \varphi_-(t)]^T$. The latter functions will be denoted by bold symbols. Having two such functions $\boldsymbol{\varphi}$,~$\boldsymbol{\psi}$, we can define
	\begin{equation}
		\langle \boldsymbol{\varphi}, \boldsymbol{\psi} \rangle = \int_0^\beta d\tau \, \varphi_e(\tau) \, \psi_e(\tau) + \int_0^T dt \, \varphi_+(t) \psi_+(t) + \int_0^T dt \, \varphi_-(t) \psi_-(t).
	\end{equation}
	It will also be natural to introduce the operation $s$ on such functions, multiplying its components by $1$,~$i$, and $-i$, i.e.
	\begin{gather}
		s\boldsymbol{\psi} = \begin{bmatrix}
			\psi_e(\tau) & i\psi_+(t) & -i \psi_-(t)
		\end{bmatrix}^T,\\
		\langle \boldsymbol{\varphi}, s\boldsymbol{\psi} \rangle = \int_0^\beta d\tau \, \varphi_e(\tau) \, \psi_e(\tau) + i \int_0^T dt \, \varphi_+(t) \psi_+(t) - i \int_0^T dt \, \varphi_-(t) \psi_-(t).
	\end{gather}
	Note that the inner product of two functions on the contour $C$ (with the operator $s$ inserted), defined by an analytic continuation similar to (\ref{eq:zero_mode cont}), equals to the usual inner product of its Euclidean parts, since the Lorentzian parts cancel, namely
	\begin{equation}
		\langle \boldsymbol{\varphi}, s\boldsymbol{\psi} \rangle = \int_0^\beta d\tau \, \varphi_e(\tau) \psi_e(\tau) = (\varphi_e, \psi_e), \qquad
		\boldsymbol{\varphi} = \begin{bmatrix}
			\varphi(-i\tau) \\ \varphi(t) \\ \varphi(t)
		\end{bmatrix}, \quad
		\boldsymbol{\psi} = \begin{bmatrix}
			\psi(-i\tau) \\ \psi(t) \\ \psi(t)
		\end{bmatrix}, 
	\end{equation}
	where $\varphi_e(\tau)=\varphi(-i\tau)$,~$\psi_e(\tau) = \psi(-i\tau)$. As a consequence, we have $\langle \boldsymbol{\varphi}, s\boldsymbol{\varphi} \rangle = \| \varphi_e \|^2 $.
%	\textcolor{red}{
%	\begin{itemize}
%		\item Inner product on analytic functions.
%	\end{itemize}}

	In terms of the inner product defined above, the gauge-fixing functional and the corresponding Faddeev-Popov determinant (\ref{eq:gf_fun_lor_bg})--(\ref{eq:fp_det_lor_bg}) can be rewritten as
	\begin{align}
		\chi[\bar{x}^{\tau_0};\bar{x}^{\tau_0}_e+\eta_e, \bar{x}^{\tau_0}+\eta_\pm] &= 
		\int_0^\beta d\tau \, \eta_{0e}^{\tau_0}(\tau) \, \eta_e(\tau) + i \int_0^T dt \, \eta_0^{\tau_0}(t) \,\bigl(\eta_+(t) - \eta_-(t)\bigr) = \langle \boldsymbol{\eta}_0^{\tau_0}, s\boldsymbol{\eta} \rangle \\
		J[\bar{x}^{\tau_0};\bar{x}^{\tau_0}_e+\eta_e, \bar{x}^{\tau_0}+\eta_\pm] &= \|\dot{\bar{x}}\| \, \biggl[ 1 - \frac1{\|\dot{\bar{x}}\|} \Bigl[\int_0^{\beta}\!\!d\tau\, \partial_{\tau} \eta_{0e}^{\tau_0}(\tau) \eta_e(\tau) + \!\int_0^T \!\! dt \, \partial_{t} \eta_0^{\tau_0}(t) \bigl(\eta_+(t) - \eta_-(t)\bigr) \Bigr]  \biggr] \nonumber \\ &= \|\dot{\bar{x}}\| \Bigl[1 - \frac1{\|\dot{\bar{x}}\|} \bigl\langle \partial_{\tau_0} \boldsymbol{\eta}_0^{\tau_0}, s\boldsymbol{\eta} \bigr\rangle \Bigr].
	\end{align}
	where
	\begin{align}
		\boldsymbol{\eta} = \begin{bmatrix}
			\eta_e(\tau) \\ \eta_+(t) \\ \eta_-(t)
		\end{bmatrix}, \qquad
		\boldsymbol{\eta}_0^{\tau_0} = \begin{bmatrix*}[l]
			\eta_{0e}^{\tau_0}(\tau) \\ \eta_0^{\tau_0}(t) \\ \eta_0^{\tau_0}(t)
		\end{bmatrix*} = 
		\frac1{\|\dot{\bar{x}}\|} \begin{bmatrix*}[r]
			\partial_\tau \bar{x}_e^{\tau_0}(\tau) \\ -i\partial_t \bar{x}^{\tau_0}(t) \\  -i\partial_t \bar{x}^{\tau_0}(t)
		\end{bmatrix*}
	\end{align}
%
%\textcolor{red}{
%	\begin{enumerate}
%		\item Introduce ``inner product'' on the contour $C$.
%		\item Rewrite the generating functional in terms of it.
%	\end{enumerate}
%}
	Thus, the generating functional can be written as
	\begin{align}
		Z[j_+, j_-] &= \frac{\|\dot{\bar{x}}\|}{\sqrt{2\pi\xi}}  e^{-S_e[\bar{x}_e]} \\ &\times \int_0^\beta d\tau_0 \, e^{i \int_0^T dt\, (j_+(t) - j_-(t)) \bar{x}^{\tau_0}(t)} \int\limits_{\substack{\eta_+(0)=\eta_e(\beta)\\\eta_-(0)=\eta_e(0)\\\eta_+(T)=\eta_-(T)}} \calD \eta_e \, \calD \eta_+ \, \calD \eta_- \;
		\Bigl[1 - \frac1{\|\dot{\bar{x}}\|} \bigl\langle \partial_{\tau_0} \boldsymbol{\eta}_0^{\tau_0}, s\boldsymbol{\eta} \bigr\rangle \Bigr] \nonumber \\ 
		&\times  \exp\biggl\{ \!  - \boldsymbol{S}_\xi^{\scriptscriptstyle (2)}[\bar{x}^{\tau_0};\boldsymbol{\eta}] + \langle \boldsymbol{\eta}, s \boldsymbol{j} \rangle 
		- S^{\text{int}}_e[\bar{x}_e^{\tau_0},\eta_e] + i S^{\text{int}}[\bar{x}^{\tau_0},\eta_+] - i S^{\text{int}}[\bar{x}^{\tau_0},\eta_-]
		\biggr\} \biggr|_{j_e=0}, \nonumber
	\end{align}
	where we introduce the source $j_e$ corresponding to the perturbation $\eta_e$ for the further convenience, and $\boldsymbol{j} = [j_e(\tau), j_+(t), j_-(t)]^T$. Here the interaction actions read
	\begin{equation}
	\begin{aligned}
		S^{\text{int}}_e[\bar{x}_e^{\tau_0},\eta_e] &= \phantom{-} \sum_{k\ge3} \frac1{k!} \int_{0}^{\beta} d\tau \, V^{(k)}(\bar{x}_e^{\tau_0}(\tau)) \, \eta_e^k(\tau), \\
		S^{\text{int}}[\bar{x}^{\tau_0},\eta_\pm] &= -\sum_{k\ge3} \frac1{k!} \int_0^T dt \, V^{(k)}(\bar{x}^{\tau_0}(t)) \, \eta_\pm^k(t),
	\end{aligned} \label{eq:action_int_lor}
	\end{equation}
	whereas $\boldsymbol{S}_\xi^{\scriptscriptstyle (2)}$ is the sum of the Euclidean and the Lorentzian quadratic actions together with the gauge-fixing term, which can also be rewritten in terms of inner product~$\langle \bullet, \bullet \rangle$, and the differential operator
	\begin{equation}
		\boldsymbol{K}^{\tau_0}=\begin{bmatrix}
			-\partial_\tau^2 + V''(\bar{x}_e^{\tau_0}) & 0 & 0 \\
			0 & \partial_t^2 + V''(\bar{x}^{\tau_0}) & 0 \\
			0 & 0 & \partial_t^2 + V''(\bar{x}^{\tau_0})
		\end{bmatrix},
	\end{equation}
	so that
	\begin{align}
		\boldsymbol{S}_\xi^{\scriptscriptstyle (2)}[\bar{x}^{\tau_0};\boldsymbol{\eta}] &=
		\frac12 \langle\boldsymbol{\eta}, s \boldsymbol{K}^{\tau_0} \boldsymbol{\eta} \rangle + \frac1{2\xi} \langle \boldsymbol{\eta}_0^{\tau_0}, s\boldsymbol{\eta} \rangle^2 \nonumber \\ &+ \frac12 \left[ \eta_e \partial_\tau \eta_e\Bigr|_0^\beta + i \eta_+ \partial_t\eta_+\Bigr|_0^T - i \eta_- \partial_t\eta_-\Bigr|_0^T \right], \label{eq:action_quad_full}
	\end{align}
	where the boundary terms arise because of the integration by parts.
%	\textcolor{red}{
%	\begin{itemize}
%		\item Refer to boundary conditions (\ref{eq:pert_bc}).
%	\end{itemize}
%	}
	To calculate the integral perturbatively, one should be able to solve the saddle point equation for the quadratic part of the action, which has the following form
	\begin{equation}
		\boldsymbol{K}_\xi^{\tau_0} \boldsymbol{\eta} = \boldsymbol{j}, \qquad
		\boldsymbol{K}_\xi^{\tau_0} = \boldsymbol{K}^{\tau_0} + \frac1\xi \boldsymbol{\eta} \, \langle \boldsymbol{\eta}_0^{\tau_0} , s\bullet \rangle 
		\label{eq:quadratic_eom_reg}
	\end{equation}
%	\begin{equation}
%		\begin{bmatrix}
%			-\partial_\tau^2 + V''(\bar{x}_e^{\tau_0}) & & \\
%			& \partial_t^2 + V''(\bar{x}^{\tau_0}) & \\
%			& & \partial_t^2 + V''(\bar{x}^{\tau_0})
%		\end{bmatrix}
%		\begin{bmatrix}
%			\eta_{e}(\tau) \\ \eta_{+}(t) \\ \eta_{-}(t)
%		\end{bmatrix} + \frac1\xi 
%		\begin{bmatrix}
%			\eta_{0e}^{\tau_0}(\tau) \\ \eta_{0+}^{\tau_0}(t)  \\ \eta_{0-}^{\tau_0}(t)
%		\end{bmatrix} \langle \boldsymbol{\eta}_0^{\tau_0} , s\boldsymbol{\eta}\rangle
%		= 
%		\begin{bmatrix}
%			j_{e}(\tau) \\ j_{+}(t) \\ j_{-}(t)
%		\end{bmatrix}, \label{eq:quadratic_eom_reg}
%	\end{equation}
	where the second term on l.h.s. has the meaning of the projection operator onto zero mode, and arises as a result of the gauge-fixing procedure, in contrast to (\ref{eq:quadratic_eom}).
	The Green's function $\boldsymbol{G}_\xi^{\tau_0}$, solving the equation above as $\boldsymbol{\eta} = \boldsymbol{G}_\xi^{\tau_0} s \boldsymbol{j}$, is constructed in Appendix~\ref{sec:app_zero_mode_gf_lor}. The latter Green's function allows to formally integrate over the perturbations $\eta_e$,~$\eta_\pm$, and rewrite the generating functional in the form, useful for the construction of a diagrammatic technique, which is one of the our main aims
	\begin{align} \label{eq:gen_fun_lor_wick}
		Z&[j_+, j_-] = Z_{\text{1-loop}} \frac1\beta \int_0^\beta d\tau_0 \; e^{i \int_0^T dt\, (j_+(t) - j_-(t)) \bar{x}^{\tau_0}(t)} \; \exp\biggl\{\frac12 \Bigl\langle\frac{\delta}{\delta \boldsymbol{\eta}}, \boldsymbol{G}_\xi^{\tau_0}\frac{\delta}{\delta \boldsymbol{\eta}} \Bigr\rangle\biggr\} \;
		 \\ 
		&\times  \Bigl[1 - \frac1{\|\dot{\bar{x}}\|} \bigl\langle \partial_{\tau_0} \boldsymbol{\eta}_0^{\tau_0}, s\boldsymbol{\eta} \bigr\rangle \Bigr] \; \exp\biggl\{
		- S^{\text{int}}_e[\bar{x}_e^{\tau_0},\eta_e] + i S^{\text{int}}[\bar{x}^{\tau_0},\eta_+] - i S^{\text{int}}[\bar{x}^{\tau_0},\eta_-]
		+ \langle \boldsymbol{\eta}, s \boldsymbol{j} \rangle \biggr\} \biggr|_{\substack{j_e=0\\\,\boldsymbol{\eta}=0}}, \nonumber
	\end{align}
	where $Z_{\text{1-loop}}$ is the same as in the pure Euclidean case (\ref{eq:part_fun_1_loop}), since the Lorentzian parts cancel each other in the Gaussian integral over perturbations.
%\textcolor{red}{
%	\begin{itemize}
%		\item Write down expression for the generating functional of real-time correlation functions in terms of Green's functions.
%	\end{itemize}
%}

\section{Diagrammatic technique} \label{sec:diagrams}

	To begin with, we describe a diagrammatic technique, following from the generating functional (\ref{eq:gen_fun_eucl_wick}) of the imaginary-time correlation functions. After the explanation of the main features of the underlying diagrammatic technique which are mostly the same in Euclidean and Lorentzian cases, we generalize it to the real-time correlation functions~(\ref{eq:gen_fun_lor_wick}).
	
\subsection{Warming up: Euclidean case}
	First of all, let us note that the generating functional (\ref{eq:gen_fun_eucl_wick}) contains both background and perturbative parts since the source $j$ has the sum of background $\bar{x}^{\tau_0}$ and the perturbation $\eta$ as a multiplier. One can separate these ones so the generating functional can be rewritten as\footnote{In this subsection, we omit the subscript $e$ of background solution $\bar{x}^{\tau_0}$, zero mode $\eta_0^{\tau_0},$ etc., emphasizing its Euclidean nature, as we do in Section~\ref{sec:imaginary_time_corr}, and in contrast to Section~\ref{sec:real_time_corr}.}
	\begin{align}
		&Z[j] = Z_{\text{1-loop}} \frac1\beta \int_0^\beta d\tau_0 \; e^{\int_0^\beta d\tau j(\tau) \bar{x}^{\tau_0}(\tau)} \;  Z_{\text{pert}}^{\tau_0}[j], \\
		&Z_{\text{pert}}^{\tau_0}[j] = e^{\frac12 \int d\tau\, d\tau' \, \frac{\delta}{\delta \eta(\tau)} G^{\tau_0}_\xi(\tau,\tau') \frac{\delta}{\delta \eta(\tau')}}  \Bigl[1 - \frac1{\|\dot{\bar{x}}\|} \int_{0}^\beta d\tau \, \dot\eta_0^{\tau_0}(\tau) \eta(\tau)\Bigr] e^{- S^{\text{int}}[\bar{x}^{\tau_0}, \eta] + \int_0^\beta d\tau \, j(\tau) \eta(\tau)}\Bigr|_{\eta=0}
	\end{align}
	Comparing the result to the expression (\ref{eq:corr_fun_eucl}) for the correlation function through the generating functional, one concludes that the process of perturbative calculation of the correlation function splits into three parts: one first separates background and perturbative parts of the correlation function, then calculates the perturbative part, using the generating functional $Z_{\text{pert}}^{\tau_0}$ of correlation functions of the perturbations, and finally averages over the imaginary time shift parameter $\tau_0$. For the further convenience, let us define the correlation functions of the perturbation as
	\begin{equation}
		\langle \eta(\tau_1) \ldots \eta(\tau_n) \rangle^{\tau_0} = \frac1{Z_{\text{pert}}^{\tau_0}}\frac{\delta^n Z_{\text{pert}}^{\tau_0}[j]}{\delta j(\tau_1) \ldots \delta j(\tau_n)}\biggl|_{j=0}, \label{eq:corr_fun_eucl_pert}
	\end{equation}
	so that the algorithm above gives the following result e.g. for 1-, 2-, and 3-point correlation functions (\ref{eq:corr_fun_eucl})
	\begin{subequations}
	\begin{align}
		&D_1(\tau_1) = \frac1\beta \int_0^\beta d\tau_0 \Bigl[ \bar{x}^{\tau_0}(\tau_1) + \langle \eta(\tau_1) \rangle^{\tau_0} \Bigr], \label{eq:corr_fun_eucl_1pt}\\
		&\begin{aligned}D_2(\tau_1, \tau_2) = \frac1\beta \int_0^\beta d\tau_0 \Bigl[& \bar{x}^{\tau_0}(\tau_1) \bar{x}^{\tau_0}(\tau_2) \\ &+ \bar{x}^{\tau_0}(\tau_1) \langle \eta(\tau_2) \rangle^{\tau_0} + \bar{x}^{\tau_0}(\tau_2) \, \langle \eta(\tau_1) \rangle^{\tau_0} + \langle \eta(\tau_1) \eta(\tau_2) \rangle^{\tau_0}  \Bigr],
		\end{aligned} \\
		&\begin{aligned}
			D_3(\tau_1,{}& \tau_2, \tau_3) = \frac1\beta \int_0^\beta d\tau_0 \Bigl[ \bar{x}^{\tau_0}(\tau_1) \, \bar{x}^{\tau_0}(\tau_2) \, \bar{x}^{\tau_0}(\tau_3) \\ &+ \bar{x}^{\tau_0}(\tau_1) \bar{x}^{\tau_0}(\tau_2)\, \langle \eta(\tau_3) \rangle^{\tau_0} + \bar{x}^{\tau_0}(\tau_1) \bar{x}^{\tau_0}(\tau_3) \, \langle \eta(\tau_2) \rangle^{\tau_0}  + \bar{x}^{\tau_0}(\tau_2) \bar{x}^{\tau_0}(\tau_3) \, \langle \eta(\tau_1) \rangle^{\tau_0}\\ 
			&+ \bar{x}^{\tau_0}(\tau_1) \, \langle \eta(\tau_2) \eta(\tau_3) \rangle^{\tau_0} + \bar{x}^{\tau_0}(\tau_2) \, \langle \eta(\tau_1) \eta(\tau_3) \rangle^{\tau_0} + \bar{x}^{\tau_0}(\tau_3) \, \langle \eta(\tau_1) \eta(\tau_2) \rangle^{\tau_0} \\ &+ \langle \eta(\tau_1) \eta(\tau_2) \eta(\tau_3) \rangle^{\tau_0}  \Bigr]. \label{eq:corr_fun_eucl_3pt}
		\end{aligned}
	\end{align}
	\end{subequations}
	The generalization to $n$-point correlation functions is obvious.
	
	Now, let us focus on the calculation of the correlation functions of the perturbations (\ref{eq:corr_fun_eucl_pert}), which is the part of the total correlation functions (\ref{eq:corr_fun_eucl}). For this purpose we derive a diagram technique, following from the generating functional (\ref{eq:gen_fun_eucl_pert}). The explicit form of the $n$-point correlation function of the perturbations, following from (\ref{eq:corr_fun_eucl_pert}), reads
	\begin{multline}
		\langle \eta(\tau_1) \ldots \eta(\tau_n) \rangle^{\tau_0} = e^{\frac12 \int d\tau\, d\tau' \, \frac{\delta}{\delta \eta(\tau)} G^{\tau_0}_\xi(\tau,\tau') \frac{\delta}{\delta \eta(\tau')}} \\ \times \eta(\tau_1) \ldots  \eta(\tau_n)\, \Bigl[1 - \frac1{\|\dot{\bar{x}}\|} \int_{0}^\beta d\tau \, \dot\eta_0^{\tau_0}(\tau) \eta(\tau)\Bigr] e^{- S^{\text{int}}[\bar{x}^{\tau_0}, \eta]}\Bigr|_{\substack{\eta=0\hspace{2.5em}\\\text{non-vacuum}}}. \label{eq:corr_fun_pert_eucl}
	\end{multline}
	where we omit the denominator $Z_{\text{pert}}^{\tau_0}$, and simultaneously exclude all terms on r.h.s., having vacuum diagrams (i.e. factors independent of external time points $\tau_i$,~$i=1,\ldots n$) as a multiplier, because the standard statement on vacuum diagrams cancellation is valid due to its $\tau_0$-independence. The variational operator gives the Wick contractions of the terms dependent of $\eta$, which are represented by the lines, connecting vertices and external points, where the latter simply equal to unity as in the absence of spin
	\begin{align}
		\vcenter{\hbox{\includegraphics[scale=1.25]{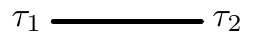}}}
		 = \; G^{\tau_0}(\tau_1, \tau_2), \qquad\qquad
		\vcenter{\hbox{\includegraphics[scale=1.25]{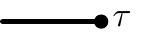}}} = \; 1.
	\end{align}
	The interaction action $S^{\text{int}}$, obtained by the expansion (\ref{eq:action_int_eucl}) of the potential $V$ about the saddle point solution $\bar{x}^{\tau_0}$, gives the vertices, which can be in principle of arbitrary order $k$, if the potential is non-polynomial. Vertices are represented as
	\begin{equation}
		\vcenter{\hbox{\includegraphics[scale=1.25]{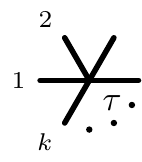}}} \;\; =  \; -
		\frac1{k!} \, V^{(k)}(\bar{x}^{\tau_0}(\tau)).
	\end{equation}
	One of the main differences compared to the conventional diagram techniques is the presence of the second term in the squared brackets in (\ref{eq:corr_fun_pert_eucl}), emerging from the Faddeev-Popov determinant and giving an additional diagram element, namely, a one-point vertex, appearing at most once in each diagram. The corresponding diagram element reads
	\begin{equation}
			\vcenter{\hbox{\includegraphics[scale=1.25]{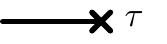}}}  = \; -\frac1{\|\dot{\bar{x}}\|} \dot \eta_0^{\tau_0}(\tau). \label{eq:1pt_vertex_eucl}
	\end{equation}
	To make things clear, let us write down some explicit expressions for the correlation functions, appearing in (\ref{eq:corr_fun_eucl_1pt})--(\ref{eq:corr_fun_eucl_3pt}), both in terms of diagrams and the corresponding integrals. For the 1-point correlation function of the perturbations we have the following diagrammatic expansion\footnote{We will formally assign one additional loop to diagrams with the one-point vertex~(\ref{eq:1pt_vertex_eucl}), since usually the one-point vertex turns out to be proportional to the coupling constant \cite{Lowe:1978ug}.}
	\begin{equation}
		\langle \eta(\tau) \rangle^{\tau_0} = 
		\begin{minipage}[c]{2.3cm}\includegraphics[scale=1.25]{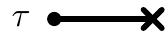}\end{minipage}  + 
		\begin{minipage}[c]{2.4cm}\includegraphics[scale=1.25]{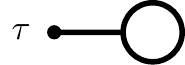}\end{minipage} \;\;+\;\; \text{higher loops},
	\end{equation}
	where
	\begin{align}
		&\begin{minipage}[c]{2.4cm}\includegraphics[scale=1.25]{1pt_1_E.pdf}\end{minipage} \;\; = \; 
		-\frac1{\|\dot{\bar{x}}\|}\int_0^\beta d\tau_1 \, G^{\tau_0}(\tau, \tau_1) \dot{\eta}^{\tau_0}_0(\tau_1), \\ 
		&\begin{minipage}[c]{2.4cm}\includegraphics[scale=1.25]{1pt_2_E.pdf}\end{minipage} \;\;= \;
		-\frac12 \int_0^\beta d\tau_1 \, V^{(3)}(\bar{x}^{\tau_0}(\tau_1)) G^{\tau_0}(\tau, \tau_1) G^{\tau_0}(\tau_1, \tau_1).
	\end{align}
	The result for the 2-point function reads
	\begin{multline}
		\langle \eta(\tau) \eta(\tau') \rangle^{\tau_0} = \begin{minipage}[t][0.14cm][b]{3.376cm}\includegraphics[scale=1.25]{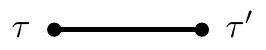} \end{minipage} + 
		\begin{minipage}[t][0.14cm][b]{3.376cm}\includegraphics[scale=1.25]{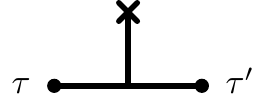}\end{minipage} + 
		\begin{minipage}[t][0.14cm][b]{3.376cm}\includegraphics[scale=1.25]{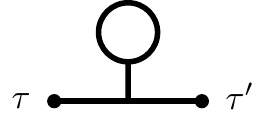} \end{minipage} \\	+ 
		\begin{minipage}[c][2cm][c]{3.376cm}\includegraphics[scale=1.25]{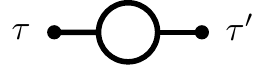}\end{minipage} + 
		\begin{minipage}[t][0.14cm][b]{3.376cm}\includegraphics[scale=1.25]{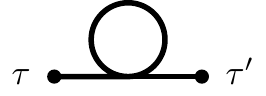}\end{minipage}+ \text{higher loops},
	\end{multline}
	where, for example
	\begin{align}
		&\begin{aligned}
			\begin{minipage}[b][1.2cm][t]{3.376cm}\includegraphics[scale=1.25]{2pt_2_E.pdf} \end{minipage} = \frac1{\|\dot{\bar{x}}\|} \int_0^\beta d\tau_1 \int_0^\beta d\tau_2 \;& V^{(3)}(\bar{x}^{\tau_0}(\tau_1)) \dot{\eta}^{\tau_0}_0(\tau_2) \\ &\times  G^{\tau_0}(\tau, \tau_1) G^{\tau_0}(\tau', \tau_1) G^{\tau_0}(\tau_1,\tau_2),
		\end{aligned}\\
		&\begin{minipage}[t][0.14cm][b]{3.376cm}\includegraphics[scale=1.25]{2pt_5_E.pdf}\end{minipage} = -\frac12 \int_0^\beta d\tau_1 \; V^{(4)}(\bar{x}^{\tau_0}(\tau_1)) \, G^{\tau_0}(\tau, \tau_1) G^{\tau_0}(\tau', \tau_1) G^{\tau_0}(\tau_1,\tau_1).
	\end{align}
	To sum up, let us emphasize the following differences in our construction compared to the conventional diagrammatic techniques, namely, the appearance  of the one-point vertex (\ref{eq:1pt_vertex_eucl}), and contribution of the background solution together with the subsequent averaging over imaginary time parameter of the background shift (\ref{eq:corr_fun_eucl_1pt})--(\ref{eq:corr_fun_eucl_3pt}).
	
\subsection{Lorentzian case} \label{sec:sub_diagr_lor}
	In full analogy to the Euclidean case we start our derivation of the diagrammatic technique for the real-time correlation functions by splitting the generating functional into background and perturbative parts. Specifically, we rewrite the generating functional (\ref{eq:gen_fun_lor_wick}) as
	\begin{align}
		&Z[j_+, j_-] = Z_{\text{1-loop}} \frac1\beta \int_0^\beta d\tau_0 \; e^{i \int_0^T dt\, (j_+(t) - j_-(t)) \bar{x}^{\tau_0}(t)} \; Z_{\text{pert}}^{\tau_0}[j_+, j_-], \\
		&Z_{\text{pert}}^{\tau_0}[j_+, j_-] = \exp\biggl\{ \frac12 \Bigl\langle\frac{\delta}{\delta \boldsymbol{\eta}}, \boldsymbol{G}_\xi^{\tau_0}\frac{\delta}{\delta \boldsymbol{\eta}} \Bigr\rangle\biggr\} \;
		\label{eq:gen_fun_lor_pert}\\ 
		&\quad\times  \Bigl[1 - \frac1{\|\dot{\bar{x}}\|} \bigl\langle \partial_{\tau_0} \boldsymbol{\eta}_0^{\tau_0}, s\boldsymbol{\eta} \bigr\rangle \Bigr] \; \exp\biggl\{
		- S^{\text{int}}_e[\bar{x}_e^{\tau_0},\eta_e] + i S^{\text{int}}[\bar{x}^{\tau_0},\eta_+] - i S^{\text{int}}[\bar{x}^{\tau_0},\eta_-]
		+ \langle \boldsymbol{\eta}, s \boldsymbol{j} \rangle \biggr\} \biggr|_{\substack{j_e=0\\\,\boldsymbol{\eta}=0}}, \nonumber
	\end{align}
	where $Z_{\text{pert}}^{\tau_0}$ allows us to define the correlation functions of the perturbations as
		\begin{align}
		\langle \eta_-(t_1^-) \ldots \eta_-(t_m^-) \eta_+(t_1^+) \ldots \eta_+(t_n^+) \rangle^{\tau_0} = \frac{i^{m-n}}{Z_{\text{pert}}^{\tau_0}} \, \frac{\delta^{n+m} Z_{\text{pert}}^{\tau_0}[j_+,j_-]}{\delta j_+(t_1^+) \ldots \delta j_+(t_n^+) \, \delta j_-(t_1^-) \ldots \delta j_-(t_m^-)}\biggl|_{j_\pm=0}. \label{eq:corr_fun_lor_pert}
	\end{align}
	Thus, from the definition (\ref{eq:corr_fun_lor}) of the real-time correlation functions, we obtain the following expressions, e.g. for 1- and 2-point functions, which now have several different types
	\begin{subequations}
	\begin{align}
		&D_{1,0}(t_1^+) = \frac1\beta \int_0^\beta d\tau_0 \Bigl[ \bar{x}^{\tau_0}(t_1^+) + \langle \eta_+(t_1^+) \rangle^{\tau_0} \Bigr],  \label{eq:1pt_lor_1} \\
		&D_{0,1}(t_1^-) = \frac1\beta \int_0^\beta d\tau_0 \Bigl[ \bar{x}^{\tau_0}(t_1^-) + \langle \eta_-(t_1^-) \rangle^{\tau_0} \Bigr], \\
		&\begin{aligned}D_{1,1}(t_1^+;t_1^-) = \frac1\beta \int_0^\beta & d\tau_0 \Bigl[ \bar{x}^{\tau_0}(t_1^+) \bar{x}^{\tau_0}(t_1^-) \\ &+ \bar{x}^{\tau_0}(t_1^+) \langle \eta_-(t_1^-) \rangle^{\tau_0} + \bar{x}^{\tau_0}(t_1^-) \, \langle \eta_+(t_1^+) \rangle^{\tau_0} + \langle \eta_-(t_1^-) \eta_+(t_1^+) \rangle^{\tau_0}  \Bigr],
		\end{aligned} \\
		&\begin{aligned}D_{2,0}(t_1^+,t_2^+) = \frac1\beta \int_0^\beta & d\tau_0 \Bigl[ \bar{x}^{\tau_0}(t_1^+) \bar{x}^{\tau_0}(t_2^+) \\ &+ \bar{x}^{\tau_0}(t_1^+) \langle \eta_+(t_2^+) \rangle^{\tau_0} + \bar{x}^{\tau_0}(t_2^+) \, \langle \eta_+(t_1^+) \rangle^{\tau_0} + \langle \eta_+(t_1^+) \eta_+(t_2^+) \rangle^{\tau_0}  \Bigr],
		\end{aligned}\\
		&\begin{aligned}D_{0,2}(t_1^-,t_2^-) = \frac1\beta \int_0^\beta & d\tau_0 \Bigl[ \bar{x}^{\tau_0}(t_1^-) \bar{x}^{\tau_0}(t_2^-) \\ &+ \bar{x}^{\tau_0}(t_1^-) \langle \eta_-(t_2^-) \rangle^{\tau_0} + \bar{x}^{\tau_0}(t_2^-) \, \langle \eta_-(t_1^-) \rangle^{\tau_0} + \langle \eta_-(t_1^-) \eta_-(t_2^-) \rangle^{\tau_0}  \Bigr]. \label{eq:2pt_lor_3}
		\end{aligned}
	\end{align}
	\end{subequations}
	
	To calculate the correlation functions (\ref{eq:corr_fun_lor_pert}), we propose a diagrammatic technique, which follows from the generating functional (\ref{eq:gen_fun_lor_pert}). Similar to the Euclidean case, described in the previous subsection, we use the definition (\ref{eq:corr_fun_lor_pert}) to write correlation functions in the following form
	\begin{multline}
		\langle  \eta_-(t_1^-) \ldots \eta_-(t_m^-)  \eta_+(t_1^+) \ldots \eta_+(t_n^+)  \rangle^{\tau_0} = 
		\exp\biggl\{ \frac12 \Bigl\langle\frac{\delta}{\delta \boldsymbol{\eta}}, \boldsymbol{G}_\xi^{\tau_0}\frac{\delta}{\delta \boldsymbol{\eta}} \Bigr\rangle\biggr\}\\
		 \times \eta_-(t_1^-) \ldots \eta_-(t_m^-) \eta_+(t_1^+) \ldots \eta_+(t_1^+) 
		\Bigl[1 - \frac1{\|\dot{\bar{x}}\|} \bigl\langle \partial_{\tau_0} \boldsymbol{\eta}_0^{\tau_0}, s\boldsymbol{\eta} \bigr\rangle \Bigr] \label{eq:corr_fun_lor_expl} \\ 
		\times \exp\biggl\{
		- S^{\text{int}}_e[\bar{x}_e^{\tau_0},\eta_e] + i S^{\text{int}}[\bar{x}^{\tau_0},\eta_+] - i S^{\text{int}}[\bar{x}^{\tau_0},\eta_-] \biggr\} \biggr|_{\substack{\boldsymbol{\eta}=0\hspace{2.5em}\\\text{non-vacuum}}},
	\end{multline}
	where we omit the denominator, and exclude the vacuum diagrams on r.h.s. Now, we have 9 different types of contractions, coming from the variational term, because of a tripled set of fields, namely $\eta_e$,~$\eta_+$, and $\eta_-$. In fact, at least 3 of them are not independent due to transposition symmetry of $\boldsymbol{G}_\xi^{\tau_0}$, i.e.
	\begin{equation}
		G_{\xi+e}^{\tau_0}(t, \tau') = G_{\xi e+}^{\tau_0}(\tau', t), \qquad
		G_{\xi -e}^{\tau_0}(t, \tau') = G_{\xi e-}^{\tau_0}(\tau', t), \qquad G_{\xi -+}^{\tau_0}(t, t') = G_{\xi +-}^{\tau_0}(t', t).
	\end{equation}
	We will denote diagrammatic elements, corresponding to $\eta_e$ by dotted lines, whereas the elements, corresponding to $\eta_+$ and $\eta_-$ will be denoted by solid lines with an inward and outward arrow, respectively. Thus, we have the following types of internal lines 
	\begin{equation}
		\begin{aligned}
			\vcenter{\hbox{\includegraphics[scale=1]{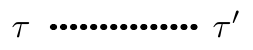}}}
			&= \; G_{\xi ee}^{\tau_0}(\tau, \tau'), & \vcenter{\hbox{\includegraphics[scale=1]{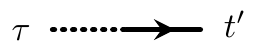}}}
			&= \; G_{\xi e+}^{\tau_0}(\tau, t'), & \vcenter{\hbox{\includegraphics[scale=1]{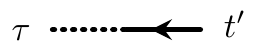}}}
			&= \; G_{\xi e-}^{\tau_0}(\tau, t'), \\
			\vcenter{\hbox{\includegraphics[scale=1]{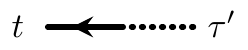}}}
			&= \; G_{\xi+e}^{\tau_0}(t, \tau'), & \vcenter{\hbox{\includegraphics[scale=1]{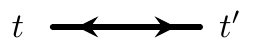}}}
			&= \; G_{\xi++}^{\tau_0}(t, t'), & \vcenter{\hbox{\includegraphics[scale=1]{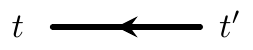}}}
			&= \; G_{\xi+-}^{\tau_0}(t, t'), \\
			\vcenter{\hbox{\includegraphics[scale=1]{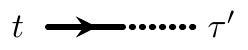}}}
			&= \; G_{\xi-e}^{\tau_0}(t, \tau'), & \vcenter{\hbox{\includegraphics[scale=1]{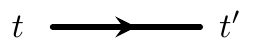}}}
			&= \; G_{\xi-+}^{\tau_0}(t, t'), & \vcenter{\hbox{\includegraphics[scale=1]{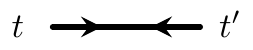}}}
			&= \; G_{\xi--}^{\tau_0}(t, t').
		\end{aligned}
	\end{equation}
	The explicit form of the components of $\boldsymbol{G}_\xi^{\tau_0}$ is given in Appendix~\ref{sec:app_zero_mode_gf_lor}. The external points simply equal to identity
	\begin{align}
%		\vcenter{\hbox{\includegraphics[scale=1.25]{outer_e_lor.pdf}}} &= 1, &
		\vcenter{\hbox{\includegraphics[scale=1.25]{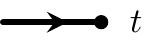}}} &= 1, &
		\vcenter{\hbox{\includegraphics[scale=1.25]{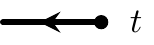}}} &= 1.
	\end{align}
	The term in square brackets in (\ref{eq:corr_fun_lor_expl}) give rise to 1-point vertices, having the following form	
	\begin{align}
		\vcenter{\hbox{\includegraphics[scale=1]{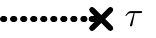}}}  &=  -\frac1{\|\dot{\bar{x}}\|} \partial_{\tau_0} \eta_{0e}^{\tau_0}(\tau), & 
		\vcenter{\hbox{\includegraphics[scale=1]{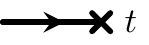}}}  &=  -\frac{i}{\|\dot{\bar{x}}\|} \partial_{\tau_0} \eta_{0}^{\tau_0}(t),  & 
		\vcenter{\hbox{\includegraphics[scale=1]{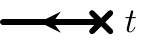}}}  &=  +\frac{i}{\|\dot{\bar{x}}\|} \partial_{\tau_0} \eta_{0}^{\tau_0}(t).
	\end{align}
	Finally, interaction actions (\ref{eq:action_int_lor}) lead to three types of $k$-point vertices
	\begin{equation}
	\begin{aligned}
		\vcenter{\hbox{\includegraphics[scale=1]{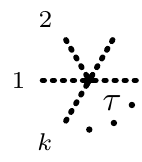}}} \; &= -
		\frac1{k!} \, V^{(k)}(\bar{x}_e^{\tau_0}(\tau)), &
		\vcenter{\hbox{\includegraphics[scale=1]{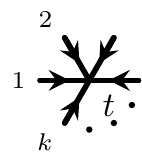}}} \; &= -
		\frac{i}{k!} \, V^{(k)}(\bar{x}^{\tau_0}(t)), &
		\vcenter{\hbox{\includegraphics[scale=1]{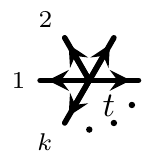}}} \; &= +
		\frac{i}{k!} \, V^{(k)}(\bar{x}^{\tau_0}(t)).	
	\end{aligned}
	\end{equation}
	As an example, we provide the first few terms of diagrammatic expansion, for 1-point correlation function of perturbations $\langle\eta_+(t) \rangle^{\tau_0}$, appearing in (\ref{eq:1pt_lor_1})--(\ref{eq:2pt_lor_3})
	\begin{align}
		\langle \eta_+(t) \rangle^{\tau_0} = 
		\begin{minipage}[c]{2.4cm}\includegraphics[scale=1.25]{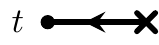}\end{minipage} &+ 
		\begin{minipage}[c]{2.4cm}\includegraphics[scale=1.25]{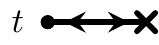}\end{minipage} +
		\begin{minipage}[c]{2.4cm}\includegraphics[scale=1.25]{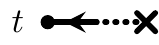}\end{minipage} \nonumber\\ +
		\begin{minipage}[c]{2.4cm}\includegraphics[scale=1.25]{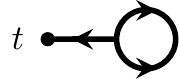}\end{minipage} &+ 
		\begin{minipage}[c]{2.4cm}\includegraphics[scale=1.25]{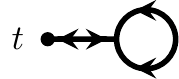}\end{minipage} + 
		\begin{minipage}[c]{2.4cm}\includegraphics[scale=1.25]{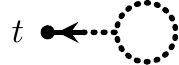}\end{minipage} 
		+ \; \text{higher loops},
	\end{align}
	together with the corresponding explicit expressions, e.g. for
	\begin{align}
		\begin{minipage}[c]{2.4cm}\includegraphics[scale=1.25]{1pt_1_lor.pdf}\end{minipage} &= +\frac{i}{\|\dot{\bar{x}}\|} \int_0^T dt_1 \, \partial_{\tau_0} \eta_{0}^{\tau_0}(t_1) \, G_{\xi+-}^{\tau_0}(t, t_1), \\
		\begin{minipage}[c]{2.4cm}\includegraphics[scale=1.25]{1pt_2_lor.pdf}\end{minipage} &= -\frac{i}{\|\dot{\bar{x}}\|} \int_0^T dt_1 \, \partial_{\tau_0} \eta_{0}^{\tau_0}(t_1) \, G_{\xi++}^{\tau_0}(t, t_1),\\
		\begin{minipage}[c]{2.4cm}\includegraphics[scale=1.25]{1pt_6_lor.pdf}\end{minipage} &= -\frac12\int_0^\beta d\tau_1 \, V^{(3)}(\bar{x}_e^{\tau_0}(\tau_1)) \, G_{\xi+e}^{\tau_0}(t,\tau_1) \, G_{\xi ee}^{\tau_0}(\tau_1,\tau_1).
	\end{align}

	The expression for 2-point correlation function of perturbations, such as $\langle \eta_+(t_1^+) \eta_+(t_2^+) \rangle^{\tau_0}$, has 31 diagrams in the expansion up to one loop, which is rather cumbersome. In principle, there is a special change of variables, referred as the Keldysh rotation \cite{Keldysh:1964ud,Arseev:2015}, exploiting the fact that some combinations of the Green's functions, namely, $G_{\xi e+}^{\tau_0} - G_{\xi e-}^{\tau_0}$ and $G_{\xi++}^{\tau_0}-G_{\xi +-}^{\tau_0}-G_{\xi -+}^{\tau_0}+G_{\xi --}^{\tau_0}$, vanish, that potentially makes diagram technique much simpler. However, this transformation simultaneously enlarge the number of vertex types. We describe the corresponding diagrammatic technique in Appendix~\ref{sec:app_keldysh_rot}.

\section{Discussion} \label{sec:discussion}
%	\textcolor{red}{ \begin{enumerate}
%		\item Resonant states and false vacuum decay
%		\item Real-time approach to false vacuum decay and Lefschets thimble approach 
%		\item Possible obstructions and modifications in non-analytic cases
%		\item Thermalization and two-component technique
%		\item OTOCs and chaos in instantonic systems.
%		\item Solid state system, admitting instantons.
%	\end{enumerate}}
%	
	In this paper, we have constructed the perturbation theory for in-in correlation functions for thermal states in the presence of instantons, working directly in real time. The perturbative expansion is encoded in the generating functional (\ref{eq:gen_fun_lor_wick}), and can be recapitulated in terms of the diagram technique, described in Section~\ref{sec:diagrams}. This is the main result of our work. The diagrammatic technique constructed is three-component, since we have a tripled set of fields in the path integral representation of the generating functional. The main difficulty was to appropriately take into account the zero mode, because the Schwinger-Keldysh closed time contour (at least naively) spoils the invariance under imaginary time translations, responsible for this zero mode. We performed an appropriate identical transformation of the generating functional, namely we averaged it over the gluing point of the Lorentzian part of the Schwinger-Keldysh contour to the Euclidean part, so that the mentioned invariance became manifest. As a side effect, we got that such averaging makes the path integral and correlation functions saturated with complex backgrounds, which is rather unusual for real-time correlation functions.
	
	Although we worked in the case of the one mechanical degree of freedom, it is not difficult to extend the underlying computations to many-particle and even field systems (avoiding the divergence issue in the latter case). The only point in our construction, that cannot be generalized to many degrees of freedom in a straightforward way, is our method of the Green's function construction (see Appendix~\ref{sec:app_zero_mode_gf_lor}), which significantly uses the fact that our system is one-dimensional. However, for the many-particle systems, admitting instantons, and having a simple tensor structure, such as a matrix mechanics \cite{Marino:2015yie}, our considerations are still valid.
	
	The diagram technique constructed can be applied to any system in a thermal state, having degenerate minima of the Hamiltonian, thus exhibiting tunneling effects. This situation frequently arises in solid state and spin systems \cite{chudnovsky1997first,garanin1997thermally,liang1998periodic,Zhang:2001wa}.
	Such systems may exhibit a quantum to classical phase transition, first observed in \cite{affleck1981quantum}, in which the trivial saddle becomes dominant over the nontrivial (instantonic) one above the critical temperature.
	Thus, our construction can be applied to the calculation of real-time correlation functions, which are the measures of the critical behavior in the vicinity of the above mentioned phase transitions. At the same time, potentially the most important applications correspond to the cases, where the standard analytic continuation methods from the imaginary time \cite{baym1961determination,Baier:1993yh,Evans:1991ky,Kobes:1990kr} may break down, such as calculation of many-point correlation functions.
	
	Evidently, the technique constructed can be extended without significant changes to the case of out-of-time-order-correlation functions (OTOCs) \cite{larkin1969quasiclassical,Haehl:2017qfl}, which are the common measure of quantum chaos \cite{Rozenbaum:2016mmv,Akutagawa:2020qbj,Kolganov:2022mpe}. However, OTOCs may give a wrong signature of the quantum chaos for potentials having local maxima \cite{Xu:2019lhc,pilatowsky2020positive,Kidd:2020mtu}. At the same time, these maxima serve as a barrier, so the tunnelling effects should be taken into account. Thus, the generalization of our construction in this direction can help in understanding quantum chaos.
	
	Finally, application of our technique to the unstable systems, exhibiting e.g. false vacuum decay \cite{coleman1977fate,*callan1977fate} can be inconsistent, since the underlying process is essentially non-equilibrium, that contradicts to the thermality of the state, though our scheme of calculation is formally the same in this case. Nevertheless, a slight modification can potentially give an alternative direction in modern attempts to treat false vacuum decay in real time \cite{Ai:2019fri}.
	
\section*{Acknowledgments}
	The author would like to express his special thanks to Andrei~O.~Barvinsky for the inspiration of this study and stimulating discussions during all stages of the project. Also the author is indebted to Dmitrii Trunin, Dmitry Galakhov, and Andrew G. Semenov for very useful conversations. This work was supported by the RFBR grant No.20-02-00297.

\appendix
\renewcommand{\thesection}{\Alph{section}}
\renewcommand{\theequation}{\Alph{section}.\arabic{equation}}

%\section{Elliptic functions}
%	\begin{align}
%		&\begin{aligned}
%			K(k) &= \frac{\pi}2\sum_{m=0}^{\infty} \frac{(\tfrac12)_m(\tfrac12)_m}{m! \, m!} k^{2m}\\ &= \sum_{m=0}^{\infty} \frac{(\tfrac12)_m(\tfrac12)_m}{m! \, m!} {k'}^{2m} \Bigl(\ln\frac1{k'}+\psi(1+m) - \psi(\tfrac12+m)\Bigr), \qquad k'=\sqrt{1-k^2}
%		\end{aligned} \\
%		&\sn\bigl( \tfrac{2K(k)}{\pi}\bt,k\bigr) = \frac{2\pi}{k K(k)} \sum_{n=0}^{\infty} \frac{q^{n+1/2}}{1-q^{2n+1}} \sin\bigl((2n+1)\bt\bigr), \qquad q = \exp(-\pi K(k') / K(k))
%	\end{align}
	
\section{Euclidean Green's functions and zero modes} \label{sec:app_zero_mode_gf_eucl}
	In this appendix we consider 1-dimensional mechanical system, admitting a periodic motion. We focus on the linear perturbations about the periodic solution and construct the Green's function for them, carefully resolving a zero mode issue. We mostly follow the construction, presented in \cite{Barvinsky:2011hv}.
	
	Let us consider a theory defined by the Euclidean action
	\begin{equation}
		S[x] = \int_0^\beta d\tau \left[\frac{\dot x^2}2 + V(x) \right]
	\end{equation}
	where a potential $V$ has extrema such that the equations of motion
	\begin{equation}
		\ddot x - V'(x) = 0 \label{eq:eom_nlin}
	\end{equation}
	admit periodic solutions $\bar{x}(\tau)$, i.e. such that
	\begin{equation}
		\bar{x}(0) = \bar{x}(\beta), \qquad \dot{\bar{x}}(0) = \dot{\bar{x}}(\beta), \label{eq:per_bc}
	\end{equation}
	where $\beta$ is a period. Consider a perturbation $\eta$ about $\bar{x}$. It satisfies linear a equation of motion
	\begin{equation}
		\bigl[ -\partial_\tau^2 + V''(\bar{x}(\tau))  \bigr] \eta(\tau) = 0. \label{eq:eom_lin}
	\end{equation}
	Differentiating (\ref{eq:eom_nlin}) evaluated on the solution $\bar{x}$ one obtains
	\begin{equation}
		\bigl[-\partial_\tau^2 + V''(\bar{x}(\tau))\bigr]\dot{\bar{x}}(\tau) = 0,
	\end{equation}
	and finds that $\eta_0 \propto \dot{\bar{x}}$ is the periodic solution for the equation of motion (\ref{eq:eom_lin}) for the perturbation. Moreover, because of periodic boundary conditions (\ref{eq:per_bc}) one finds that $\eta_0$ is also periodic, so that it is a zero mode of the differential operator $K$, defining the equation (\ref{eq:eom_lin}) for perturbation $\eta$
	\begin{equation}
		(K \eta_0)(\tau) = \int_0^\beta d\tau' \, K(\tau, \tau') \eta_0(\tau') = 0, \qquad
		K(\tau, \tau') = \bigl[ -\partial_\tau^2 + V''(\bar{x}(\tau))  \bigr] \delta(\tau-\tau').
	\end{equation}
	Thus, we define the normalized zero mode as
	\begin{equation}
		\eta_0(\tau) = \frac1{\|\dot{\bar{x}}\|} \dot{\bar{x}}(\tau), \qquad
		\|\dot{\bar{x}}\|^2 = \int_0^\beta d\tau \, \bigl(\dot{\bar{x}}(\tau)\bigr)^2.
	\end{equation}
	
	We will construct the Green's function of the operator $K$ using the technique, equivalent to the variation of constants method. For this purpose, we need for the second independent solution $\zeta_{0}$ of the equation (\ref{eq:eom_lin}), which can be obtained from the constancy of the Wronskian
	\begin{equation}
		W[\eta_0, \zeta_0] = \eta_0 \partial_\tau \zeta_0 - \zeta_0 \partial_\tau \eta_0 = 1,
	\end{equation}
	so that 
	\begin{equation}
		\zeta_0(\tau) =  \eta_0(\tau) \int^\tau_0  \frac{d\tau'}{(\eta_0(\tau'))^2}. \label{eq:nonzero_mode}
	\end{equation}
	Note that $\zeta_0$ is finite if $\eta_0$ has only simple zeros.
	
	While the zero mode $\eta_0$ is periodic, the second solution $\zeta_0$ is generally not. Nevertheless, non-periodicity of the latter can be handled in a systematic way. Namely, since $\zeta_0(\tau + \beta)$ is also solution of the homogeneous equation, it equals to a linear combination of $\eta_0(\tau)$ and $\zeta_0(\tau)$
	\begin{equation}
		\zeta_0(\tau+\beta) = \zeta_0(\tau) + \Delta \, \eta_0(\tau),
	\end{equation}
	where $\Delta$ is a constant called monodromy, and the unit coefficient near $\zeta_0$ can be explained as follows. Suppose $\zeta_0(\tau+\beta) = \tilde \Delta \, \zeta_0(\tau) + \Delta \, \eta_0(\tau)$ and apply a chain of identical transformations to the Wronskian
	\begin{equation}
		W[\eta_0(\tau), \zeta_0(\tau)] = W[\eta_0(\tau+\beta), \zeta_0(\tau+\beta)] = W[\eta_0(\tau), \zeta_0(\tau+\beta)] = \tilde \Delta \, W[\eta_0(\tau), \zeta_0(\tau)],
	\end{equation}
	where we use the constancy of $W$, periodicity of $\eta_0$, and vanishing of $W[\eta_0, \eta_0]$ in order. Since $W[\eta_0, \zeta_0] = 1 \ne 0$, we conclude that $\tilde \Delta = 1$.

\subsection{Subtracting zero mode.}
	Consider the inhomogeneous equation with the source term $j$
	\begin{equation}
		\bigl[ -\partial_\tau^2 + V''(\bar{x}(\tau))  \bigr] \eta(\tau) = j(\tau), \label{eq:eom_lin_nh}
	\end{equation}
	supplemented by the periodic boundary conditions
	\begin{equation}
		\eta(0) = \eta(\beta), \qquad \partial_\tau \eta(0) = \partial_\tau \eta(\beta). \label{eq:eom_lin_bc}
	\end{equation}
	Multiplying both sides of the equation by the zero mode $\eta_0$ and integrating from $0$ to $\beta$, one obtains
	\begin{align}
		0 &= \|\dot{\bar{x}}\|\int_0^\beta d\tau \, \bigl[ -\partial_\tau^2 \eta(\tau) + V''(\bar{x}(\tau)) \eta(\tau) - j(\tau)  \bigr] \eta_0(\tau)\\&= 
		\int_0^\beta d\tau \, \bigl[\ddot{\bar{x}}(\tau) - V'(\bar{x}(\tau))\bigr] \partial_\tau \eta(\tau) + \bigl[-\dot{\bar{x}}(\tau) \, \partial_\tau \eta(\tau) + V'(\bar{x}(\tau)) \eta(\tau)\bigr]\Bigr|_0^\beta - \int_0^\beta d\tau \, j(\tau) \, \eta_0(\tau).  \nonumber 
	\end{align}
	The first term on the r.h.s.~vanishes due to the background e.o.m. (\ref{eq:eom_nlin}), the second term is zero due to the periodicity of $\eta$ and $\bar{x}$, so that the source $j$ should be orthogonal to the zero mode
	\begin{equation}
		\int_0^\beta d\tau \, j(\tau) \, \eta_0(\tau) = 0.
	\end{equation}
	Thus, the Green's function for the problem (\ref{eq:eom_lin_nh})--(\ref{eq:eom_lin_bc}) should satisfy the equation
	\begin{equation}
		\bigl[-\partial_\tau^2 + V''(\bar{x})\bigr] G(\tau,\tau') = \delta(\tau-\tau') - \eta_0(\tau) \eta_0(\tau') \label{eq:green_eq}
 	\end{equation}
	with zero mode subtracted.
	
	We will find the Green's function in the following form
	\begin{equation}
		G(\tau, \tau') = G_0(\tau, \tau') + \Omega(\tau, \tau') + \alpha H_{\eta\eta}(\tau, \tau') + \beta H_{\eta\zeta}(\tau, \tau') + \gamma H_{\zeta\zeta}(\tau, \tau')  \label{eq:green_ans}
	\end{equation}
	where $\alpha$, $\beta$, $\gamma$ are the constants to be defined from the boundary conditions, $G_0(\tau, \tau')$ is the Green's function, satisfying
	\begin{equation}
		\bigl[-\partial_\tau^2 + V''(\bar{x})\bigr] G_0(\tau,\tau') = \delta(\tau-\tau')
	\end{equation}
	and is taken to be
	\begin{align}
		&G_0(\tau, \tau') =  G_0^>(\tau,\tau') \, \theta(\tau-\tau') + G_0^<(\tau,\tau') \, \theta(\tau'-\tau), \\
		&G_0^>(\tau,\tau') = \frac12 \bigl(\eta_0(\tau)\zeta_0(\tau') - \zeta_0(\tau) \eta_0(\tau')\bigr) = -G_0^<(\tau,\tau') \label{eq:gf_eucl_wight}
	\end{align}
	whereas the other functions involved read
	\begin{equation} \label{eq:gf_eucl_cons}
		\begin{aligned}
			&\Omega(\tau, \tau') = \omega(\tau) \eta_0(\tau') + \eta_0(\tau) \omega(\tau'), \\
			&H_{\eta\eta}(\tau, \tau') = \eta_0(\tau) \eta_0(\tau'), \\
			&H_{\eta\zeta}(\tau, \tau') = \zeta_0(\tau) \eta_0(\tau') + \eta_0(\tau) \zeta_0(\tau'), \\
			&H_{\zeta\zeta}(\tau, \tau') =  \zeta_0(\tau) \zeta_0(\tau').
		\end{aligned}
	\end{equation}
	Here $\omega(\tau)$ is defined as follows. Substitution of (\ref{eq:green_ans}) to (\ref{eq:green_eq}) gives the following equation on $\omega(\tau)$
	\begin{equation}
		\bigl[-\partial_\tau^2 + V''(\bar{x})\bigr] \omega(\tau) = -\eta_0(\tau).
	\end{equation}
	It can be resolved in the following way. We multiply the equation by $\eta_0$ and subtract from the identity $\omega(\tau) [-\partial_\tau^2 + V''(\bar{x})] \eta_0(\tau) = 0$, and obtain the following equation on
	\begin{equation}
		\frac{d}{d\tau} W[\eta_0, \omega] = (\eta_0(\tau))^2 \quad \leftrightarrow\quad 
		\frac{d}{d\tau}\biggl(\frac{\omega(\tau)}{\eta_0(\tau)}\biggr) = \int_0^\tau d\tau' \, (\eta_0(\tau'))^2,
	\end{equation}
	which can be integrated as
	\begin{align}
		\omega(\tau) &= \eta_0(\tau) \int_0^\tau d\tau' \, \frac1{(\eta_0(\tau'))^2} \int_0^{\tau'} d\tau'' \, (\eta_0(\tau''))^2 \nonumber\\
		&=
		\zeta_0(\tau)\int_0^\tau d\tau' \, (\eta_0(\tau'))^2 - \eta_0(\tau) \int_0^\tau d\tau' \eta_0(\tau') \zeta_0(\tau'),
	\end{align}
	where we use integration by parts in the latter equality.
	
	As noted, the coefficients $\alpha$, $\beta$, $\gamma$ of the ansatz (\ref{eq:green_ans}) should be determined from the boundary conditions. However, there is an additional freedom, since any solution of (\ref{eq:green_eq}) can be shifted by the term, proportional the zero mode with an arbitrary coefficient, namely $\eta_0(\tau) \eta_0(\tau')$. Thus, one should impose the gauge condition, which we choose as
	\begin{equation}
		\int_0^\beta dt \, \eta_0(\tau) \, G(\tau, \tau') = 0. \label{eq:gauge_cond}
	\end{equation}
	
\subsection{Periodic boundary conditions.}
	Let us impose the periodic boundary conditions to find the unknown coefficients in (\ref{eq:green_ans})
	\begin{equation}
		G(\tau+\beta, \tau') = G(\tau), \qquad \partial_\tau G(\tau+\beta, \tau') = \partial_\tau G(\tau).
	\end{equation}
	In order to find the explicit equations, we examine the periodic properties of the constituents of $G$. Namely, for each function of time $F(\tau)$ we define its monodromy $\hat{\Delta} F$ as
	\begin{equation}
		\hat{\Delta} F(\tau) = F(\tau + \beta) - F(\tau).
	\end{equation}
	Thus, we have
	\begin{equation}
		\begin{aligned}
			&\hat{\Delta} G_0(\tau, \tau') = -2 \zeta_0(\tau) \eta_0(\tau') + H_{\eta\zeta}(\tau, \tau') - \frac\Delta2 H_{\eta\eta}(\tau, \tau'), \\
			&\hat{\Delta} \Omega(\tau, \tau') = \zeta_0(\tau) \eta_0(\tau') + \bigl(\Delta - (\eta_0, \zeta_0)\bigr) H_{\eta\eta}(\tau, \tau'), \\
			&\hat{\Delta} H_{\eta\eta}(\tau, \tau') = 0, \\
			&\hat{\Delta} H_{\eta\zeta}(\tau, \tau') = \Delta \, H_{\eta\eta}(\tau, \tau'), \\
			&\hat{\Delta} H_{\zeta\zeta}(\tau, \tau') = \Delta \, H_{\eta\zeta}(\tau, \tau') - \Delta \, \zeta_0(\tau) \eta_0(\tau'), 			
		\end{aligned}
	\end{equation}
	where $(\bullet, \bullet)$ is $L_2$ inner product
	\begin{equation}
		(\eta_0, \zeta_0) = \int_0^\beta d\tau \, \eta_0(\tau) \zeta_0(\tau).
	\end{equation}
	Substituting the monodromies above to the periodicity condition $\Delta G(\tau, \tau')=0$, one fixes $\beta$, $\gamma$ as
	\begin{equation}
		\beta = \frac1{\Delta} (\eta_0, \zeta_0) - \frac12, \qquad 
		\gamma = -\frac1{\Delta}. \label{eq:beta_gamma}
	\end{equation}
	The coefficient $\alpha$ should be fixed from the gauge condition (\ref{eq:gauge_cond}). The corresponding integrals for the constituents of $G$ are as follows
	\begin{equation}
		\begin{aligned}
			&\int_0^\beta d\tau\, \eta_0(\tau) \, G_0(\tau, \tau') = \frac12 \, \zeta_0(\tau') - \frac12 (\eta_0, \zeta_0) \, \eta_0(\tau') - \omega(\tau'), \\
			&\int_0^\beta d\tau\, \eta_0(\tau) \, \Omega(\tau, \tau') = (\eta_0, \omega) \, \eta_0(\tau') + \omega(\tau'),	\\
			&\int_0^\beta d\tau\, \eta_0(\tau) \, H_{\eta\eta}(\tau, \tau') = \eta_0(\tau'),  \\
			&\int_0^\beta d\tau\, \eta_0(\tau) \, H_{\eta\zeta}(\tau, \tau') = \zeta_0(\tau') + (\eta_0, \zeta_0) \, \eta_0(\tau'),  \\
			&\int_0^\beta d\tau\, \eta_0(\tau) \, H_{\zeta\zeta}(\tau, \tau') = (\eta_0, \zeta_0) \, \zeta_0(\tau'),
		\end{aligned}
	\end{equation}
	Substitution to (\ref{eq:gauge_cond}) together with (\ref{eq:beta_gamma}) gives
	\begin{equation}
		\alpha =  (\eta_0, \zeta_0) - (\eta_0, \omega) - \frac1{\Delta} (\eta_0, \zeta_0)^2. \label{eq:alpha}
	\end{equation}

	To solve the regularized version (\ref{eq:gf_eq_eucl_reg}) of the equation (\ref{eq:green_eq}), which we reproduce here for the convenience 
	\begin{equation}
		\bigl[-\partial_\tau^2 + V''(\bar{x})\bigr] G_\xi(\tau, \tau') + \frac1{\xi} \eta_0(\tau) \int_0^\beta d\tau'' \, \eta_0(\tau'') G_\xi(\tau'', \tau') = \delta(\tau-\tau'),
	\end{equation}
	one should shift $G$ by the projection operator onto $\eta_0$ (multiplied by the coefficient $\xi$), which is absent in $G$, i.e.\footnote{Green's functions $G_0$ and $G_{\xi=0}$ should not be confused.}
	\begin{equation}
		G_\xi(\tau, \tau') = G(\tau, \tau') + \xi \eta_0(\tau) \eta_0(\tau').
	\end{equation}

\section{Three-component Green's functions for instantonic systems} \label{sec:app_zero_mode_gf_lor}
	 In Appendix~\ref{sec:app_zero_mode_gf_eucl} we coherently construct the imaginary-time Green's function subject to periodic boundary conditions, using, one might say, ``bottom-up approach''. Instead of doing the same for the many-component Green's function, solving the equation (\ref{eq:quadratic_eom_reg}), we will use the ``top-down approach'', namely we will provide the answer for it, and prove that it is correct.
	  
	 In fact, we will find the many-component Green's function\footnote{Hereinafter in this appendix we omit the superscript $\tau_0$ of the transformed background solution $\bar{x}^{\tau_0}$, corresponding zero mode $\boldsymbol{\eta}_0^{\tau_0}$, operator $\boldsymbol{K}^{\tau_0}$, Green's function $\boldsymbol{G}^{\tau_0}$, etc. for the brevity reasons.}
	 \begin{equation}
	 	\boldsymbol{G} = 
	 	\begin{bmatrix}
	 		G_{ee}(\tau,\tau') & G_{e+}(\tau, t') & G_{e-}(\tau, t') \\
	 		G_{+e}(t,\tau') & G_{++}(t, t') & G_{+-}(t, t') \\
	 		G_{-e}(t,\tau') & G_{-+}(t, t') & G_{--}(t, t') \\
	 	\end{bmatrix}
	 \end{equation}
	 satisfying the equation
	 \begin{equation} \label{eq:gf_many_comp_eq}
	 	s\boldsymbol{K} \boldsymbol{G} = \boldsymbol{1}
	 	 - s \boldsymbol{\eta}_0 \langle \boldsymbol{\eta}_0,  \bullet \rangle,
	\end{equation}
	where
	\begin{equation}
		\boldsymbol{1} =
		\begin{bmatrix}
			\delta(\tau-\tau') & 0 & 0 \\
			0 & \delta(t-t') & 0 \\
			0 & 0 & \delta(t-t').
		\end{bmatrix}
	\end{equation}
	It also should satisfy the boundary conditions (\ref{eq:pert_bc})
 	and the gauge-fixing condition
 	\begin{equation} \label{eq:gf_lor_gauge_cond}
 		\langle \boldsymbol{\eta}_0, s \boldsymbol{G} \, \bullet \rangle = 0.
 	\end{equation}
 	Once $\boldsymbol{G}$ is found, one can define a shifted Green's function $\boldsymbol{G}_\xi$ as 
 	\begin{equation}
 		\boldsymbol{G}_\xi = \boldsymbol{G} + \xi \boldsymbol{\eta}_0  \langle \boldsymbol{\eta}_0, \bullet \rangle, \label{eq:green_lor_split}
 	\end{equation}
 	that solves the equation (\ref{eq:quadratic_eom_reg}) as
 	\begin{equation}
 		\boldsymbol{\eta} = \boldsymbol{G}_\xi s \boldsymbol{j} = \boldsymbol{G} s \boldsymbol{j} + \xi \boldsymbol{\eta}_0  \langle \boldsymbol{\eta}_0, \boldsymbol{j} \rangle. \label{eq:sol_inhom_lor}
 	\end{equation}
 	
 	Let us describe the constituents of the Green's function that we are constructing. The main building block is the zero-mode $\eta_0(z)$ defined in (\ref{eq:zero_mode_full}) and satisfying the equation (\ref{eq:zero_mode_eq_compl}), which we reproduce here for the convenience
 	\begin{equation}
 		\bigl[\partial_z^2 + V''(\bar{x}) \bigr] \eta_0(z) = 0, \qquad \eta_0(z) = -\frac{i}{\|\dot{\bar{x}}\|}\partial_z\bar{x}(z).
 	\end{equation}
 	The second linearly independent solution of this equation can be constructed as
 	\begin{equation}
 		\zeta_0(z) = i \eta_0(z) \int_0^z \frac{dz'}{(\eta_0(z'))^2},
 	\end{equation}
 	Note that $\eta_0(-i\tau)$ is just Euclidean zero mode of the previous section.
 	I will be useful (but not necessary) to assume that $\eta_0$ has an analytic structure, such that $\zeta_0$ is independent of the contour, connecting $z'=0$ and $z'=z$ (e.g. $\eta_0(z)$ is a meromorphic function of $z$). In fact, we only need the existence of the integral over straight lines parallel to real and imaginary axes. Next, we define the function $\omega(z)$, satisfying the equation
 	\begin{eqnarray}
 		\bigl[\partial_z^2 + V''(\bar{x}) \bigr] \omega(z) = -\eta_0(z),
 	\end{eqnarray} 
	which will be needed to obtain $\eta_0$ on r.h.s. of (\ref{eq:gf_many_comp_eq}). The explicit expression for it reads
 	\begin{align}
 		\omega(z) &= -\eta_0(z) \int_0^z dz' \, \frac1{(\eta_0(z'))^2} \int_0^{z'} dz'' \, (\eta_0(z''))^2 \nonumber\\
 		&=i
 		\zeta_0(z)\int_0^z dz' \, (\eta_0(z'))^2 - i\eta_0(z) \int_0^z dz' \eta_0(z') \zeta_0(z'), \label{eq:omega_lor}
 	\end{align}
 	where assumptions on analytic properties are the same as in the $\zeta_0$ case.
 	Next, we introduce the following functions of two complex numbers $z$,~$z'$ as\footnote{Almost the same quantities (\ref{eq:gf_eucl_cons}),~(\ref{eq:gf_eucl_wight}) differ from the ones introduced here by the factor $i$ in all the arguments.}
 	\begin{equation}
	 	\begin{aligned}
	 		&\Omega(z, z') = \omega(z) \eta_0(z') + \eta_0(z) \omega(z') \\
	 		&H(z, z') = \alpha H_{\eta\eta}(z, z') + \beta H_{\eta\zeta}(z, z')+ \gamma H_{\zeta\zeta}(z,z') \\
	 		&H_{\eta\eta}(z, z') = \eta_0(z) \eta_0(z') \\
	 		&H_{\eta\zeta}(z, z') = \eta_0(z) \zeta_0(z') + \zeta_0(z) \eta_0(z') \\
	 		&H_{\zeta\zeta}(z,z') = \zeta_0(z) \zeta_0(z') \\
	 		&G_0^>(z, z') = \frac12 \bigl(\eta_0(z)\zeta_0(z') - \zeta_0(z) \eta_0(z')\bigr) = -G_0^<(z,z')
	 	\end{aligned} \label{eq:gf_cons_def}
 	\end{equation}
 	where $\alpha$,~$\beta$, and $\gamma$ are the same as in the purely Euclidean case, i.e. (\ref{eq:alpha}),~(\ref{eq:beta_gamma}). In terms of the quantities above, the Euclidean Green's function (\ref{eq:green_ans}), which we identify with $G_{ee}$, appearing in (\ref{eq:gf_many_comp_eq}), reads
 	\begin{multline}
 		G_{ee}(\tau,\tau') = G_0^>(-i\tau, -i\tau') \theta(\tau-\tau') +  G_0^<(-i\tau, -i\tau') \theta(\tau'-\tau)\\ + \Omega(-i\tau, -i\tau') + H(-i\tau, -i\tau').
 	\end{multline}
 	We will formally obtain the remaining components of the Green's function, using the continuation of the arguments of $G_{ee}(\tau,\tau')$ from the imaginary axis to the real one. A possible obstacle in this program is the presence of the functions $\theta(\tau-\tau')$,~$\theta(\tau'-\tau)$, which are not analytic. To overcome this difficulty, we replace the difference of imaginary times in its arguments by the difference of natural parameters of the whole contour $C$. Using this prescription, one obtains the Green's function in terms of three components
 	\begin{align}
 		\boldsymbol{G} = \boldsymbol{G}_0 + \boldsymbol{\Omega} + \boldsymbol{H} \label{eq:gf_lor_sum}
 	\end{align}
 	that are defined as
 	\begin{gather}
 		\boldsymbol{G}_0 = \left[\begin{smallmatrix}
 			G_0^>(-i\tau, -i\tau') \theta(\tau-\tau') +  G_0^<(-i\tau, -i\tau') \theta(\tau'-\tau)\hspace{-1em}  & G_0^>(-i\tau, t') & G_0^>(-i\tau, t')\\
 			G_0^<(t, -i\tau') & \hspace{-1em}G_0^>(t, t') \theta(t-t') +  G_0^<(t, t') \theta(t'-t)\hspace{-1em} & G_0^<(t, t') \\
 			G_0^<(t, -i\tau') & G_0^>(t, t') & \hspace{-1em}G_0^>(t, t') \theta(t'-t) +  G_0^<(t, t') \theta(t-t')
 		\end{smallmatrix}\right] \nonumber \\
	 	\boldsymbol{\Omega} = \left[
	 	\begin{smallmatrix}
	 		\Omega(-i\tau,-i\tau') & \Omega(-i\tau, t') & \Omega(-i\tau, t') \\
	 		\Omega(t,-i\tau') & \Omega(t,t') & \Omega(t,t') \\
	 		\Omega(t,-i\tau') & \Omega(t,t') & \Omega(t,t') \\
	 	\end{smallmatrix}
	 	\right],\qquad
	 	\boldsymbol{H} = \left[
	 	\begin{smallmatrix}
	 		H(-i\tau,-i\tau') & H(-i\tau, t') & H(-i\tau, t') \\
	 		H(t,-i\tau') & H(t,t') & H(t,t') \\
	 		H(t,-i\tau') & H(t,t') & H(t,t') \\
	 	\end{smallmatrix}
	 	\right] \label{eq:gf_lor_sum_comp}
 	\end{gather}
	and satisfy the following equations
 	\begin{equation}
 		s\boldsymbol{K} \boldsymbol{G}_0 = \boldsymbol{1}, \qquad 
 		\boldsymbol{K} \boldsymbol{\Omega} = \boldsymbol{\eta}_0 \langle \boldsymbol{\eta}_0, \bullet \rangle, \qquad 
 		\boldsymbol{K} \boldsymbol{H} = 0,
 	\end{equation}
 	which follow from the definition (\ref{eq:gf_cons_def}). The boundary conditions (\ref{eq:pert_bc}) are satisfied by the solution (\ref{eq:sol_inhom_lor}), constructed with the use of Green's function obtained, because of our analytic continuation prescription. At the same time, the gauge condition (\ref{eq:gf_lor_gauge_cond}) needs to be checked. For this purpose, we act by this equality on some test function $\boldsymbol{\varphi}$, namely
 	\begin{align}
 		\langle \boldsymbol{\eta}_0, s \boldsymbol{G} \boldsymbol{\varphi} \rangle = {} &\int_0^\beta d\tau' \biggl[ \int_0^\beta d\tau \, \eta_{0e}(\tau) G_{ee}(\tau,\tau') + i\int_0^T dt \, \eta_0(t) \bigl(G_{e+}(t, \tau') - G_{e-}(t,\tau')\bigr) \biggr] \varphi_e(\tau') \nonumber \\
 		+{}&\int_0^T dt' \biggl[ \int_0^\beta d\tau \, \eta_{0e}(\tau) G_{e+}(\tau,t') + i\int_0^T dt \, \eta_0(t) \bigl(G_{++}(t, \tau') - G_{-+}(t,\tau')\bigr) \biggr] \varphi_+(\tau') \nonumber \\
 		+{}&\int_0^T dt' \biggl[ \int_0^\beta d\tau \, \eta_{0e}(\tau) G_{e-}(\tau,t') + i\int_0^T dt \, \eta_0(t) \bigl(G_{+-}(t, \tau') - G_{--}(t,\tau')\bigr) \biggr] \varphi_-(\tau'),
 	\end{align}
 	and show that coefficients near its components $\varphi_e$,~$\varphi_+$, and $\varphi_-$ vanish. First term in the first line equals to zero because of gauge condition (\ref{eq:gauge_cond}), while the second term vanishes identically, since $G_{e+}=G_{e-}$. The first term in the second line equals to $\omega(t)$, whereas the second one reads
 	\begin{multline}
 		i\int_0^T dt \, \eta_0(t) \bigl(G_{++}(t, \tau') - G_{-+}(t,\tau')\bigr) = i\int_0^{t'} dt \, \eta_0(t) \bigl(G_0^<(t,t') - G_0^>(t,t')\bigr) \\ = 
 		-i \zeta_0(t')\int_0^{t'} dt \, (\eta_0(t))^2 + i \eta_0(t')\int_0^{t'} dt \, \eta_0(t) \zeta_0(t).
 	\end{multline}
 	Comparing with the definition (\ref{eq:omega_lor}), one finds that it is nothing but $-\omega(t)$, that exactly cancel the first term in the first line. The same considerations show that the third line also vanishes. Thus, Green's function $\boldsymbol{G}$ indeed satisfies the gauge condition (\ref{eq:gf_lor_gauge_cond}).
 	
 	Finally, let us write down the quadratic action (\ref{eq:action_quad_full}), with the solution (\ref{eq:sol_inhom_lor}) substituted
 	\begin{equation}
 		\boldsymbol{S}_\xi^{\scriptscriptstyle (2)}[\bar{x};\boldsymbol{\eta}] = \frac12 \langle s\boldsymbol{j}, \boldsymbol{G}_\xi s \boldsymbol{j}  \rangle,
 	\end{equation}
 	where all the boundary terms vanish, since (\ref{eq:sol_inhom_lor}) obeys the boundary conditions (\ref{eq:pert_bc}).
 	It also worth noting that though our construction of the Green's function uses an analytic continuation procedure as an intermediate step, it is done explicitly anywhere, so that the final expression does not imply any analytic continuation. The only requirement is that zero mode is known explicitly at whole complex time plane and has the above mentioned analytic properties.
 	
\section{Keldysh rotation} \label{sec:app_keldysh_rot}

	In Section~\ref{sec:diagrams} we have constructed the diagram technique for real-time correlation functions for instantonic systems. This technique is three-component, having nine kinds of internal lines expressed through components of Green's function, three types of vertices for each term in the power expansion of the interaction action, and three additional 1-point vertices, coming from the functional integration measure. Such a variety of the diagrammatic elements makes the underlying diagram technique very cumbersome. At the same time, not all of the nine components of the Green's function are independent. Transposition symmetry immediately rules out three dependent components. The causal structure of the Green's function predicts that two more linear combinations of its components vanish. To get rid of redundant components, one should perform a linear transformation in the space of fields, called Keldysh rotation \cite{Keldysh:1964ud,Arseev:2015}, after which one lefts with four independent Green's function components, corresponding to the same number of internal line types. As we will see, this transformation simultaneously enlarge the number of vertex types.
	
	Let us perform the following change of variables, referred as the Keldysh rotation, in the space of fields $x_e$,~$x_+$, and $x_-$
	\begin{equation}
		x_c(t) = \frac12 \bigl(x_+(t) + x_-(t)\bigr), \qquad x_q(t) = x_+(t) - x_-(t), \label{eq:keldysh_rot}
	\end{equation}
	so that the new independent set of fields consists of $x_e$,~$x_c$, and $x_q$. Here $x_c$ is called the ``classical field'', whereas $x_q$ bears the name the ``quantum field''. Next, we rewrite the generating functional (\ref{eq:gen_fun_lor_wick}) in terms of new variables, and derive the corresponding diagrammatic technique.
	
	Substitution of the original fields perturbations (\ref{eq:pert_def_lor}) into the change of variables gives
	\begin{equation}
		x_c(t) = \bar{x}^{\tau_0}(t) + \eta_c(t), \qquad x_q(t) = \eta_q(t),
	\end{equation}
	where $\eta_c$,~$\eta_q$ are expressed through the perturbations of the original fields in exactly the same way (\ref{eq:keldysh_rot}) as $x_c$,~$x_q$
	\begin{equation}
		\eta_c(t) = \frac12 \bigl(\eta_+(t) + \eta_-(t)\bigr), \qquad \eta_q = \eta_+(t) - \eta_-(t). \label{eq:pert_lor_kr}
	\end{equation}
	Thus, together with the perturbation $\eta_e$ of the Euclidean field, it constitute the new independent set of perturbations $\tilde{\boldsymbol{\eta}}$
	\begin{equation}
		\tilde{\boldsymbol{\eta}} = \begin{bmatrix} \eta_e(\tau) \\ \eta_c(t) \\ \eta_q(t) \end{bmatrix}, \label{eq:pert_keldysh_rot}
	\end{equation}
	which connected to the initial set $\boldsymbol{\eta} = [\eta_e(\tau), \eta_+(t), \eta_-(t)]^T$ through the equality
	\begin{equation}	
		\boldsymbol{\eta} = v \tilde{\boldsymbol{\eta}}, \qquad 
		v = \begin{bmatrix*}[r]
			1 \hspace{0.8em}& 0 & 0 \\
			0 \hspace{0.8em}& 1 & \frac12 \\
			0 \hspace{0.8em}& 1 & -\frac12
		\end{bmatrix*}.
	\end{equation}
	Now, let us rewrite the variational operator in (\ref{eq:gen_fun_lor_wick}) in terms of $\tilde{\boldsymbol{\eta}}$. We obtain
	\begin{equation}
		\Bigl\langle\frac{\delta}{\delta \boldsymbol{\eta}}, \boldsymbol{G}_\xi^{\tau_0}\frac{\delta}{\delta \boldsymbol{\eta}} \Bigr\rangle = \Bigl\langle\frac{\delta}{\delta \tilde{\boldsymbol{\eta}}}, \tilde{\boldsymbol{G}}_\xi^{\tau_0}\frac{\delta}{\delta \tilde{\boldsymbol{\eta}}} \Bigr\rangle, \qquad \tilde{\boldsymbol{G}}_\xi^{\tau_0} = v^{-1} \,\boldsymbol{G}_\xi^{\tau_0} \, (v^{-1})^T.
	\end{equation}
	Using the splitting (\ref{eq:green_lor_split}) of $\boldsymbol{G}_\xi^{\tau_0}$ into two terms
	\begin{equation}
		\tilde{\boldsymbol{G}}_\xi^{\tau_0} = \tilde{\boldsymbol{G}}^{\tau_0} + \xi  \tilde{\boldsymbol{\eta}}_0^{\tau_0}  \langle  \tilde{\boldsymbol{\eta}}_0^{\tau_0}, \bullet \rangle, \qquad \tilde{\boldsymbol{G}}^{\tau_0} = v^{-1} \,\boldsymbol{G}^{\tau_0} \, (v^{-1})^T, \quad \tilde{\boldsymbol{\eta}}_0^{\tau_0} = v^{-1} \boldsymbol{\eta}_0^{\tau_0}\label{eq:green_lor_split_kr}
	\end{equation}
	we obtain the following answer for the first term
	\begin{equation}
		\tilde{\boldsymbol{G}}^{\tau_0} = 
		\begin{bmatrix}
			G^{\tau_0}_{ee}(\tau,\tau') & G^{\tau_0}_{ec}(\tau, t') & 0 \\
			G^{\tau_0}_{ce}(t,\tau') & G^{\tau_0}_{cc}(t, t') & G^{\tau_0}_{cq}(t, t') \\
			0 & G^{\tau_0}_{qc}(t, t') & 0 \\
		\end{bmatrix}.
	\end{equation}
	Two off-diagonal terms equal to $G^{\tau_0}_{e+}-G^{\tau_0}_{e-}=0$ and $G^{\tau_0}_{+e}-G^{\tau_0}_{-e}=0$, whereas the diagonal component is proportional to $G_{++}^{\tau_0}-G_{+-}^{\tau_0}-G_{-+}^{\tau_0}+G_{--}^{\tau_0}=0$. Vanishing of these combinations can be verified by either the explicit expressions (\ref{eq:gf_lor_sum}),~(\ref{eq:gf_lor_sum_comp}), or the causal structure of two-point correlation functions, which these Green's function components correspond to. The explicit expressions for the non-vanishing components read
	\begin{subequations} \label{eq:greens_fun_expl_kr}
	\begin{align}
		G_{ec}(\tau, t') &= \frac12 \bigl(G_{e+}(\tau, t') + G_{e-}(\tau, t') \bigr) = G_0^>(-i\tau, t') + H(-i\tau, t') + \Omega(-i\tau, t'), \\
		G_{ce}(t, \tau') &= \frac12 \bigl(G_{+e}(t, \tau') + G_{-e}(t, \tau') \bigr) = G_0^<(t, -i\tau') + H(t, -i\tau') + \Omega(t, -i\tau'), \\
		G_{cq}(t,t') &= \frac12 \bigl(G_{++}(t,t') - G_{+-}(t,t') + G_{-+}(t,t') - G_{--}(t,t')\bigr) = 2G_0^>(t,t') \theta(t-t'), \!\!\\
		G_{qc}(t,t') &= \frac12 \bigl(G_{++}(t,t') + G_{+-}(t,t') - G_{-+}(t,t') - G_{--}(t,t')\bigr) = 2G_0^<(t,t') \theta(t'-t), \!\!\!\\
		G_{cc}(t,t') &= \frac12 \bigl(G_{++}(t,t') + G_{+-}(t,t') + G_{-+}(t,t') + G_{--}(t,t')\bigr) = H(t,t') + \Omega(t,t'),
	\end{align}
	\end{subequations}
	where we omit the superscripts $\tau_0$ for the readability reasons, and $G_{ee}$ remains untransformed under the Keldysh rotation.
	The second term in (\ref{eq:green_lor_split_kr}) does not spoil the matrix structure of $\tilde{\boldsymbol{G}}^{\tau_0}$, since
	\begin{equation}
		\tilde{\boldsymbol{\eta}}_0^{\tau_0} = v^{-1} \boldsymbol{\eta}_0^{\tau_0} = \begin{bmatrix}
			\eta_{0e}^{\tau_0}(\tau) \\ \eta_{0}^{\tau_0}(t) \\ 0
		\end{bmatrix},
	\end{equation}
	so $\tilde{\boldsymbol{G}}^{\tau_0}_\xi$ has the same vanishing components as $\tilde{\boldsymbol{G}}^{\tau_0}$. 
	
	Next, let us rewrite the sum of the Lorentzian interaction actions (\ref{eq:action_int_lor}) in terms of the new variables. Using the Newton's binomial formula and the relation (\ref{eq:pert_lor_kr}) between the original and Keldysh-rotated variables, one obtains
	\begin{multline}
		\tilde{S}^{\text{int}}[\bar{x}^{\tau_0},\eta_c,\eta_q]= S^{\text{int}}[\bar{x}^{\tau_0},\eta_+] - S^{\text{int}}[\bar{x}^{\tau_0},\eta_-] = -\sum_{k\ge3} \frac1{k!} \int_0^T dt \, V^{(k)}(\bar{x}^{\tau_0}(t)) \, \bigl(\eta_+^k(t) - \eta_-^k(t)) \\ =
		-\sum_{k\ge3} \sum_{r=0}^{\lfloor\frac{k-1}2 \rfloor} \frac1{2^{2r} (2r+1)!(k-2r-1)!} \int_0^T dt \, V^{(k)}(\bar{x}^{\tau_0}(t)) \, \eta_q^{2r+1}(t) \eta_c^{k-2r-1}(t), \label{eq:action_int_kr}
	\end{multline}
	where $\lfloor\bullet\rfloor$ denotes rounding down. Thus, we have $\lfloor\frac{k-1}2\rfloor+1$ different $k$-point Lorentzian interaction vertices for each non-vanishing term in the sum over $k$, so the number of vertices increases after Keldysh rotation for $k>4$. Note that each term in the sum has odd (so, at least one) power of the quantum field.
	
	We just have to express the terms in the generating functional (\ref{eq:gen_fun_lor_wick}), containing the inner products, namely $\bigl\langle \partial_{\tau_0} \boldsymbol{\eta}_0^{\tau_0}, s\boldsymbol{\eta} \bigr\rangle$ and $\langle \boldsymbol{j}, s \boldsymbol{\eta}\rangle$, in terms of the Keldysh-rotated variables.
	Since the Keldysh rotation does not preserve a bilinear form, defined as $\boldsymbol{\psi},\boldsymbol{\varphi} \mapsto \langle \boldsymbol{\psi}, s \boldsymbol{\varphi} \rangle$, there are two different Keldysh-rotated bases, which we denote by tilde and check symbols, respectively
	\begin{equation}
		\tilde{\boldsymbol{\psi}} = v^{-1} \boldsymbol{\psi}, \qquad \check{\boldsymbol{\psi}} = \tilde s^{-1} v^T s \, \boldsymbol{\psi},
	\end{equation}
	where $\tilde s \tilde{\boldsymbol{\psi}} = \tilde s \, [\psi_e, \psi_c, \psi_q]^T = [\psi_e, i\psi_c, i\psi_q]^T$, so that $\tilde{\boldsymbol{\psi}}$ and $\check{\boldsymbol{\psi}}$ differ by the order of the classical and quantum components
	\begin{equation}
		\tilde{\boldsymbol{\psi}} = \begin{bmatrix}
			\psi_e(\tau) \\ \psi_c(t) \\ \psi_q(t)
		\end{bmatrix}, \qquad
		\check{\boldsymbol{\psi}} = \tilde{s}^{-1} v^T s \,v \, \tilde{\boldsymbol{\psi}} = \begin{bmatrix}
			\psi_e(\tau) \\ \psi_q(t) \\ \psi_c(t)
		\end{bmatrix},
	\end{equation}
	so that the bilinear form takes the following equivalent forms
	\begin{equation*}
		\langle \boldsymbol{\psi}, s \boldsymbol{\varphi} \rangle = \langle \check{\boldsymbol{\psi}}, \tilde s \tilde{\boldsymbol{\varphi}} \rangle = \langle \tilde{\boldsymbol{\psi}}, \tilde s \check{\boldsymbol{\varphi}} \rangle = \int_0^\beta d\tau \, \psi_e(\tau) \varphi_e(\tau) + i \int_0^T dt \, \bigl[ \psi_c(t) \varphi_q(t) + \psi_q(t) \varphi_c(t) \bigr].
	\end{equation*}
	Thus, the second term in the square brackets in (\ref{eq:gen_fun_lor_wick}) can be rewritten in terms of $\tilde{\boldsymbol{\eta}}$ as
	\begin{equation}
		\bigl\langle \partial_{\tau_0} \boldsymbol{\eta}_0^{\tau_0}, s\boldsymbol{\eta} \bigr\rangle = \bigl\langle \partial_{\tau_0} \check{\boldsymbol{\eta}}_0^{\tau_0}, \tilde s\tilde{\boldsymbol{\eta}} \bigr\rangle, \qquad \check{\boldsymbol{\eta}}_0^{\tau_0} = \tilde s^{-1} v^T s \, \boldsymbol{\eta}_0^{\tau_0} = \begin{bmatrix}
			\eta_{0e}^{\tau_0}(\tau) \\
			0 \\ \eta_{0}^{\tau_0}(t),
		\end{bmatrix}
	\end{equation}
	where $\tilde s\tilde{\boldsymbol{\eta}} = [\eta_e, i \eta_c, i\eta_q]^T$, so that we have no 1-point vertex, including the classical field. Similarly, we can express the source term in terms the Keldysh basis as
	\begin{equation}
		 \langle \boldsymbol{j}, s \boldsymbol{\eta}\rangle = \langle \check{\boldsymbol{j}}, \tilde s \tilde{\boldsymbol{\eta}} \rangle = \int_0^\beta d\tau \, j_e(\tau) \eta_e(\tau) + i \int_0^T dt \, \bigl[ j_c(t) \eta_q(t) + j_q(t) \eta_c(t) \bigr], \label{eq:src_term_kr}
	\end{equation}
	where the explicit expression for  $\check{\boldsymbol{j}}$ reads
	\begin{equation}
		\check{\boldsymbol{j}} = \tilde s^{-1} v^T s \, \boldsymbol{j} = 
		\begin{bmatrix}
			j_e(\tau) \\ j_+(t) - j_-(t) \\ \frac12 (j_+(t) + j_-(t))
		\end{bmatrix} = \begin{bmatrix}
		j_e(\tau) \\ j_q(t) \\ j_c(t)
	\end{bmatrix}.
	\end{equation}
	Thus, from (\ref{eq:src_term_kr}) we see that $j_c$ is the source for the field $\eta_q$, whereas $j_q$ is the source for $\eta_c$, that may look rather non-intuitive for the first time.
	
	Combining all the results together, we rewrite the generating functional (\ref{eq:gen_fun_lor_wick}) in terms of the Keldysh rotated fields as
	\begin{align}
		\tilde Z[j_c&, j_q] = Z^{\text{1-loop}} \frac1\beta \int_0^\beta d\tau_0 \; e^{i \int_0^T dt\, j_q(t) \bar{x}^{\tau_0}(t)} \; \exp\biggl\{\frac12 \Bigl\langle\frac{\delta}{\delta \tilde{\boldsymbol{\eta}}}, \tilde{\boldsymbol{G}}_\xi^{\tau_0}\frac{\delta}{\delta \tilde{\boldsymbol{\eta}}} \Bigr\rangle\biggr\} \;
		\\ 
		&\times  \Bigl[1 - \frac1{\|\dot{\bar{x}}\|} \bigl\langle \partial_{\tau_0} \check{\boldsymbol{\eta}}_0^{\tau_0}, \tilde s\tilde{\boldsymbol{\eta}} \bigr\rangle \Bigr] \; \exp\biggl\{
		- S^{\text{int}}_e[\bar{x}_e^{\tau_0},\eta_e] + i \tilde{S}^{\text{int}}[\bar{x}^{\tau_0},\eta_c,\eta_q]
		+ \langle \tilde{\boldsymbol{\eta}}, \tilde s \check{\boldsymbol{j}} \rangle \biggr\} \biggr|_{\substack{j_e=0\\\,\tilde{\boldsymbol{\eta}}=0}}, \nonumber
	\end{align}
	and define the corresponding correlation functions as its variations
	\begin{align}
		\tilde D_{nm}(t_{1}^c, \ldots t_{n}^c;t_{1}^q, \ldots t_{m}^q) = (-i)^{m+n}
		\frac{1}{\tilde Z} \frac{\delta^{n+m} \tilde Z[j_c, j_q]}{\delta j_q(t_1^c) \ldots \delta j_q(t_n^c) \, \delta j_c(t_1^q) \ldots \delta j_c(t_m^q)}\biggl|_{j_q=j_c=0}, \label{eq:corr_fun_lor_kr}
	\end{align}
	that can be expressed as a linear combination of the original correlation functions (\ref{eq:corr_fun_lor}).
	Note that in the Keldysh-rotated basis we have the background contribution only to the classical fields, i.e. when varying the source $j_q$.
	Now, we can split the generating functional $\tilde Z$ into background and perturbative parts
	\begin{align}
		&\tilde Z[j_c, j_q] = Z^{\text{1-loop}} \frac1\beta \int_0^\beta d\tau_0 \; e^{i \int_0^T dt\, j_q(t) \bar{x}^{\tau_0}(t)} \; \tilde{Z}_{\text{pert}}^{\tau_0}[j_c,j_q], \\
		&\tilde{Z}_{\text{pert}}^{\tau_0}[j_c,j_q] = 	
		\exp\biggl\{\frac12 \Bigl\langle\frac{\delta}{\delta \tilde{\boldsymbol{\eta}}}, \tilde{\boldsymbol{G}}_\xi^{\tau_0}\frac{\delta}{\delta \tilde{\boldsymbol{\eta}}} \Bigr\rangle\biggr\} \label{eq:gen_fun_lor_pert_kr} \\ & \qquad\quad\qquad \times \Bigl[1 - \frac1{\|\dot{\bar{x}}\|} \bigl\langle \partial_{\tau_0} \check{\boldsymbol{\eta}}_0^{\tau_0}, \tilde s\tilde{\boldsymbol{\eta}} \bigr\rangle \Bigr] \exp\biggl\{
		- S^{\text{int}}_e[\bar{x}_e^{\tau_0},\eta_e] + i \tilde{S}^{\text{int}}[\bar{x}^{\tau_0},\eta_c,\eta_q]
		+ \langle \tilde{\boldsymbol{\eta}}, \tilde s \check{\boldsymbol{j}} \rangle \biggr\} \biggr|_{\substack{j_e=0\\\,\tilde{\boldsymbol{\eta}}=0}}. \nonumber
	\end{align}
	The perturbative part of the generating functional $\tilde{Z}_{\text{pert}}^{\tau_0}$ allows us to define the correlation functions of the perturbations in the Keldysh-rotated basis as
	\begin{align}
		\langle \eta_q(t_1^q) \ldots \eta_q(t_m^q) \eta_c(t_1^c) \ldots \eta_c(t_n^c) \rangle^{\tau_0} = \frac{(-i)^{m+n}}{\tilde{Z}_{\text{pert}}^{\tau_0}} \, \frac{\delta^{n+m} \tilde{Z}_{\text{pert}}^{\tau_0}[j_c,j_q]}{\delta j_q(t_1^c) \ldots \delta j_q(t_n^c) \, \delta j_c(t_1^q) \ldots \delta j_c(t_m^q)}\biggl|_{j_q=j_c=0}. \label{eq:corr_fun_lor_kr_pert}
	\end{align}
	Now, one can express the Keldysh rotated correlation functions (\ref{eq:corr_fun_lor_kr}) in terms of the background solution ${\bar x}^{\tau_0}$ and the correlation functions of the perturbations. For 1- and 2-point functions we obtain the following expressions
	\begin{subequations}
		\begin{align}
			&\tilde{D}_{1,0}(t_1^c) = \frac1\beta \int_0^\beta d\tau_0 \Bigl[ \bar{x}^{\tau_0}(t_1^c) + \langle \eta_c(t_1^c) \rangle^{\tau_0} \Bigr], \\
			&\tilde{D}_{0,1}(t_1^q) = \frac1\beta \int_0^\beta d\tau_0 \, \langle \eta_q(t_1^q) \rangle^{\tau_0}, \\
			&\tilde{D}_{1,1}(t_1^c;t_1^q) = \frac1\beta \int_0^\beta  d\tau_0 \Bigl[\bar{x}^{\tau_0}(t_1^c) \langle \eta_q(t_1^q) \rangle^{\tau_0} + \langle \eta_c(t_1^c) \eta_q(t_1^q) \rangle^{\tau_0}  \Bigr], \\
			&\begin{aligned}\tilde{D}_{2,0}(t_1^c,t_2^c) = \frac1\beta \int_0^\beta & d\tau_0 \Bigl[ \bar{x}^{\tau_0}(t_1^c) \bar{x}^{\tau_0}(t_2^c) \\ &+ \bar{x}^{\tau_0}(t_1^c) \langle \eta_c(t_2^c) \rangle^{\tau_0} + \bar{x}^{\tau_0}(t_2^c) \langle \eta_c(t_1^c) \rangle^{\tau_0} + \langle \eta_c(t_1^c) \eta_c(t_2^c) \rangle^{\tau_0}  \Bigr],
			\end{aligned}\\
			&\tilde{D}_{0,2}(t_1^q,t_2^q) = \frac1\beta \int_0^\beta d\tau_0 \, \langle \eta_q(t_1^q) \eta_q(t_2^q) \rangle^{\tau_0}.
		\end{align}
	\end{subequations}
	One can see from the operator formalism, that $\tilde D_{0,n}$, and $\tilde{D}_{0,1}$,~$\tilde{D}_{0,2}$ in particular, must vanish.

	In order to calculate the correlation functions of the perturbations in the Keldysh-rotated basis (\ref{eq:corr_fun_lor_kr_pert}), we will construct a diagrammatic technique, which follows from the generating functional (\ref{eq:gen_fun_lor_pert_kr}). Just like without the Keldysh rotation, we use the definition (\ref{eq:corr_fun_lor_kr_pert}) to write correlation functions of the perturbations in the following explicit form
	\begin{multline}
		\langle \eta_q(t_1^q) \ldots \eta_q(t_m^q) \eta_c(t_1^c) \ldots \eta_c(t_n^c) \rangle^{\tau_0} = 	
		\exp\biggl\{\frac12 \Bigl\langle\frac{\delta}{\delta \tilde{\boldsymbol{\eta}}}, \tilde{\boldsymbol{G}}_\xi^{\tau_0}\frac{\delta}{\delta \tilde{\boldsymbol{\eta}}} \Bigr\rangle\biggr\} \\  \times \eta_q(t_1^q) \ldots \eta_q(t_m^q) \eta_c(t_1^c) \ldots \eta_c(t_n^c) \Bigl[1 - \frac1{\|\dot{\bar{x}}\|} \bigl\langle \partial_{\tau_0} \check{\boldsymbol{\eta}}_0^{\tau_0}, \tilde s\tilde{\boldsymbol{\eta}} \bigr\rangle \Bigr] \\ \times \exp\biggl\{
		- S^{\text{int}}_e[\bar{x}_e^{\tau_0},\eta_e] + i \tilde{S}^{\text{int}}[\bar{x}^{\tau_0},\eta_c,\eta_q]  \biggr\} \biggr|_{\substack{\tilde{\boldsymbol{\eta}}=0\hspace{2.5em}\\\text{non-vacuum}}}. \label{eq:corr_fun_lor_kr_pert_expl}
	\end{multline}
	where the denominator cancels vacuum diagrams, so we exclude them on r.h.s. As we have seen, we left with 6 different types of contractions, coming from the variational term above, whereas only 4 of them are truly independent. We will denote diagrammatic elements, corresponding to $\eta_e$ by dotted lines as before, whereas the elements, corresponding to $\eta_c$ and $\eta_q$ will be denoted by solid and dashed lines, respectively. Thus, we have the following types internal lines	
	\begin{equation}
		\begin{aligned}
			\vcenter{\hbox{\includegraphics[scale=1.25]{propagator_ee_lor.pdf}}}
			&= \; G_{\xi ee}^{\tau_0}(\tau, \tau'), & & & \vcenter{\hbox{\includegraphics[scale=1.25]{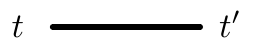}}}
			&= \; G_{\xi cc}^{\tau_0}(t, t') \\
			\vcenter{\hbox{\includegraphics[scale=1.25]{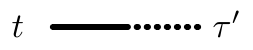}}}
			&= \; G_{\xi ce}^{\tau_0}(t, \tau'), & & & \vcenter{\hbox{\includegraphics[scale=1.25]{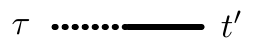}}}
			&= \; G_{\xi ec}^{\tau_0}(\tau, t'),\\
			\vcenter{\hbox{\includegraphics[scale=1.25]{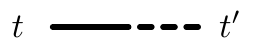}}}
			&= \; G_{\xi cq}^{\tau_0}(t, t'), & & & \vcenter{\hbox{\includegraphics[scale=1.25]{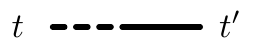}}}
			&= \; G_{\xi qc}^{\tau_0}(t, t'). 
		\end{aligned}
	\end{equation}
	Similarly, the external points for the perturbations of the classical and quantum fields are represented as
	\begin{align}
		\vcenter{\hbox{\includegraphics[scale=1.25]{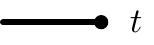}}} &= 1, &
		\vcenter{\hbox{\includegraphics[scale=1.25]{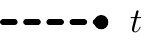}}} &= 1.
	\end{align}
	The term in the square brackets in the second line of (\ref{eq:corr_fun_lor_kr_pert_expl}) gives rise to the following one-point vertices for perturbations of the Euclidean and quantum fields
	\begin{align}
		\vcenter{\hbox{\includegraphics[scale=1.25]{stub_e_lor.pdf}}}  &=  -\frac1{\|\dot{\bar{x}}\|} \partial_{\tau_0} \eta_{0e}^{\tau_0}(\tau), & 
		\vcenter{\hbox{\includegraphics[scale=1.25]{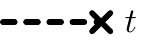}}}  &=  -\frac{i}{\|\dot{\bar{x}}\|} \partial_{\tau_0} \eta_{0}^{\tau_0}(t).
	\end{align}
	Finally, the interaction terms in the third line of (\ref{eq:corr_fun_lor_kr_pert_expl}) lead to the various types of vertices. Expansion of the Euclidean interaction action gives $k$-point vertices, corresponding to the self-coupling of the Euclidean field perturbations, as before
	\begin{equation}
			\vcenter{\hbox{\includegraphics[scale=1]{vertex_e_lor.pdf}}} \; = -
		\frac1{k!} \, V^{(k)}(\bar{x}_e^{\tau_0}(\tau)).
	\end{equation}
	At the same time, the expression for the interaction actions (\ref{eq:action_int_kr}) of the Lorentzian field perturbations leads to vertices that couple $2r+1$ quantum and $k-2r-1$ classical field perturbations for each appropriate $k$ and $r$
	\begin{equation}
		\vcenter{\hbox{\includegraphics[scale=1]{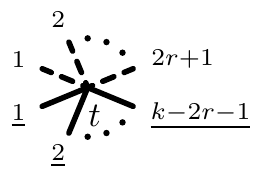}}} =  -\frac{i}{2^{2r} (2r+1)!(k-2r-1)!}  V^{(k)}(\bar{x}^{\tau_0}(t)).
	\end{equation}

	As an example, we provide the explicit diagrammatic expressions for particular 1- and 2-point correlation functions of perturbations. For 1-point correlation function of the classical field perturbations, we have the following first few terms of the diagrammatic expansion
	\begin{align}
		\langle \eta_c(t) \rangle^{\tau_0} = 
		\begin{minipage}[c]{2.4cm}\includegraphics[scale=1.25]{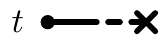}\end{minipage} &+ 
		\begin{minipage}[c]{2.4cm}\includegraphics[scale=1.25]{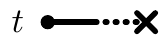}\end{minipage} \nonumber\\ +
		\begin{minipage}[c]{2.4cm}\includegraphics[scale=1.25]{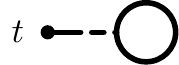}\end{minipage} &+ 
		\begin{minipage}[c]{2.4cm}\includegraphics[scale=1.25]{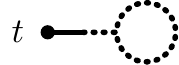}\end{minipage} + \; \text{higher loops},
	\end{align}
	where, e.g. the explicit expressions for the second and third diagrams read
	\begin{align}
		\begin{minipage}[c]{2.4cm}\includegraphics[scale=1.25]{1pt_2_kr.pdf}\end{minipage} &= -\frac1{\|\dot{\bar{x}}\|} \int_0^\beta d\tau_1 \, \partial_{\tau_0} \eta_{0e}^{\tau_0}(\tau_1) \, G^{\tau_0}_{\xi ce}(t, \tau_1), \\
		\begin{minipage}[c]{2.4cm}\includegraphics[scale=1.25]{1pt_3_kr.pdf}\end{minipage} &= -\frac{i}2 \int_0^T dt_1 \, V^{(3)}(\bar{x}^{\tau_0}(t_1)) \, G^{\tau_0}_{\xi cq}(t, t_1) G^{\tau_0}_{\xi cc}(t_1, t_1).
	\end{align}
	The diagrammatic expansion for 2-point correlation function of the classical and quantum field perturbations, is as follows
	\begin{align}
		\langle \eta_c(t) \eta_q(t') \rangle^{\tau_0} = \begin{minipage}[t][0.14cm][b]{3.376cm}\includegraphics[scale=1.25]{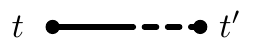} \end{minipage} &+ 
		\begin{minipage}[t][0.14cm][b]{3.376cm}\includegraphics[scale=1.25]{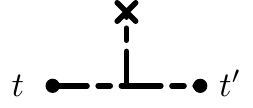}\end{minipage} + 
		\begin{minipage}[t][0.14cm][b]{3.376cm}\includegraphics[scale=1.25]{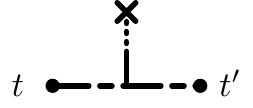} \end{minipage} \nonumber \\ &\nonumber \\+ 
		\begin{minipage}[t][0.14cm][b]{3.376cm}\includegraphics[scale=1.25]{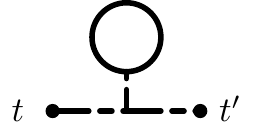} \end{minipage} &+ 
		\begin{minipage}[t][0.14cm][b]{3.376cm}\includegraphics[scale=1.25]{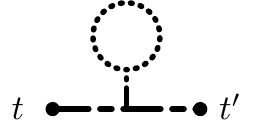} \end{minipage} \\ +
		\begin{minipage}[c][2cm][c]{3.376cm}\includegraphics[scale=1.25]{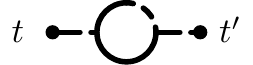}\end{minipage} &+ 
		\begin{minipage}[t][0.18cm][b]{3.376cm}\includegraphics[scale=1.25]{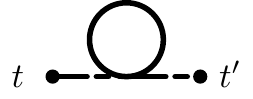}\end{minipage}+ \; \text{higher loops}. \nonumber
	\end{align}
	Let us provide the explicit expressions for some of the diagrams involved in the expansion
	\begin{align}
		&\nonumber \\
		&\begin{aligned}
			\begin{minipage}[t][0.14cm][b]{3.376cm}\includegraphics[scale=1.25]{2pt_4_kr.pdf} \end{minipage} = -\frac12 \int_0^T & dt_1 \int_0^T dt_2 \, V^{(3)}(\bar{x}^{\tau_0}(t_1)) V^{(3)}(\bar{x}^{\tau_0}(t_2)) \\ &\times G^{\tau_0}_{\xi cq}(t, t_1) G^{\tau_0}_{\xi cq}(t_1, t') G^{\tau_0}_{\xi cq}(t_1, t_2) G^{\tau_0}_{\xi cc}(t_2, t_2),
		\end{aligned}\\
		&\begin{aligned}
			\begin{minipage}[c][2cm][c]{3.376cm}\includegraphics[scale=1.25]{2pt_6_kr.pdf}\end{minipage} = -\int_0^T & dt_1 \int_0^T dt_2 \, V^{(3)}(\bar{x}^{\tau_0}(t_1)) V^{(3)}(\bar{x}^{\tau_0}(t_2)) \\ &\times G^{\tau_0}_{\xi cq}(t, t_1) G^{\tau_0}_{\xi cq}(t_1, t_2) G^{\tau_0}_{\xi cc}(t_1, t_2) G^{\tau_0}_{\xi cq}(t_2, t').
		\end{aligned}
	\end{align}
	Note that in the above expansions we exclude the diagrams, that vanish identically due to the causal structure of the Green's functions $G^{\tau_0}_{\xi cq}$,~and $G^{\tau_0}_{\xi cq}$, which can be deduced from the explicit expressions (\ref{eq:greens_fun_expl_kr}).
	
	Let us summarize the outcome of the Keldysh rotation (\ref{eq:keldysh_rot}) and oppose the corresponding diagrammatic technique to those of Section~\ref{sec:diagrams}. After the Keldysh rotation we have 4 independent internal lines, and 2 one-point vertices, in contrast to 6 internal lines and 3 one-point vertices before the rotation. Moreover, some diagrams vanish due to causal properties of Green's functions, namely $G^{\tau_0}_{\xi cq}(t, t') = G^{\tau_0}_{\xi qc}(t', t) = 0$ for $t \le t'$. At the same time, the Keldysh rotation enlarges the number of vertices, so that one have $\lfloor \frac{k-1}2\rfloor+2$ types of $k$-point vertices instead of 3 types of $k$-points vertices before the Keldysh rotation.

\bibliography{main}

%apsrev4-2.bst 2019-01-14 (MD) hand-edited version of apsrev4-1.bst
%Control: key (0)
%Control: author (8) initials jnrlst
%Control: editor formatted (1) identically to author
%Control: production of article title (0) allowed
%Control: page (0) single
%Control: year (1) truncated
%Control: production of eprint (0) enabled
\begin{thebibliography}{40}%
\makeatletter
\providecommand \@ifxundefined [1]{%
 \@ifx{#1\undefined}
}%
\providecommand \@ifnum [1]{%
 \ifnum #1\expandafter \@firstoftwo
 \else \expandafter \@secondoftwo
 \fi
}%
\providecommand \@ifx [1]{%
 \ifx #1\expandafter \@firstoftwo
 \else \expandafter \@secondoftwo
 \fi
}%
\providecommand \natexlab [1]{#1}%
\providecommand \enquote  [1]{``#1''}%
\providecommand \bibnamefont  [1]{#1}%
\providecommand \bibfnamefont [1]{#1}%
\providecommand \citenamefont [1]{#1}%
\providecommand \href@noop [0]{\@secondoftwo}%
\providecommand \href [0]{\begingroup \@sanitize@url \@href}%
\providecommand \@href[1]{\@@startlink{#1}\@@href}%
\providecommand \@@href[1]{\endgroup#1\@@endlink}%
\providecommand \@sanitize@url [0]{\catcode `\\12\catcode `\$12\catcode
  `\&12\catcode `\#12\catcode `\^12\catcode `\_12\catcode `\%12\relax}%
\providecommand \@@startlink[1]{}%
\providecommand \@@endlink[0]{}%
\providecommand \url  [0]{\begingroup\@sanitize@url \@url }%
\providecommand \@url [1]{\endgroup\@href {#1}{\urlprefix }}%
\providecommand \urlprefix  [0]{URL }%
\providecommand \Eprint [0]{\href }%
\providecommand \doibase [0]{https://doi.org/}%
\providecommand \selectlanguage [0]{\@gobble}%
\providecommand \bibinfo  [0]{\@secondoftwo}%
\providecommand \bibfield  [0]{\@secondoftwo}%
\providecommand \translation [1]{[#1]}%
\providecommand \BibitemOpen [0]{}%
\providecommand \bibitemStop [0]{}%
\providecommand \bibitemNoStop [0]{.\EOS\space}%
\providecommand \EOS [0]{\spacefactor3000\relax}%
\providecommand \BibitemShut  [1]{\csname bibitem#1\endcsname}%
\let\auto@bib@innerbib\@empty
%</preamble>
\bibitem [{\citenamefont {Keldysh}(1964)}]{Keldysh:1964ud}%
  \BibitemOpen
  \bibfield  {author} {\bibinfo {author} {\bibfnamefont {L.~V.}\ \bibnamefont
  {Keldysh}},\ }\bibfield  {title} {\bibinfo {title} {{Diagram technique for
  nonequilibrium processes}},\ }\href@noop {} {\bibfield  {journal} {\bibinfo
  {journal} {Zh. Eksp. Teor. Fiz.}\ }\textbf {\bibinfo {volume} {47}},\
  \bibinfo {pages} {1515} (\bibinfo {year} {1964})}\BibitemShut {NoStop}%
\bibitem [{\citenamefont {Schwinger}(1961)}]{schwinger1961brownian}%
  \BibitemOpen
  \bibfield  {author} {\bibinfo {author} {\bibfnamefont {J.~S.}\ \bibnamefont
  {Schwinger}},\ }\bibfield  {title} {\bibinfo {title} {{Brownian motion of a
  quantum oscillator}},\ }\href {https://doi.org/10.1063/1.1703727} {\bibfield
  {journal} {\bibinfo  {journal} {J. Math. Phys.}\ }\textbf {\bibinfo {volume}
  {2}},\ \bibinfo {pages} {407} (\bibinfo {year} {1961})}\BibitemShut {NoStop}%
\bibitem [{\citenamefont {Arseev}(2015)}]{Arseev:2015}%
  \BibitemOpen
  \bibfield  {author} {\bibinfo {author} {\bibfnamefont {P.~I.}\ \bibnamefont
  {Arseev}},\ }\bibfield  {title} {\bibinfo {title} {On the nonequilibrium
  diagram technique: derivation, some features and applications},\ }\href
  {https://doi.org/10.3367/UFNe.0185.201512b.1271} {\bibfield  {journal}
  {\bibinfo  {journal} {Phys. Usp.}\ }\textbf {\bibinfo {volume} {58}},\
  \bibinfo {pages} {1159} (\bibinfo {year} {2015})}\BibitemShut {NoStop}%
\bibitem [{\citenamefont {Matsubara}(1955)}]{matsubara1955new}%
  \BibitemOpen
  \bibfield  {author} {\bibinfo {author} {\bibfnamefont {T.}~\bibnamefont
  {Matsubara}},\ }\bibfield  {title} {\bibinfo {title} {A new approach to
  quantum-statistical mechanics},\ }\href@noop {} {\bibfield  {journal}
  {\bibinfo  {journal} {Progress of theoretical physics}\ }\textbf {\bibinfo
  {volume} {14}},\ \bibinfo {pages} {351} (\bibinfo {year} {1955})}\BibitemShut
  {NoStop}%
\bibitem [{\citenamefont {Abrikosov}\ \emph {et~al.}(2012)\citenamefont
  {Abrikosov}, \citenamefont {Gorkov},\ and\ \citenamefont
  {Dzyaloshinski}}]{abrikosov2012methods}%
  \BibitemOpen
  \bibfield  {author} {\bibinfo {author} {\bibfnamefont {A.~A.}\ \bibnamefont
  {Abrikosov}}, \bibinfo {author} {\bibfnamefont {L.~P.}\ \bibnamefont
  {Gorkov}},\ and\ \bibinfo {author} {\bibfnamefont {I.~E.}\ \bibnamefont
  {Dzyaloshinski}},\ }\href@noop {} {\emph {\bibinfo {title} {Methods of
  quantum field theory in statistical physics}}}\ (\bibinfo  {publisher}
  {Courier Corporation},\ \bibinfo {year} {2012})\BibitemShut {NoStop}%
\bibitem [{\citenamefont {Baym}\ and\ \citenamefont
  {Mermin}(1961)}]{baym1961determination}%
  \BibitemOpen
  \bibfield  {author} {\bibinfo {author} {\bibfnamefont {G.}~\bibnamefont
  {Baym}}\ and\ \bibinfo {author} {\bibfnamefont {N.~D.}\ \bibnamefont
  {Mermin}},\ }\bibfield  {title} {\bibinfo {title} {Determination of
  thermodynamic green's functions},\ }\href {https://doi.org/10.1063/1.1703704}
  {\bibfield  {journal} {\bibinfo  {journal} {Journal of Mathematical Physics}\
  }\textbf {\bibinfo {volume} {2}},\ \bibinfo {pages} {232} (\bibinfo {year}
  {1961})}\BibitemShut {NoStop}%
\bibitem [{\citenamefont {Evans}(1992)}]{Evans:1991ky}%
  \BibitemOpen
  \bibfield  {author} {\bibinfo {author} {\bibfnamefont {T.~S.}\ \bibnamefont
  {Evans}},\ }\bibfield  {title} {\bibinfo {title} {{N point finite temperature
  expectation values at real times}},\ }\href
  {https://doi.org/10.1016/0550-3213(92)90357-H} {\bibfield  {journal}
  {\bibinfo  {journal} {Nucl. Phys. B}\ }\textbf {\bibinfo {volume} {374}},\
  \bibinfo {pages} {340} (\bibinfo {year} {1992})}\BibitemShut {NoStop}%
\bibitem [{\citenamefont {Kobes}(1990)}]{Kobes:1990kr}%
  \BibitemOpen
  \bibfield  {author} {\bibinfo {author} {\bibfnamefont {R.}~\bibnamefont
  {Kobes}},\ }\bibfield  {title} {\bibinfo {title} {{A Correspondence Between
  Imaginary Time and Real Time Finite Temperature Field Theory}},\ }\href
  {https://doi.org/10.1103/PhysRevD.42.562} {\bibfield  {journal} {\bibinfo
  {journal} {Phys. Rev. D}\ }\textbf {\bibinfo {volume} {42}},\ \bibinfo
  {pages} {562} (\bibinfo {year} {1990})}\BibitemShut {NoStop}%
\bibitem [{\citenamefont {Baier}\ and\ \citenamefont
  {Niegawa}(1994)}]{Baier:1993yh}%
  \BibitemOpen
  \bibfield  {author} {\bibinfo {author} {\bibfnamefont {R.}~\bibnamefont
  {Baier}}\ and\ \bibinfo {author} {\bibfnamefont {A.}~\bibnamefont
  {Niegawa}},\ }\bibfield  {title} {\bibinfo {title} {{Analytic continuation of
  thermal N point functions from imaginary to real energies}},\ }\href
  {https://doi.org/10.1103/PhysRevD.49.4107} {\bibfield  {journal} {\bibinfo
  {journal} {Phys. Rev. D}\ }\textbf {\bibinfo {volume} {49}},\ \bibinfo
  {pages} {4107} (\bibinfo {year} {1994})},\ \Eprint
  {https://arxiv.org/abs/hep-ph/9307362} {arXiv:hep-ph/9307362} \BibitemShut
  {NoStop}%
\bibitem [{\citenamefont {Wang}\ and\ \citenamefont
  {Heinz}(2002)}]{Wang:1998wg}%
  \BibitemOpen
  \bibfield  {author} {\bibinfo {author} {\bibfnamefont {E.}~\bibnamefont
  {Wang}}\ and\ \bibinfo {author} {\bibfnamefont {U.~W.}\ \bibnamefont
  {Heinz}},\ }\bibfield  {title} {\bibinfo {title} {{A Generalized fluctuation
  dissipation theorem for nonlinear response functions}},\ }\href
  {https://doi.org/10.1103/PhysRevD.66.025008} {\bibfield  {journal} {\bibinfo
  {journal} {Phys. Rev. D}\ }\textbf {\bibinfo {volume} {66}},\ \bibinfo
  {pages} {025008} (\bibinfo {year} {2002})},\ \Eprint
  {https://arxiv.org/abs/hep-th/9809016} {arXiv:hep-th/9809016} \BibitemShut
  {NoStop}%
\bibitem [{\citenamefont {Kugler}\ \emph {et~al.}(2021)\citenamefont {Kugler},
  \citenamefont {Lee},\ and\ \citenamefont {von Delft}}]{kugler2021multipoint}%
  \BibitemOpen
  \bibfield  {author} {\bibinfo {author} {\bibfnamefont {F.~B.}\ \bibnamefont
  {Kugler}}, \bibinfo {author} {\bibfnamefont {S.-S.~B.}\ \bibnamefont {Lee}},\
  and\ \bibinfo {author} {\bibfnamefont {J.}~\bibnamefont {von Delft}},\
  }\bibfield  {title} {\bibinfo {title} {Multipoint correlation functions:
  spectral representation and numerical evaluation},\ }\href@noop {} {\bibfield
   {journal} {\bibinfo  {journal} {Physical Review X}\ }\textbf {\bibinfo
  {volume} {11}},\ \bibinfo {pages} {041006} (\bibinfo {year} {2021})},\
  \Eprint {https://arxiv.org/abs/2101.00707} {arXiv:2101.00707
  [cond-mat.str-el]} \BibitemShut {NoStop}%
\bibitem [{\citenamefont {Tripolt}\ \emph {et~al.}(2019)\citenamefont
  {Tripolt}, \citenamefont {Gubler}, \citenamefont {Ulybyshev},\ and\
  \citenamefont {Von~Smekal}}]{Tripolt:2018xeo}%
  \BibitemOpen
  \bibfield  {author} {\bibinfo {author} {\bibfnamefont {R.-A.}\ \bibnamefont
  {Tripolt}}, \bibinfo {author} {\bibfnamefont {P.}~\bibnamefont {Gubler}},
  \bibinfo {author} {\bibfnamefont {M.}~\bibnamefont {Ulybyshev}},\ and\
  \bibinfo {author} {\bibfnamefont {L.}~\bibnamefont {Von~Smekal}},\ }\bibfield
   {title} {\bibinfo {title} {{Numerical analytic continuation of Euclidean
  data}},\ }\href {https://doi.org/10.1016/j.cpc.2018.11.012} {\bibfield
  {journal} {\bibinfo  {journal} {Comput. Phys. Commun.}\ }\textbf {\bibinfo
  {volume} {237}},\ \bibinfo {pages} {129} (\bibinfo {year} {2019})},\ \Eprint
  {https://arxiv.org/abs/1801.10348} {arXiv:1801.10348 [hep-ph]} \BibitemShut
  {NoStop}%
\bibitem [{\citenamefont {Lowe}\ and\ \citenamefont
  {Stone}(1978)}]{Lowe:1978ug}%
  \BibitemOpen
  \bibfield  {author} {\bibinfo {author} {\bibfnamefont {M.}~\bibnamefont
  {Lowe}}\ and\ \bibinfo {author} {\bibfnamefont {M.}~\bibnamefont {Stone}},\
  }\bibfield  {title} {\bibinfo {title} {{A TWO LOOP CALCULATION ABOUT A
  QUANTUM MECHANICAL INSTANTON}},\ }\href
  {https://doi.org/10.1016/0550-3213(78)90021-4} {\bibfield  {journal}
  {\bibinfo  {journal} {Nucl. Phys. B}\ }\textbf {\bibinfo {volume} {136}},\
  \bibinfo {pages} {177} (\bibinfo {year} {1978})}\BibitemShut {NoStop}%
\bibitem [{\citenamefont {Escobar-Ruiz}\ \emph
  {et~al.}(2015{\natexlab{a}})\citenamefont {Escobar-Ruiz}, \citenamefont
  {Shuryak},\ and\ \citenamefont {Turbiner}}]{Escobar-Ruiz:2015nsa}%
  \BibitemOpen
  \bibfield  {author} {\bibinfo {author} {\bibfnamefont {M.~A.}\ \bibnamefont
  {Escobar-Ruiz}}, \bibinfo {author} {\bibfnamefont {E.}~\bibnamefont
  {Shuryak}},\ and\ \bibinfo {author} {\bibfnamefont {A.~V.}\ \bibnamefont
  {Turbiner}},\ }\bibfield  {title} {\bibinfo {title} {{Three-loop Correction
  to the Instanton Density. I. The Quartic Double Well Potential}},\ }\href
  {https://doi.org/10.1103/PhysRevD.92.025046} {\bibfield  {journal} {\bibinfo
  {journal} {Phys. Rev. D}\ }\textbf {\bibinfo {volume} {92}},\ \bibinfo
  {pages} {025046} (\bibinfo {year} {2015}{\natexlab{a}})},\ \Eprint
  {https://arxiv.org/abs/1501.03993} {arXiv:1501.03993 [hep-th]} \BibitemShut
  {NoStop}%
\bibitem [{\citenamefont {Escobar-Ruiz}\ \emph
  {et~al.}(2015{\natexlab{b}})\citenamefont {Escobar-Ruiz}, \citenamefont
  {Shuryak},\ and\ \citenamefont {Turbiner}}]{Escobar-Ruiz:2015rfa}%
  \BibitemOpen
  \bibfield  {author} {\bibinfo {author} {\bibfnamefont {M.~A.}\ \bibnamefont
  {Escobar-Ruiz}}, \bibinfo {author} {\bibfnamefont {E.}~\bibnamefont
  {Shuryak}},\ and\ \bibinfo {author} {\bibfnamefont {A.~V.}\ \bibnamefont
  {Turbiner}},\ }\bibfield  {title} {\bibinfo {title} {{Three-loop Correction
  to the Instanton Density. II. The Sine-Gordon potential}},\ }\href
  {https://doi.org/10.1103/PhysRevD.92.025047} {\bibfield  {journal} {\bibinfo
  {journal} {Phys. Rev. D}\ }\textbf {\bibinfo {volume} {92}},\ \bibinfo
  {pages} {025047} (\bibinfo {year} {2015}{\natexlab{b}})},\ \Eprint
  {https://arxiv.org/abs/1505.05115} {arXiv:1505.05115 [hep-th]} \BibitemShut
  {NoStop}%
\bibitem [{\citenamefont {Callan}\ \emph {et~al.}(1978)\citenamefont {Callan},
  \citenamefont {Dashen},\ and\ \citenamefont {Gross}}]{Callan:1977gz}%
  \BibitemOpen
  \bibfield  {author} {\bibinfo {author} {\bibfnamefont {C.~G.}\ \bibnamefont
  {Callan}, \bibfnamefont {Jr.}}, \bibinfo {author} {\bibfnamefont {R.~F.}\
  \bibnamefont {Dashen}},\ and\ \bibinfo {author} {\bibfnamefont {D.~J.}\
  \bibnamefont {Gross}},\ }\bibfield  {title} {\bibinfo {title} {{Toward a
  Theory of the Strong Interactions}},\ }\href
  {https://doi.org/10.1103/PhysRevD.17.2717} {\bibfield  {journal} {\bibinfo
  {journal} {Phys. Rev. D}\ }\textbf {\bibinfo {volume} {17}},\ \bibinfo
  {pages} {2717} (\bibinfo {year} {1978})}\BibitemShut {NoStop}%
\bibitem [{\citenamefont {Callan}\ \emph {et~al.}(1976)\citenamefont {Callan},
  \citenamefont {Dashen},\ and\ \citenamefont {Gross}}]{Callan:1976je}%
  \BibitemOpen
  \bibfield  {author} {\bibinfo {author} {\bibfnamefont {C.~G.}\ \bibnamefont
  {Callan}, \bibfnamefont {Jr.}}, \bibinfo {author} {\bibfnamefont {R.~F.}\
  \bibnamefont {Dashen}},\ and\ \bibinfo {author} {\bibfnamefont {D.~J.}\
  \bibnamefont {Gross}},\ }\bibfield  {title} {\bibinfo {title} {{The Structure
  of the Gauge Theory Vacuum}},\ }\href
  {https://doi.org/10.1016/0370-2693(76)90277-X} {\bibfield  {journal}
  {\bibinfo  {journal} {Phys. Lett. B}\ }\textbf {\bibinfo {volume} {63}},\
  \bibinfo {pages} {334} (\bibinfo {year} {1976})}\BibitemShut {NoStop}%
\bibitem [{\citenamefont {Coleman}(1977)}]{coleman1977fate}%
  \BibitemOpen
  \bibfield  {author} {\bibinfo {author} {\bibfnamefont {S.~R.}\ \bibnamefont
  {Coleman}},\ }\bibfield  {title} {\bibinfo {title} {{The Fate of the False
  Vacuum. 1. Semiclassical Theory}},\ }\href
  {https://doi.org/10.1103/PhysRevD.16.1248} {\bibfield  {journal} {\bibinfo
  {journal} {Phys. Rev. D}\ }\textbf {\bibinfo {volume} {15}},\ \bibinfo
  {pages} {2929} (\bibinfo {year} {1977})}\BibitemShut {NoStop}%
\bibitem [{\citenamefont {Callan}\ and\ \citenamefont
  {Coleman}(1977)}]{callan1977fate}%
  \BibitemOpen
  \bibfield  {author} {\bibinfo {author} {\bibfnamefont {C.~G.}\ \bibnamefont
  {Callan}, \bibfnamefont {Jr.}}\ and\ \bibinfo {author} {\bibfnamefont
  {S.~R.}\ \bibnamefont {Coleman}},\ }\bibfield  {title} {\bibinfo {title}
  {{The Fate of the False Vacuum. 2. First Quantum Corrections}},\ }\href
  {https://doi.org/10.1103/PhysRevD.16.1762} {\bibfield  {journal} {\bibinfo
  {journal} {Phys. Rev. D}\ }\textbf {\bibinfo {volume} {16}},\ \bibinfo
  {pages} {1762} (\bibinfo {year} {1977})}\BibitemShut {NoStop}%
\bibitem [{\citenamefont {Linde}(1983)}]{Linde:1981zj}%
  \BibitemOpen
  \bibfield  {author} {\bibinfo {author} {\bibfnamefont {A.~D.}\ \bibnamefont
  {Linde}},\ }\bibfield  {title} {\bibinfo {title} {{Decay of the False Vacuum
  at Finite Temperature}},\ }\href
  {https://doi.org/10.1016/0550-3213(83)90072-X} {\bibfield  {journal}
  {\bibinfo  {journal} {Nucl. Phys. B}\ }\textbf {\bibinfo {volume} {216}},\
  \bibinfo {pages} {421} (\bibinfo {year} {1983})}\BibitemShut {NoStop}%
\bibitem [{\citenamefont {Chudnovsky}\ and\ \citenamefont
  {Garanin}(1997)}]{chudnovsky1997first}%
  \BibitemOpen
  \bibfield  {author} {\bibinfo {author} {\bibfnamefont {E.~M.}\ \bibnamefont
  {Chudnovsky}}\ and\ \bibinfo {author} {\bibfnamefont {D.~A.}\ \bibnamefont
  {Garanin}},\ }\bibfield  {title} {\bibinfo {title} {{First- and Second-Order
  Transitions between Quantum and Classical Regimes for the Escape Rate of a
  Spin System}},\ }\href {https://doi.org/10.1103/PhysRevLett.79.4469}
  {\bibfield  {journal} {\bibinfo  {journal} {Phys. Rev. Lett.}\ }\textbf
  {\bibinfo {volume} {79}},\ \bibinfo {pages} {4469} (\bibinfo {year}
  {1997})},\ \Eprint {https://arxiv.org/abs/cond-mat/9805060}
  {arXiv:cond-mat/9805060} \BibitemShut {NoStop}%
\bibitem [{\citenamefont {Liang}\ \emph {et~al.}(1998)\citenamefont {Liang},
  \citenamefont {Muller-Kirsten}, \citenamefont {Park},\ and\ \citenamefont
  {Zimmershied}}]{liang1998periodic}%
  \BibitemOpen
  \bibfield  {author} {\bibinfo {author} {\bibfnamefont {J.~Q.}\ \bibnamefont
  {Liang}}, \bibinfo {author} {\bibfnamefont {H.~J.~W.}\ \bibnamefont
  {Muller-Kirsten}}, \bibinfo {author} {\bibfnamefont {D.~K.}\ \bibnamefont
  {Park}},\ and\ \bibinfo {author} {\bibfnamefont {F.}~\bibnamefont
  {Zimmershied}},\ }\bibfield  {title} {\bibinfo {title} {{Periodic instantons
  and quantum classical transitions in spin systems}},\ }\href
  {https://doi.org/10.1103/PhysRevLett.81.216} {\bibfield  {journal} {\bibinfo
  {journal} {Phys. Rev. Lett.}\ }\textbf {\bibinfo {volume} {81}},\ \bibinfo
  {pages} {216} (\bibinfo {year} {1998})},\ \Eprint
  {https://arxiv.org/abs/cond-mat/9805209} {arXiv:cond-mat/9805209}
  \BibitemShut {NoStop}%
\bibitem [{\citenamefont {Polyakov}(1977)}]{Polyakov:1976fu}%
  \BibitemOpen
  \bibfield  {author} {\bibinfo {author} {\bibfnamefont {A.~M.}\ \bibnamefont
  {Polyakov}},\ }\bibfield  {title} {\bibinfo {title} {{Quark Confinement and
  Topology of Gauge Groups}},\ }\href
  {https://doi.org/10.1016/0550-3213(77)90086-4} {\bibfield  {journal}
  {\bibinfo  {journal} {Nucl. Phys. B}\ }\textbf {\bibinfo {volume} {120}},\
  \bibinfo {pages} {429} (\bibinfo {year} {1977})}\BibitemShut {NoStop}%
\bibitem [{\citenamefont {Dotdaev}\ \emph {et~al.}(2021)\citenamefont
  {Dotdaev}, \citenamefont {Rodionov},\ and\ \citenamefont
  {Tikhonov}}]{titov2016korshunov}%
  \BibitemOpen
  \bibfield  {author} {\bibinfo {author} {\bibfnamefont {A.~S.}\ \bibnamefont
  {Dotdaev}}, \bibinfo {author} {\bibfnamefont {Y.~I.}\ \bibnamefont
  {Rodionov}},\ and\ \bibinfo {author} {\bibfnamefont {K.~S.}\ \bibnamefont
  {Tikhonov}},\ }\bibfield  {title} {\bibinfo {title} {{Instantons in the
  out-of-equilibrium Coulomb blockade}},\ }\href
  {https://doi.org/10.1016/j.physleta.2021.127736} {\bibfield  {journal}
  {\bibinfo  {journal} {Phys. Lett. A}\ }\textbf {\bibinfo {volume} {419}},\
  \bibinfo {pages} {127736} (\bibinfo {year} {2021})},\ \Eprint
  {https://arxiv.org/abs/2012.04390} {arXiv:2012.04390 [cond-mat.mes-hall]}
  \BibitemShut {NoStop}%
\bibitem [{\citenamefont {Konstantinov}\ and\ \citenamefont
  {Perel}(1961)}]{konstantinov1961diagram}%
  \BibitemOpen
  \bibfield  {author} {\bibinfo {author} {\bibfnamefont {O.}~\bibnamefont
  {Konstantinov}}\ and\ \bibinfo {author} {\bibfnamefont {V.}~\bibnamefont
  {Perel}},\ }\bibfield  {title} {\bibinfo {title} {A diagram technique for
  evaluating transport quantities},\ }\href@noop {} {\bibfield  {journal}
  {\bibinfo  {journal} {SOVIET PHYSICS JETP-USSR}\ }\textbf {\bibinfo {volume}
  {12}},\ \bibinfo {pages} {142} (\bibinfo {year} {1961})}\BibitemShut
  {NoStop}%
\bibitem [{\citenamefont {Kadanoff}\ and\ \citenamefont
  {Baym}(2018)}]{kadanoff2018quantum}%
  \BibitemOpen
  \bibfield  {author} {\bibinfo {author} {\bibfnamefont {L.~P.}\ \bibnamefont
  {Kadanoff}}\ and\ \bibinfo {author} {\bibfnamefont {G.}~\bibnamefont
  {Baym}},\ }\href@noop {} {\emph {\bibinfo {title} {Quantum statistical
  mechanics: Green’s function methods in equilibrium and nonequilibrium
  problems}}}\ (\bibinfo  {publisher} {CRC Press},\ \bibinfo {year}
  {2018})\BibitemShut {NoStop}%
\bibitem [{\citenamefont {Mari\~no}(2015)}]{Marino:2015yie}%
  \BibitemOpen
  \bibfield  {author} {\bibinfo {author} {\bibfnamefont {M.}~\bibnamefont
  {Mari\~no}},\ }\href@noop {} {\emph {\bibinfo {title} {{Instantons and Large
  N}: {An Introduction to Non-Perturbative Methods in Quantum Field Theory}}}}\
  (\bibinfo  {publisher} {Cambridge University Press},\ \bibinfo {year}
  {2015})\BibitemShut {NoStop}%
\bibitem [{\citenamefont {Garanin}\ and\ \citenamefont
  {Chudnovsky}(1997)}]{garanin1997thermally}%
  \BibitemOpen
  \bibfield  {author} {\bibinfo {author} {\bibfnamefont {D.}~\bibnamefont
  {Garanin}}\ and\ \bibinfo {author} {\bibfnamefont {E.}~\bibnamefont
  {Chudnovsky}},\ }\bibfield  {title} {\bibinfo {title} {Thermally activated
  resonant magnetization tunneling in molecular magnets: Mn 12 ac and others},\
  }\href@noop {} {\bibfield  {journal} {\bibinfo  {journal} {Physical Review
  B}\ }\textbf {\bibinfo {volume} {56}},\ \bibinfo {pages} {11102} (\bibinfo
  {year} {1997})},\ \Eprint {https://arxiv.org/abs/cond-mat/9805057}
  {arXiv:cond-mat/9805057 [cond-mat.stat-mech]} \BibitemShut {NoStop}%
\bibitem [{\citenamefont {Zhang}\ and\ \citenamefont
  {Muller-Kirsten}(2001)}]{Zhang:2001wa}%
  \BibitemOpen
  \bibfield  {author} {\bibinfo {author} {\bibfnamefont {Y.-b.}\ \bibnamefont
  {Zhang}}\ and\ \bibinfo {author} {\bibfnamefont {H.~J.~W.}\ \bibnamefont
  {Muller-Kirsten}},\ }\bibfield  {title} {\bibinfo {title} {{Instanton
  approach to Josephson tunneling between trapped condensates}},\ }\href
  {https://doi.org/10.1007/s100530170010} {\bibfield  {journal} {\bibinfo
  {journal} {Eur. Phys. J. D}\ }\textbf {\bibinfo {volume} {17}},\ \bibinfo
  {pages} {351} (\bibinfo {year} {2001})},\ \Eprint
  {https://arxiv.org/abs/cond-mat/0110054} {arXiv:cond-mat/0110054}
  \BibitemShut {NoStop}%
\bibitem [{\citenamefont {Affleck}(1981)}]{affleck1981quantum}%
  \BibitemOpen
  \bibfield  {author} {\bibinfo {author} {\bibfnamefont {I.}~\bibnamefont
  {Affleck}},\ }\bibfield  {title} {\bibinfo {title} {{Quantum Statistical
  Metastability}},\ }\href {https://doi.org/10.1103/PhysRevLett.46.388}
  {\bibfield  {journal} {\bibinfo  {journal} {Phys. Rev. Lett.}\ }\textbf
  {\bibinfo {volume} {46}},\ \bibinfo {pages} {388} (\bibinfo {year}
  {1981})}\BibitemShut {NoStop}%
\bibitem [{\citenamefont {Larkin}\ and\ \citenamefont
  {Ovchinnikov}(1969)}]{larkin1969quasiclassical}%
  \BibitemOpen
  \bibfield  {author} {\bibinfo {author} {\bibfnamefont {A.}~\bibnamefont
  {Larkin}}\ and\ \bibinfo {author} {\bibfnamefont {Y.~N.}\ \bibnamefont
  {Ovchinnikov}},\ }\bibfield  {title} {\bibinfo {title} {Quasiclassical method
  in the theory of superconductivity},\ }\href@noop {} {\bibfield  {journal}
  {\bibinfo  {journal} {Sov Phys JETP}\ }\textbf {\bibinfo {volume} {28}},\
  \bibinfo {pages} {1200} (\bibinfo {year} {1969})}\BibitemShut {NoStop}%
\bibitem [{\citenamefont {Haehl}\ \emph {et~al.}(2019)\citenamefont {Haehl},
  \citenamefont {Loganayagam}, \citenamefont {Narayan},\ and\ \citenamefont
  {Rangamani}}]{Haehl:2017qfl}%
  \BibitemOpen
  \bibfield  {author} {\bibinfo {author} {\bibfnamefont {F.~M.}\ \bibnamefont
  {Haehl}}, \bibinfo {author} {\bibfnamefont {R.}~\bibnamefont {Loganayagam}},
  \bibinfo {author} {\bibfnamefont {P.}~\bibnamefont {Narayan}},\ and\ \bibinfo
  {author} {\bibfnamefont {M.}~\bibnamefont {Rangamani}},\ }\bibfield  {title}
  {\bibinfo {title} {{Classification of out-of-time-order correlators}},\
  }\href {https://doi.org/10.21468/SciPostPhys.6.1.001} {\bibfield  {journal}
  {\bibinfo  {journal} {SciPost Phys.}\ }\textbf {\bibinfo {volume} {6}},\
  \bibinfo {pages} {001} (\bibinfo {year} {2019})},\ \Eprint
  {https://arxiv.org/abs/1701.02820} {arXiv:1701.02820 [hep-th]} \BibitemShut
  {NoStop}%
\bibitem [{\citenamefont {Rozenbaum}\ \emph {et~al.}(2017)\citenamefont
  {Rozenbaum}, \citenamefont {Ganeshan},\ and\ \citenamefont
  {Galitski}}]{Rozenbaum:2016mmv}%
  \BibitemOpen
  \bibfield  {author} {\bibinfo {author} {\bibfnamefont {E.~B.}\ \bibnamefont
  {Rozenbaum}}, \bibinfo {author} {\bibfnamefont {S.}~\bibnamefont
  {Ganeshan}},\ and\ \bibinfo {author} {\bibfnamefont {V.}~\bibnamefont
  {Galitski}},\ }\bibfield  {title} {\bibinfo {title} {{Lyapunov Exponent and
  Out-of-Time-Ordered Correlator\textquoteright{}s Growth Rate in a Chaotic
  System}},\ }\href {https://doi.org/10.1103/PhysRevLett.118.086801} {\bibfield
   {journal} {\bibinfo  {journal} {Phys. Rev. Lett.}\ }\textbf {\bibinfo
  {volume} {118}},\ \bibinfo {pages} {086801} (\bibinfo {year} {2017})},\
  \Eprint {https://arxiv.org/abs/1609.01707} {arXiv:1609.01707
  [cond-mat.dis-nn]} \BibitemShut {NoStop}%
\bibitem [{\citenamefont {Akutagawa}\ \emph {et~al.}(2020)\citenamefont
  {Akutagawa}, \citenamefont {Hashimoto}, \citenamefont {Sasaki},\ and\
  \citenamefont {Watanabe}}]{Akutagawa:2020qbj}%
  \BibitemOpen
  \bibfield  {author} {\bibinfo {author} {\bibfnamefont {T.}~\bibnamefont
  {Akutagawa}}, \bibinfo {author} {\bibfnamefont {K.}~\bibnamefont
  {Hashimoto}}, \bibinfo {author} {\bibfnamefont {T.}~\bibnamefont {Sasaki}},\
  and\ \bibinfo {author} {\bibfnamefont {R.}~\bibnamefont {Watanabe}},\
  }\bibfield  {title} {\bibinfo {title} {{Out-of-time-order correlator in
  coupled harmonic oscillators}},\ }\href
  {https://doi.org/10.1007/JHEP08(2020)013} {\bibfield  {journal} {\bibinfo
  {journal} {JHEP}\ }\textbf {\bibinfo {volume} {08}},\ \bibinfo {pages}
  {013}},\ \Eprint {https://arxiv.org/abs/2004.04381} {arXiv:2004.04381
  [hep-th]} \BibitemShut {NoStop}%
\bibitem [{\citenamefont {Kolganov}\ and\ \citenamefont
  {Trunin}(2022)}]{Kolganov:2022mpe}%
  \BibitemOpen
  \bibfield  {author} {\bibinfo {author} {\bibfnamefont {N.}~\bibnamefont
  {Kolganov}}\ and\ \bibinfo {author} {\bibfnamefont {D.~A.}\ \bibnamefont
  {Trunin}},\ }\bibfield  {title} {\bibinfo {title} {{Classical and quantum
  butterfly effect in nonlinear vector mechanics}},\ }\href
  {https://doi.org/10.1103/PhysRevD.106.025003} {\bibfield  {journal} {\bibinfo
   {journal} {Phys. Rev. D}\ }\textbf {\bibinfo {volume} {106}},\ \bibinfo
  {pages} {025003} (\bibinfo {year} {2022})},\ \Eprint
  {https://arxiv.org/abs/2205.05663} {arXiv:2205.05663 [hep-th]} \BibitemShut
  {NoStop}%
\bibitem [{\citenamefont {Xu}\ \emph {et~al.}(2020)\citenamefont {Xu},
  \citenamefont {Scaffidi},\ and\ \citenamefont {Cao}}]{Xu:2019lhc}%
  \BibitemOpen
  \bibfield  {author} {\bibinfo {author} {\bibfnamefont {T.}~\bibnamefont
  {Xu}}, \bibinfo {author} {\bibfnamefont {T.}~\bibnamefont {Scaffidi}},\ and\
  \bibinfo {author} {\bibfnamefont {X.}~\bibnamefont {Cao}},\ }\bibfield
  {title} {\bibinfo {title} {{Does scrambling equal chaos?}},\ }\href
  {https://doi.org/10.1103/PhysRevLett.124.140602} {\bibfield  {journal}
  {\bibinfo  {journal} {Phys. Rev. Lett.}\ }\textbf {\bibinfo {volume} {124}},\
  \bibinfo {pages} {140602} (\bibinfo {year} {2020})},\ \Eprint
  {https://arxiv.org/abs/1912.11063} {arXiv:1912.11063 [cond-mat.stat-mech]}
  \BibitemShut {NoStop}%
\bibitem [{\citenamefont {Pilatowsky-Cameo}\ \emph {et~al.}(2020)\citenamefont
  {Pilatowsky-Cameo}, \citenamefont {Ch\'avez-Carlos}, \citenamefont
  {Bastarrachea-Magnani}, \citenamefont {Str\'ansk\'y}, \citenamefont
  {Lerma-Hern\'andez}, \citenamefont {Santos},\ and\ \citenamefont
  {Hirsch}}]{pilatowsky2020positive}%
  \BibitemOpen
  \bibfield  {author} {\bibinfo {author} {\bibfnamefont {S.}~\bibnamefont
  {Pilatowsky-Cameo}}, \bibinfo {author} {\bibfnamefont {J.}~\bibnamefont
  {Ch\'avez-Carlos}}, \bibinfo {author} {\bibfnamefont {M.~A.}\ \bibnamefont
  {Bastarrachea-Magnani}}, \bibinfo {author} {\bibfnamefont {P.}~\bibnamefont
  {Str\'ansk\'y}}, \bibinfo {author} {\bibfnamefont {S.}~\bibnamefont
  {Lerma-Hern\'andez}}, \bibinfo {author} {\bibfnamefont {L.~F.}\ \bibnamefont
  {Santos}},\ and\ \bibinfo {author} {\bibfnamefont {J.~G.}\ \bibnamefont
  {Hirsch}},\ }\bibfield  {title} {\bibinfo {title} {{Positive quantum Lyapunov
  exponents in experimental systems with a regular classical limit}},\ }\href
  {https://doi.org/10.1103/PhysRevE.101.010202} {\bibfield  {journal} {\bibinfo
   {journal} {Phys. Rev. E}\ }\textbf {\bibinfo {volume} {101}},\ \bibinfo
  {pages} {010202} (\bibinfo {year} {2020})},\ \Eprint
  {https://arxiv.org/abs/1909.02578} {arXiv:1909.02578 [cond-mat.stat-mech]}
  \BibitemShut {NoStop}%
\bibitem [{\citenamefont {Kidd}\ \emph {et~al.}(2021)\citenamefont {Kidd},
  \citenamefont {Safavi-Naini},\ and\ \citenamefont {Corney}}]{Kidd:2020mtu}%
  \BibitemOpen
  \bibfield  {author} {\bibinfo {author} {\bibfnamefont {R.~A.}\ \bibnamefont
  {Kidd}}, \bibinfo {author} {\bibfnamefont {A.}~\bibnamefont {Safavi-Naini}},\
  and\ \bibinfo {author} {\bibfnamefont {J.~F.}\ \bibnamefont {Corney}},\
  }\bibfield  {title} {\bibinfo {title} {{Saddle-point scrambling without
  thermalization}},\ }\href {https://doi.org/10.1103/PhysRevA.103.033304}
  {\bibfield  {journal} {\bibinfo  {journal} {Phys. Rev. A}\ }\textbf {\bibinfo
  {volume} {103}},\ \bibinfo {pages} {033304} (\bibinfo {year} {2021})},\
  \Eprint {https://arxiv.org/abs/2010.08093} {arXiv:2010.08093 [quant-ph]}
  \BibitemShut {NoStop}%
\bibitem [{\citenamefont {Ai}\ \emph {et~al.}(2019)\citenamefont {Ai},
  \citenamefont {Garbrecht},\ and\ \citenamefont {Tamarit}}]{Ai:2019fri}%
  \BibitemOpen
  \bibfield  {author} {\bibinfo {author} {\bibfnamefont {W.-Y.}\ \bibnamefont
  {Ai}}, \bibinfo {author} {\bibfnamefont {B.}~\bibnamefont {Garbrecht}},\ and\
  \bibinfo {author} {\bibfnamefont {C.}~\bibnamefont {Tamarit}},\ }\bibfield
  {title} {\bibinfo {title} {{Functional methods for false vacuum decay in real
  time}},\ }\href {https://doi.org/10.1007/JHEP12(2019)095} {\bibfield
  {journal} {\bibinfo  {journal} {JHEP}\ }\textbf {\bibinfo {volume} {12}},\
  \bibinfo {pages} {095}},\ \Eprint {https://arxiv.org/abs/1905.04236}
  {arXiv:1905.04236 [hep-th]} \BibitemShut {NoStop}%
\bibitem [{\citenamefont {Barvinsky}\ and\ \citenamefont
  {Nesterov}(2012)}]{Barvinsky:2011hv}%
  \BibitemOpen
  \bibfield  {author} {\bibinfo {author} {\bibfnamefont {A.~O.}\ \bibnamefont
  {Barvinsky}}\ and\ \bibinfo {author} {\bibfnamefont {D.~V.}\ \bibnamefont
  {Nesterov}},\ }\bibfield  {title} {\bibinfo {title} {{Monodromies and
  functional determinants in the CFT driven quantum cosmology}},\ }\href
  {https://doi.org/10.1103/PhysRevD.85.064006} {\bibfield  {journal} {\bibinfo
  {journal} {Phys. Rev. D}\ }\textbf {\bibinfo {volume} {85}},\ \bibinfo
  {pages} {064006} (\bibinfo {year} {2012})},\ \Eprint
  {https://arxiv.org/abs/1111.4474} {arXiv:1111.4474 [hep-th]} \BibitemShut
  {NoStop}%
\end{thebibliography}%
	
\end{document}